\documentclass[12pt]{iopart}
\input{epsf}
\usepackage[dvips]{graphicx}
\usepackage{longtable}
\usepackage{bm}
\usepackage{amssymb}
\usepackage{dcolumn}
\usepackage{latexsym}
\usepackage{color}
\usepackage[colorlinks]{hyperref}

\def\be{\begin{equation}}
\def\ee{\end{equation}}
\def\ba{\begin{eqnarray}}
\def\ea{\end{eqnarray}}

\def\lsim{\mathrel{\vcenter{\hbox{$<$}\nointerlineskip\hbox{$\sim$}}}}

\begin{document}

\title[Dark Matter and Dark Energy Interactions]
{Dark Matter and Dark Energy Interactions: Theoretical
Challenges, Cosmological Implications and Observational Signatures.}

\author{
B. Wang$^{1}$\footnote{wang\_b@sjtu.edu.cn}
E. Abdalla$^{2}$\footnote{eabdalla@usp.br},
F. Atrio-Barandela$^{3}$\footnote{atrio@usal.es},
D. Pav\'on$^{4}$\footnote{diego.pavon@uab.es},
}

\address{$^{1}$Department of Physics, Shanghai Jiao Tong University, China}
\address{$^{2}$Instituto de Fisica, Universidade de S\^ao Paulo, Brazil}
\address{$^{3}$F{\'\i}sica Te\'orica, Universidad de Salamanca, Spain}
\address{$^{4}$Departamento de F{\'\i}sica, Universidad Aut\'onoma de Barcelona, Spain}

\begin{abstract}
Models where Dark Matter and Dark Energy interact with each other
have been proposed to solve the coincidence problem. We review the
motivations underlying the need to introduce such interaction, its
influence on the background dynamics and how it modifies the evolution
of linear perturbations. We test models using the most recent
observational data and we find that the interaction is compatible with the
current astronomical and cosmological data. Finally, we describe the
forthcoming data sets from current and future facilities that are being
constructed or designed that will allow a clearer understanding of the
physics of the dark sector.
\end{abstract}
\maketitle
\section{Introduction.} \label{sec:sec1}

The first observational evidence that the Universe had entered a period of accelerated
expansion was obtained when Supernovae Type Ia (SNIa) were found to be fainter
than expected \cite{perlmutter_98f,Perlmutter_1999f,Riess_1998f,Riess_1999f}.
This fact has been confirmed
by many independent observations such as temperature anisotropies of the
Cosmic Microwave Background (CMB) \cite{Spergel_03f,Planck_I_14f,Planck_I_15f},
inhomogeneities in the matter distribution \cite{tegmark_04f,cole_05f},
the integrated Sachs--Wolfe (ISW) effect \cite{Boughn_04f}, Baryon
Acoustic Oscillations (BAO) \cite{Eisenstein_2005a}, weak lensing (WL)
\cite{contaldi_03f}, and gamma-ray bursts \cite{kodama_08f}. Within the
framework of General Relativity (GR), the accelerated expansion is driven
by a new energy density component with negative pressure, termed
Dark Energy (DE). The nature of this unknown matter field
has given rise to a great scientific effort in order to understand
its properties.

The observational evidence is consistent with a cosmological constant $\Lambda$
being the possible origin of the Dark Energy (DE) driving the present epoch of the accelerated expansion of our universe and a
Dark Matter (DM) component giving rise to galaxies and their large
scale structures distributions \cite{liddle_04f,Planck_XVI_14f,Planck_I_15f}.
The DM is assumed to have negligible pressure and temperature and is
termed Cold. Thanks to the agreement with observations the model is
commonly known as $\Lambda$CDM, to indicate the nature of its main components.
While favored by the observations, the model is not satisfactory from
the theoretical point of view: the value of the cosmological constant
is many orders of magnitude smaller than what it was estimated in
Particle Physics \cite{weinberg_89f}. It was suggested soon that DE could be
dynamic, evolving with time \cite{caldwell_98f,liddle_98f,peebles_03f}.
This new cosmological model also suffers from a severe fine-tune problem
known as {\em coincidence problem} \cite{zlatev_99f} that can be
expressed with the following simple terms: if the time variation
of matter and DE are very different why are their current values so similar?
Cosmological models where DM and DE do not evolve separately but
interact with each other were first introduced to justify the currently
small value of the cosmological constant \cite{wetterich_88f,wetterich_95f}
but they were found to be very useful to alleviate the coincidence problem.
In this review we will summarize the theoretical developments of this field
and the observational evidence that constrains  the  DM/DE interaction and
could, eventually, lead to its detection.

The emergence of galaxies and Large Scale Structure is driven by the
growth of matter density perturbations which themselves are connected
to the anisotropies of the CMB \cite{peebles_80f}. An interaction between
the components of the dark sector will affect the overall evolution of the
Universe and its expansion history, the growth matter and baryon density
perturbations, the pattern of temperature anisotropies of the CMB and the
evolution of the gravitational potential at late times would be different
than in the concordance model. These observables are directly linked to
the underlying theory of gravity \cite{weinberg_13fa,joyce_15f} and,
consequently, the interaction could be constrained with observations of
the background evolution and the emergence of large scale structure.

This review is organized as follows: In this introduction we describe the
concordance model and we discuss
some of its shortcomings that motivates considering interactions within
the dark sector. Since the nature of DE and DM are currently unknown,
in  Sec.~\ref{sec:sec2} we introduce two possible and different approaches
to describe the DE and the DM: fluids and scalar fields.
Based on general considerations like the holographic principle, we discuss
why the interaction within the dark sector is to be expected.
In Sec.~\ref{sec:sec3} we review the influence of the interaction on the background
dynamics. We find that a DM/DE interaction could solve the coincidence
problem and satisfy the second law of thermodynamics. In Sec.~\ref{sec:sec4} the
evolution of matter density perturbations is described for the phenomenological fluid
interacting models. In Sec.~\ref{sec:sec5} we discuss how the interaction modifies the
non-linear evolution and the subsequent collapse of density perturbations.
In Sec.~\ref{sec:sec6} we describe the main observables that are used in Sec.~\ref{sec:sec7}
to constrain the interaction.  Finally, in Sec.~\ref{sec:sec8} we
describe the present and future observational facilities and their
prospects to measure or constrain the interaction. In Table~\ref{table:acronyms}
we list the acronyms commonly used in this review.

\begin{table}
\begin{indented}
\item[]
\hspace*{-1cm}\begin{tabular}{@{}llll}
\br
Acronym & Meaning & Acronym & Meaning \\
\mr
A-P     & Alcock-Packzynki & KSZ     & Kinematic Sunyaev-Zeldovich  \\
BAO     & Baryon Accoustic Oscillations & LBG     & Lyman Break Galaxies \\
CDM     & Cold Dark Matter & LHS  & Left Hand Side (of an equation)\\
CL      & Confidence Level & LISW   & Late Integrated Sachs-Wolfe \\
CMB     & Cosmic Microwave Background & LSS & Large Scale Structure \\
DE      & Dark Energy & MCMC    & Monte Carlo Markov Chain\\
DETF    & Dark Energy Task Force & RHS     & Right Hand Side (of an equation)\\
DM      & Dark Matter &  RSD     & Redshift Space Distortions\\
EoS     & Equation of State & SL      & Strong Lensing\\
EISW    & Early Integrated Sachs-Wolfe & SNIa    & Supernova Type Ia\\
FRW     & Friedman-Robertson-Walker & TSZ     & Thermal Sunyaev-Zeldovich \\
ISW     & Integrated Sachs-Wolfe & WL      & Weak Lensing\\

\br
\end{tabular}
\end{indented}
\caption{\label{table:acronyms}List of commonly used acronyms.}
\end{table}

\subsection{The Concordance Model.} \label{subsec:concordance}

The current cosmological model is described by the Friedmann-Robertson-Walker
(FRW) metric, valid for a homogeneous and isotropic Universe \cite{weinberg_72f}
\begin{equation}
ds^2=-dt^2+a^2(t)\left[\frac{dr^2}{1-Kr^2}+r^2d\theta^2+r^2\sin^2\theta d^2\phi\right],
\label{eq:frw-canonical}
\end{equation}
where $a(t)$ is the scale factor at time $t$,
the present time is denoted by $t_0$ and the scale factor is normalized to
$a(t_0)=1$; $K$ is the Gaussian curvature of the space-time. We have chosen
units $c=1$ but we will reintroduce the speed of light when needed.
A commonly used reparametrization is the {\it conformal time},
defined implicitly as $dt=a(\tau)d\tau$. In terms of this
coordinate, the line element is
\begin{equation}
ds^2=a^2(\tau)\left[-d\tau^2
    +\frac{dr^2}{1-Kr^2}+r^2d\theta^2+r^2\sin^2\theta d^2\phi\right].
\label{eq:frw-conformal}
\end{equation}
If we describe the matter content of the Universe as a perfect fluid with
mean energy density $\rho$ and pressure $p$, Friedmann's equations are
\cite{kolb_90f}
\begin{eqnarray}
\label{eq:friedmann1a}
&&H^2+\frac{K}{a^2}=\frac{8\pi G}{3}\sum \rho_i +\frac{\Lambda}{3} , \\
\label{eq:friedmann1b}
&& \frac{\ddot{a}}{a} =-\frac{4\pi G}{3}\sum(\rho_i + 3p_i)+ \frac{\Lambda}{3} ,
\end{eqnarray}
where $H=\dot{a}/a$ is the Hubble function and
$\rho_i, p_i$ are the energy density and pressure of the
different matter components, related by an equation
of state (EoS) parameter $\omega_i=p_i/\rho_i$.
In terms of the conformal time, the expression ${\cal H}=a^{-1}(da/d\tau)=aH$ is used.
Usually densities are measured in units of the critical
density: $\Omega=\rho/\rho_{cr}$ with $\rho_{cr}=3H^2/(8\pi G)$.
The curvature term can be brought to the right hand side (RHS)
by defining $\rho_K=-3K/(8\pi Ga^2)$.
As a matter of convention, a sub-index ``0'' denotes the
current value of any given quantity. Due to the historically
uncertain value of the Hubble constant, its value is usually quoted as
$H_0=100h$kms$^{-1}$Mpc$^{-1}$ so the parameter $h$ encloses the
observational uncertainty on the value of the Hubble constant.

\begin{table*}[tbp]
\centering\begin{tabular}{cc}
    \hline
$H_0$/[kms$^{-1}$Mpc$^{-1}$] & $67.74\pm 0.46$ \\
$\Omega_{b,0}h^2$ & $0.02230\pm 0.00014$ \\
$\Omega_{c,0}h^2$ & $0.1188\pm 0.0010$ \\
$\Omega_{\Lambda,0}$ & $0.6911\pm 0.0062$ \\
    \hline
$\Omega_{K,0}$	&	$0.0008^{+0.0040}_{-0.0039}$ \\
$\omega_d$ &  $-1.019^{+0.075}_{-0.080}$ \\
\hline
\end{tabular}
\caption{Cosmological parameters of the $\Lambda$CDM model, derived from
the CMB temperature fluctuations measured by Planck with the addition of external
data sets. Error bars are given at the 68\% confidence level. The data for curvature and EoS parameter are constraints on 1-parameter extensions to the base $\Lambda$CDM model for combinations of Planck power spectra, Planck lensing and external data.  The errors are at the 95\% confidence
level. Data taken from \cite{Planck_XIII_15f}.}
\label{table:parameters}
\end{table*}

The cosmological constant provides the simplest explanation of the
present period of accelerated expansion. When $\Lambda$ is
positive and dominates the RHS of eq.~(\ref{eq:friedmann1b})
then $\ddot{a}>0$ and the expansion is accelerated. The accelerated expansion can also be described by the deceleration parameter $q=-\ddot{a}/(aH^2)<0$.
If we set the cosmological constant to zero
in eqs.~(\ref{eq:friedmann1a},\ref{eq:friedmann1b}),
it can be reintroduced as a fluid with energy density
$\rho_\Lambda=\Lambda/(8\pi G)$ and an EoS parameter $\omega_\Lambda=-1$.
In addition to the cosmological constant and the curvature terms
the concordance model includes other energy density
components: Baryons (b), Cold DM (c), and Radiation (r), characterized by
the EoS parameters $\omega_b=\omega_{c}=0$ and $\omega_r=1/3$, respectively.
Then, eq.~(\ref{eq:friedmann1b}) can be
expressed as $\sum\Omega_i=1$ where the sum extends over
all energy densities, $i=(b,c,\Lambda,K,r)$.

If $\Lambda\equiv 0$ and the source of the accelerated expansion is
a fluid, known as DE (d), then such fluid would need to have a pressure negative
enough to violate the strong energy condition, i.e., $(\rho_d+3p_d)<0$.
The DE EoS parameter could be constant or it could vary with time.
Hereafter in this review, DE models will refer to models
where the dominant DE component has an EoS $\omega_d<-1/3$ while
concordance and $\Lambda$CDM will refer to the specific case when
$w_d=\omega_\Lambda=-1$.

In Table~\ref{table:parameters} we present the most recent values given
by the Planck Collaboration derived by fitting the $\Lambda$CDM model to the
measured CMB anisotropies and other external data sets \cite{Planck_I_15f}.
The quoted errors are given at the 68\% confidence level (CL).
When more general models with 1-parameter extensions to the base $\Lambda$CDM model are fit to the same data, it is
possible to derive constraints on the curvature and a
constant DE EoS parameter. In these two cases, the quoted
error bars are at the 95\% CL. The table shows that
in the $\Lambda$CDM model the energy density budget is dominated by
$\rho_\Lambda$ and $\rho_c$. Other components like
massive neutrinos or the curvature $\rho_K$ are not dynamically
important and will not be considered in this review.

\subsection{Observational magnitudes.}\label{sec:distances}

The first evidence of the present accelerated expansion
came when comparing the measured brightness of SNIa at redshifts
$z\ge 0.4$ to their flux expected in different cosmological
models \cite{perlmutter_98f,Riess_1998f}.
The method relies on measuring distances using {\it standard
candles}, sources with well known intrinsic properties.
In Cosmology, distances are measured very differently than in the
Minkowski space-time, they are parametrized in terms of
the time travelled by the radiation from the source to the observer
by magnitudes such as the redshift and look-back time.
Depending on the observational technique,
distances are numerically different and their comparison provides
important information on the parameters defining the metric. The most
commonly used distance estimators are luminosity and angular diameter distances.

\subsubsection*{Redshift $z$.}
If $\nu_e$ and $\nu_0$ are the frequencies of a line at the source and at
the observer, the redshift of the source is defined as
$z=(\nu_e/\nu_0)-1$. In Cosmology, the redshift is
directly related to the expansion factor at the time of emission $t_e$
and observation $t_0$ as \cite{weinberg_72f}
\begin{equation}
1+z=\frac{a(t_0)}{a(t_e)} .
\label{eq:redshift}
\end{equation}
Due to the expansion of the Universe, spectral lines are shifted to longer
wavelengths from the value measured in the laboratory. The redshift measures
the speed at which galaxies recede from the observer but it is not
a measure of distance; the inhomogeneities in the matter distribution
generate peculiar velocities that add to the velocity due to the Hubble
expansion.  It needs to be noted that the relative velocities have no unambiguous implications in curved spacetime, they cannot be unwarily related to the cosmological redshift, more so at large redshifts.  Objects with the same redshift could be at different
distances from the observer if they are not comoving with the Hubble flow.
The redshift can be used to define the time variation of cosmological magnitudes.
For instance, eq.~(\ref{eq:friedmann1a}) can be written in terms of the
EoS parameter as
\begin{equation}
E(z)=\frac{H(z)}{H_0}=(\sum_i\Omega_{i,0}f_i(z))^{1/2} ,\quad
f_i(z) =\exp\left[{3\int_0^z \frac{1+\omega_i(z^\prime)}{1+z^\prime}dz^\prime}\right] ,
\label{Eofz}
\end{equation}
where the sum extends over all energy density components, $i=(b,c,d,r)$.

\subsubsection*{Luminosity Distance $D_L$.}
The distance obtained by comparing the luminosity $L$ of a standard
candle to its measured flux $F$ is known as luminosity distance
$D_L=\sqrt{L/4\pi F}$. For the flat universe, the luminosity distance is given in terms of the  cosmological parameters in the form  \cite{weinberg_72f}
\begin{equation}
D_L=(1+z)cH_0^{-1}\int _0^z\frac{dz^\prime}{E(z^\prime)} ,
\label{eq:DL}
\end{equation}
The Hubble function (see Eq.~(\ref{Eofz})) encodes the information on the
time evolution of the different energy components.

\subsubsection*{Angular Diameter Distance $D_A$.}
The distance resulting from the ratio of the intrinsic size of a
standard ruler $x$ to the angle $\theta$ subtended in the sky is
$D_A=x/\theta$. For the flat universe, it can be expressed in terms of the Hubble function as
\begin{equation}
D_A=\frac{cH_0^{-1}}{(1+z)}\int _0^z\frac{dz^\prime}{E(z^\prime)} ,
\label{eq:DA}
\end{equation}
From Eqs.~(\ref{eq:DL}) and (\ref{eq:DA}) these distances verify $D_L=(1+z)^2D_A$.

\subsubsection*{Look-back time $t_L$ and age of cosmological sources.}
\label{sec:look-back}
The look-back time is defined as the difference between the age
of the Universe today and its age at some redshift $z$
\begin{equation}
t_L(z)= H_0^{-1}\int_0^z\frac{dz'}{(1+z')E(z')}\equiv t_0-t_{age}(z) -df ,
\label{eq:look-back}
\end{equation}
where $t_{age}(z)=t_L(z_F)-t_L(z)$ and $z_F$ is the redshift of the formation of a source observed. If $t_0$
is the age of the Universe today, then the look-back time is
$t_L(z)=t_L(z_F)-t_{age}(z)=t_0-t_{age}(z)-df$,
with $df=t_0-t_L(z_F)$. From stellar population synthesis
one can estimate the age of a particular galaxy and compute
its look-back time. Since the redshift of formation of the object is
not directly observable, look-back time as tests of cosmological
models can only be applied when
many similar objects are observed at different redshifts in order
to marginalize over the nuisance parameter $df$ \cite{Capozziello_2004a}.

\subsection{Problems with the Concordance Model.}

Although the concordance model fits reasonably well all the available data,
it suffers from two fine-tune initial value problems: the cosmological constant
and the coincidence problem.

\subsubsection*{The Cosmological Constant Problem.}

Table~\ref{table:parameters} shows that today $\Omega_\Lambda\sim 1$ which
implies that $\Lambda\sim 3H_0^2$. The corresponding
energy density is a constant of amplitude $\rho_\Lambda=10^{-47}(GeV)^4$.
The cosmological constant can be interpreted as the energy density
of the vacuum. At the Planck scale,  the contribution to the quantum vacuum of
the ground state of all known matter fields is $\rho_{vac}=10^{74}(GeV)^4$,
121 orders of magnitude larger \cite{weinberg_89f}. Therefore,
the initial conditions for the concordance model requires
setting a value of $\rho_\Lambda$ that is several orders of magnitude smaller
than the theoretical expectation.

\subsubsection*{The Coincidence Problem.} \label{sec:sec1.coincidence}

The energy density associated to the cosmological constant,
$\rho_\Lambda$, is constant in time but the DM density varies as
$\rho_c\propto a(t)^{-3}$. The
CMB  blackbody temperature, that today is $T_0=2.5\times 10^{-4}$eV,
scales as $T=T_0/a(t)$ can be used to relate the current ratio of the
matter to cosmological constant energy density
to its value at the Planck scale, $T_{Planck}=10^{19}$GeV, as
\begin{equation}
\frac{\rho_\Lambda}{\rho_m(t_{Planck})}=\frac{\rho_\Lambda}{\rho_{m,0}}
\left(\frac{T_{Planck}}{T_0}\right)^{-3}\simeq 10^{-95} .
\label{eq:coincidence}
\end{equation}
This expression shows that the initial values of the energy densities
associated to matter and cosmological constant would not be very likely
fixed by random processes. At the Planck time, the initial conditions are heavily
tuned 95 orders of magnitude \cite{zlatev_99f}.  Although someone may not consider it as a problem and may regard it as the values of DE and DM that ought to be in the evolving universe.

The problem of the initial conditions in the `concordance' model has led to
study different alternatives to the cosmological constant as a source of
the current period of accelerated expansion such as scalar fields with
different equations of state \cite{ratra_88f,caldwell_98f,liddle_98f,AbdallaE_2013g};
these models are termed {\it quintessence} if $\omega_d>-1$ and
{\it phantom} if $\omega_d<-1$.  Other popular alternatives are
{\it k-essence}, a scalar field with a non-canonical kinetic energy term
\cite{armendarizpicon_07f,Chiba,chimento_04f,scherrer_04f} and
the Chaplygin gas, a fluid with EoS $p \propto \rho^{-\alpha}$
($\alpha>0$) \cite{kamenschik_01f,sen_05f}. Yet,
these models suffer similar fine-tune problems \cite{chimento_00f} and do not
fit the observations better than the `concordance' model. Furthermore,
the cosmological constant enjoys a solid motivation since it can
be interpreted as the vacuum energy density while the alternative
models do not.

\subsection{Why Interacting DM/DE models?}

Most cosmological models implicitly assume that matter and DE only interact
gravitationally. If the different species do not have any other
interaction there is the energy conservation equation for each component
\begin{equation}
\frac{d\rho_i}{dt}+3H(1+\omega_i)\rho_i=0 .\\
\label{eq:friedmann2a}
\end{equation}
where $i=(b,c,d,r)$. In view of the unknown nature of both DE and DM, it
is difficult to describe these components in term of a well established
theory. Since DE and DM dominate the energy content
of the universe today, it is equally reasonable to assume
that these dark components could interact among themselves
\cite{brax_06f} and with other components. A few properties can
be derived from observations:
(A) The DE must contribute with a negative pressure to the energy budget
while the DM pressure is small, possibly zero.
(B) The DE coupling with baryons is probably
negligible, being tightly constrained by local gravity measurements
\cite{hagiwara_02f,peebles_03f}. (C) Coupling with radiation
is also very difficult since photons will no longer follow a
geodesic path and light deflection of stellar sources during solar eclipses
would contradict the observations.
(D) The coupling between DE and DM must also be small since
the concordance model, where the DE is a cosmological constant and by definition
non-interacting, is an excellent fit to the data.
Of all these possibilities, a DM/DE interaction is the most
attractive since it can either solve the coincidence problem by allowing
solutions with a constant DM/DE ratio at late times or alleviate it,
if the ratio varies more slowly than in the concordance model.

Modified gravity models can be expressed in terms of the DE/DM interaction
in the Einstein frame (see Sec.~\ref{sec:sec2_fR}). This
equivalence implies that if we determine an interaction, we will be extending
gravitational theory beyond the scope of GR, which gives further motivation
to study the interaction between the dark sectors. Unfortunately, since we
neither have a clear understanding of the nature of DM nor of DE, the nature
of the interaction between them is also an unsolved problem. There is no
clear consensus on what interaction kernel is the most adequate and different
versions, based on multiple considerations, coexist in the literature.
Ultimately, the existence of the interaction must be resolved observationally.
\section{Interacting DM/DE models.}\label{sec:sec2}

The present observational data is insufficient to determine the
nature of the DE, leaving a great freedom to construct models.
The cosmological constant can be interpreted as a fluid with an EoS
parameter $\omega_\Lambda=-1$ (see Sec.~\ref{subsec:concordance}) or,
equivalently, it can be seen as a scalar field
with a vanishing kinetic energy \cite{kolb_90f}.
Following this example, it is often assumed that the
DE is part of the field theory description of
Nature, an approach that has been extremely successful when
applied to the early Universe.
Such an effort is not just a pure theoretical attempt
of understanding, but also a step towards a general characterization
of the Dark Sector.

The lack of information on the nature and dynamics of
DM and DE makes it difficult to describe these
components from first principles, in terms of well established
physical theories. The DE can be treated as fluids, scalar fields,
vector fields, etc, and assumptions like the {\it holographic principle}
can be made to construct models. We will review these approaches
and further we will include the interactions between these DE descriptions with DM to show how they can be used to solve some of the shortcomings
of the concordance model. More details can be found in \cite{bolotin_15f}.

\begin{table}
\centering
\begin{tabular}{ccc}
\hline
Model & $Q$ & DE EoS\\
\hline
I   & $\xi_2 H\rho_d$        & $-1<\omega_d<0$\\
II  & $\xi_2 H\rho_d$        & $\omega_d< -1$ \\
III & $\xi_1 H\rho_c$        & $\omega_d< -1$ \\
IV  & $\xi H(\rho_c+\rho_d)$ & $\omega_d<-1$  \\
\lasthline
\end{tabular}
\caption{Phenomenological interacting models considered in this review.}
\label{table:models}
\end{table}

\subsection{Phenomenological Fluid Models.}\label{sec:2.1}

In the concordance model the energy density of each fluid component $i=(r,b,c,d)$,
radiation, baryons, CDM and DE, respectively, is
conserved separately: $\dot\rho_i+3H(1+\omega_i)\rho_i=0$ (eq.~\ref{eq:friedmann2a}).
In interacting models, the total energy density of the dark sector is conserved,
but the DM and DE densities evolve as
\begin{eqnarray}
&&\dot{\rho}_{c} + 3H\rho_{c} = Q, \label{eq:cons_c}\\
&&\dot{\rho}_{d} + 3H(1+\omega_d)\rho_{d}=-Q \, , \label{eq:cons_d}
\end{eqnarray}
where $Q$ represents the interaction kernel. In the absence of a fundamental
theory the quantity $Q$ cannot be derived from first principles.
We will assume that the interaction only represents a small
correction to the evolution history of the Universe; if
$|Q|\gg 0$, then either the Universe would have remained in the
matter dominated regime (if $Q>0$) or the Universe would have not
experienced a matter dominated period, altering the formation
of galaxies and large scale structure (if $Q<0$). Similarly
to how interactions behave in particle physics, one would expect the
kernel to be a function of the energy densities involved,
$\rho_{d},\rho_{c}$ and of time, $H^{-1}$. The Taylor expansion
of the interaction terms at first order would be:
$Q=H(\xi_{2}\rho_{d}+\xi_{1}\rho_{c})$, where
the coefficients $\xi_{1},\xi_{2}$ are constants
to be determined observationally. Given the lack of information,
it is convenient to use  a single parameter instead of two.
Three choices are made here: $\xi_{1}=0$, $\xi_{2}=0$ and
$\xi_{1}=\xi_{2}$. This leads to the kernels
\begin{equation}
Q=H\xi_{1}\rho_{c};\quad
Q=H\xi_{2}\rho_{d};\quad
Q=H\xi(\rho_{d}+\rho_{c}).
\label{eq:kernels}
\end{equation}
In Table~\ref{table:models} we present the phenomenological
models that will be  considered in this review. We distinguish
phantom and quintessence EoS parameters and we analyze only those
models with stable density perturbations (see Sec.~\ref{sec:sec4}).

The underlying reason why the interaction alleviates the coincidence problem
is simple to illustrate.  Due to the interaction, the ratio of energy densities
$r=\rho_{c}/\rho_{d}$ evolves with the scale factor as $r\propto a^{-\zeta}$,
where $\zeta$ is a constant parameter in the range $[0, 3]$.
The deviation of $\zeta$ from zero quantifies the severity of
the coincidence problem. When $\zeta=3$ the solution corresponds
to the $\Lambda$CDM model with $\omega_\Lambda=-1$ and $Q=0$.
If $\zeta=0$ then $r=const$ and the coincidence problem
is solved \cite{zimdahl_04d}. As examples, let us now consider
two specific kernels.

\begin{figure}
\centering
\epsfxsize=\textwidth\epsfbox{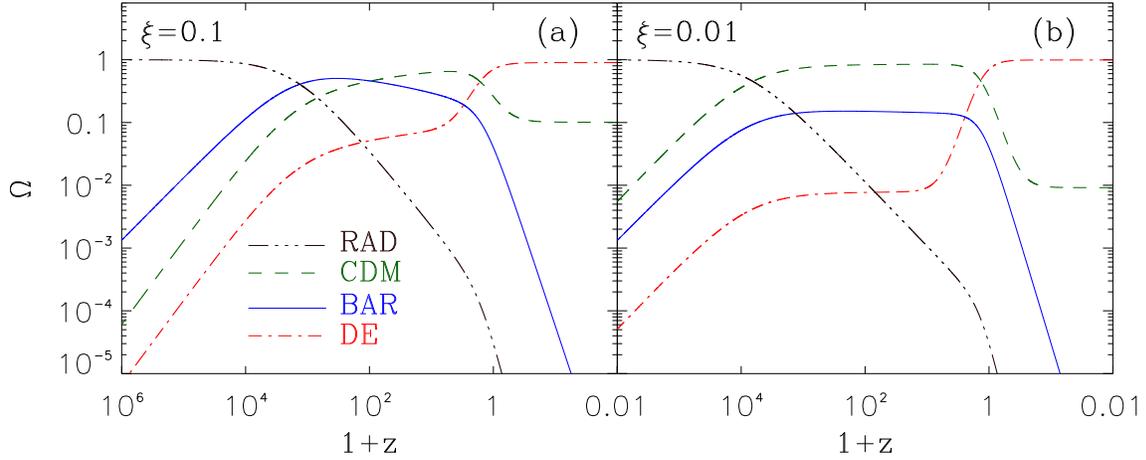}
\vspace*{-5cm}
\caption{Evolution of energy densities on a interacting DM/DE
model with kernel $Q=H\xi(\rho_{d}+\rho_{c})$. Lines
correspond to: Baryons (solid), DM (dashed), DE (dot-dashed) with
an EoS parameter $\omega_d=-1.1$ and  radiation (triple dot-dashed).
In (a) $\xi=0.1$ and in (b) $\xi=0.01$.}
\label{fig:densities}
\end{figure}

\subsubsection{A solution of the coincidence problem.}\label{sec:sec2.solution}

The kernel $Q=H\xi(\rho_{d}+\rho_{c})$ has attractor solutions
with a constant DM/DE ratio, $r=\rho_{c}/\rho_{d}=const$. In fact,
the past attractor solution
is unstable and evolves towards the future attractor solution.
To verify this behavior, we write the equation of the DM/DE ratio
\begin{equation}
\frac{dr}{dt}=-3\Gamma Hr, \qquad\qquad \Gamma=-\omega_d-
\xi^{2}\frac{(\rho_{c}+\rho_{d})^{2}}{3\rho_{c}\rho_{d}},
\label{eq:r-evolution}
\end{equation}
and stationary solutions are obtained imposing $r_{s}\Gamma(r_{s})=0$.
If, to simplify, we hold $\omega_d$ constant, then
\begin{equation}
r^{\pm}_{s}=-1+2b\pm 2\sqrt{b(b-1)} ,  \qquad b=-\frac{3\omega_d}{4\xi}>1 .
\label{r+-}
\end{equation}
Of these two stationary solutions, the past solution $r^{+}_{s}$ is unstable
while the future solution $r^{-}_{s}$ is stable \cite{Zimdahl_01f,Chimento_2003a}.
In the range $r^{-}_{s}<r<r^{+}_{s}$ the function $r(t)$ decreases monotonously.
As the Universe expands, $r(t)$ will evolve from $r_{s}^{+}$ to
the attractor solution $r^{-}_{s}$ avoiding the coincidence problem.
This DE fluid model can also be seen as a scalar field with a
power law potential at early times followed by an exponential
potential at late times \cite{Olivares_05f}.
In Fig.~\ref{fig:densities} we represent the energy densities of the
model with kernel $Q=H\xi(\rho_{c}+\rho_{d})$. In (a) the value
of the coupling constant $\xi=0.1$ was chosen to show that the Universe
undergoes a baryon domination period, altering the sequence
of cosmological eras. This value of $\xi$ would not fit
the observations since during most of the matter dominated
period baryons would dominate the formation of galaxies and
would proceed more slowly within shallower potential wells.  The
matter-radiation equality would occur after recombination so that
the anisotropies of the CMB would be altered.  In (b)
the smaller value gives rise to the correct sequence of
cosmological eras.

\begin{figure}
\centering
\epsfxsize=\textwidth\epsfbox{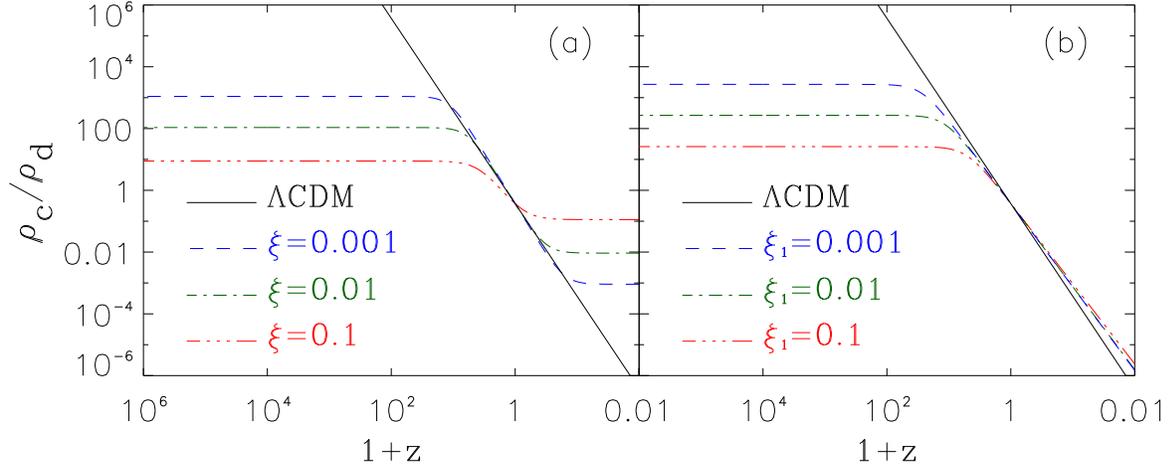}
\vspace*{-5cm}
\caption{Evolution of the ratio of DM to DE densities for different
model parameters.  In (a) the kernel is $Q=H\xi (\rho_{c}+\rho_{d})$ and in
(b) $Q=H\xi_1\rho_{c}$. The solid line, common to both panels, corresponds
to the concordance model, while the dashed lines correspond to different
interaction kernel parameters.}
\label{fig:ratios}
\end{figure}

The kernel $Q=H\xi_{1}\rho_{c}$ has also been extensively studied in the
literature \cite{campo_06f,wei_07f,Amendola_2007a,Guo_07f}. In this case
the ratio evolves as
\begin{eqnarray}
\dot{r}&=&H[\xi_{1}(1+r)+3\omega_d];\\
r&=&-(3\omega_d+\xi_{1})r_0/\{\xi_{1}r_0-(1+z)^{-(3\omega_d+\xi_{1})}
[\xi_{1}(1+r_0)+3\omega_d]\},\label{eq:r}
\end{eqnarray}
where $r_0=r(t_0)$ is the current DM/DE density ratio. Eq.~(\ref{eq:r}) does not
have future attractor solutions with $r=const$; the interaction alleviates
the coincidence problem but does not solve it.  The time evolution of
the ratio for different kernels is illustrated in Fig.~\ref{fig:ratios}
for a DE EoS $\omega_d=-1.1$.
In (a) the kernel is $Q=H\xi(\rho_{c}+\rho_{d})$ and the
ratio is constant both in the past and in the future. In (b) $Q=H\xi_1\rho_{c}$ and
the ratio is constant in the past but in the future it will evolve with time,
but the variation is $\mid(\dot{r}/r)_{0} \mid \leq H_{0}$, slower
than in $\Lambda$CDM, alleviating the coincidence problem.

\begin{figure}[tbp]
\centering
\epsfxsize=0.7\textwidth\epsfbox{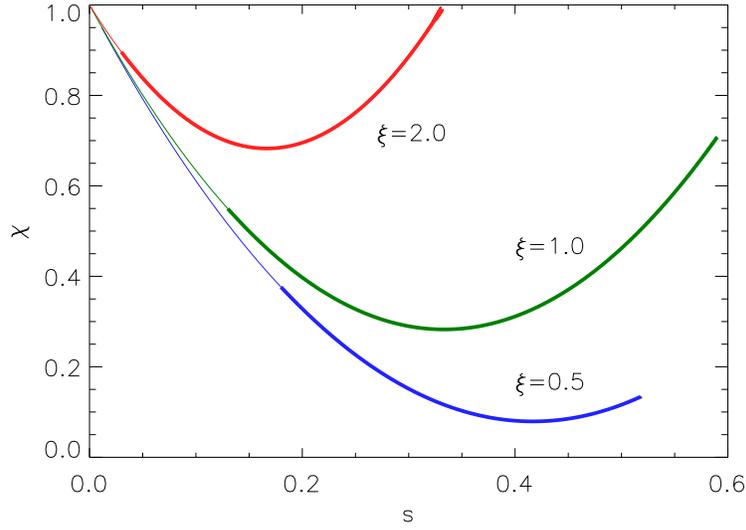}
\caption{Selected curves $\chi(s)$ for a DE EoS parameter
$\omega_d =-1.1$ and $r_{0} =3/7$ and three different values of $\xi$.
The thick lines correspond to the past evolution in the
interval $z=[0,20]$ and the thin lines to the future evolution
in $z=[-0.9,0]$.}
\label{fig:statefinder1}
\end{figure}

\subsubsection{Statefinder parameters and the coincidence problem.}

At the background level, it is possible to choose models with a
varying EoS parameter such that it reproduces the same Hubble function $H(z)$
as the DM/DE interaction models. Then, observables such as angular and
luminosity distances or look-back time can not be used to test the interaction.
One exception is when DE decays into DM, since $\omega_d(z)$
would take imaginary values \cite{campo_15f}. At the background level,
the dimensionless parameters
\begin{equation}
\chi=\frac{1}{aH^3}\frac{d^3a}{dt^3}, \quad  s=\frac{\chi-1}{3(q-\frac{1}{2})},
\end{equation}
first introduced in \cite{sahni_03d}, are
more useful to discriminate cosmological models. For instance,
If $\omega_d=const$ and the energy density ratio scales as a power
law of the scale factor, $r\propto a^{-\zeta}$ then
\begin{eqnarray}
\chi&=&1+\frac{9}{2}\frac{\omega_d}{1+r_0(1+z)^\zeta}
\left[1+\omega_d-\left(\omega_d+\frac{\zeta}{3}\right)
\frac{r_0(1+z)^\zeta}{1+r_0(1+z)^\zeta}\right]\,,
\label{eq:state_chi}\\
s&=&1+\omega_d-\left(\omega_d+\frac{\zeta}{3}\right)
	\frac{r_0(1+z)^\zeta}{1+r_0(1+z)^\zeta},
\label{eq:state_s}
\end{eqnarray}
and as indicated in Sec.~\ref{sec:2.1}, a lower value of $\zeta$ corresponds to a
model with a less severe coincidence problem. In Fig.~\ref{fig:statefinder1} we
represent the function $\chi(s)$ for three values of $\zeta$ to demonstrate that
lower values of $\zeta$ correspond to lower curves in the $s-\chi$. Hence, for
any specific model, the statefinder parameters are useful to determine the
severity of the coincidence problem. In the particular case of the
concordance model, these parameters are constants: $\chi=1$ and $s=0$
and any deviation for those values would be an observational
evidence against the concordance model. Similar conclusions have been
reached by \cite{damour_90f}.

\subsection{More forms of the interaction kernels from phenomenology}

Here we briefly consider further phenomenological
proposals of interaction kernels, not included in Table 3, that have also been discussed in the literatures:

\noindent $(i)$ $\, Q = \xi H \, \rho_{c} \, \rho_{d}/(\rho_{c}\,
+ \, \rho_{d})$. At early times ($\rho_{c} \gg \rho_{d}$) it is
seen that $r$ diverges as $a\rightarrow 0$, see Fig.23 in  \cite{campo_15f}. Further, using eq. (4) in \cite{campo_15f}
and that the above expression approaches (in that limit) to $Q\approx \xi H\rho_d$, it follows that $\xi$ is constrained to be $\xi\approx ar^{-1}(dr/da)-3\omega_d$.

\noindent $(ii)$ $\, Q = - \xi (\dot{\rho}_{c}\, + \,
\dot{\rho}_{d})$ \cite{sanchez2014}. This model interpolates
between radiation dominance and a far future de Sitter phase and
is in  good agreement with observational data; however, the  DM
component is not exactly cold, $\omega_{c}=
0.049^{+0.181}_{-0.460}$.

\noindent $(iii)$ $\,  Q = 3 (\Gamma_{d} \, \rho_{d} \, + \,
\Gamma_{c} \, \rho_{c})$  this interaction term was motivated  by
models in reheating, curvaton, and decay of DM into radiation. For
it to alleviate the coincidence problem and  the ratio $r$ be
positive and finite at early and late times the constant
coefficients $\Gamma$ must have opposite signs and satisfy
$\Gamma_{d} > \Gamma_{c}$. In this kind of models the matter
perturbations stay finite at all times. However, the models are
unable, in general, to solve the coincidence problem and,
simultaneously, ensure that $\rho_{c}$ and $\rho_{d}$ never become
negative \cite{caldera-cabral2009}. In the particular case that
$\Gamma_{c}$ vanished these constraints can be met, but the DE
dominated phase would be  transitory and the universe would revert
to DM domination in contradiction to the second law of
thermodynamics \cite{radicella2012}. Further, $r$ would diverge
irrespective of whether both $\Gamma$ coefficients were different
from zero or just one of them.

\noindent $(iv)$ $\, Q = \rho_{n0} \, a^{-3(1+\omega_{n})}
\dot{f}(\phi)$, this interaction term is not given a priori but
ascertained from the cosmic dynamics \cite{verma2010}. Here
$f(\phi)$ is a function of the scalar field, $\phi$, which
interacts with the dominant background fluid (matter or radiation)
and plays the role of a cosmological constant since the
corresponding EoS is set to -1. This function obeys $f \propto
t^{3}$, and the subindex $n$ stands for matter and radiation in
the dust and radiation eras, respectively. In this scenario $Q$ is
fixed to zero in the early inflation phase and the last de Sitter
expansion but it differs from zero in both the radiation and
matter eras.

\noindent $(v)$ In the scenario proposed in
\cite{verma2014} DE (in the form of a cosmological constant), DM and radiation arise from the action of a Higgs-like
mechanism on an underlying  tachyon field.  A small time dependent
perturbation in the EoS of the cosmological constant, so that
$\omega_{d} = -1 \, + \, \varepsilon(t)$, leads to a small shift
in the EoS of radiation and matter. The pressure of the latter
results slightly negative whence it contributes to drive the
acceleration. The three components interact non-gravitationally
with each other via two interaction terms, namely, $Q_{1} = \alpha
\dot{\bar{\rho}}_{d}$ and $Q_{2} = \beta \dot{\bar{\rho}}_{r}$
(the over-bars indicates that we are dealing with the shifted
energy densities). Different dynamics follow depending on whether
$\mid Q_{1}\mid > \mid Q_{2}\mid$ or $\mid Q_{2}\mid > \mid
Q_{1}\mid$ or $Q_{1} = Q_{2}$.

\noindent $(vi)$ In \cite{Shahalam2015} the interaction was taken in the form $Q = \alpha \dot{\rho}_{c}$,
$Q = \beta \dot{\rho}_{d}$, and also $Q = \sigma (\dot{\rho}_{c}\,
+ \, \dot{\rho}_{d})$. In all three instances no baryonic matter
is considered and $\omega_{d} > -1$. In each case the analysis of
the corresponding autonomous system reveals the existence of  a
late time stable attractor such that the ratio between both energy
densities is of the order of unity, thus solving the coincidence
problem.

Other linear and nonlinear kernels and their background evolution have been extensively studied in \cite{chimento2010}.

\subsection{Scalar Fields in Cosmology.}

From the observational point of view, phenomenological fluid models are viable
candidates of DM and DE; they fit the observational data with realistic
interaction kernels \cite{Wang_2005Gong,Wang_07f},
although they are not motivated by a dynamical principle. Alternative
formulations are usually based on a particle field approach.
 This choice not only has been very useful to describe
the physics of the early Universe, but in this context it also defines what
physical principles are involved. The situation is somewhat clearer for the DM,
with several candidates defined in terms of extensions of the Standard model.
The first DM candidate were massive neutrinos, ruled out since they
failed to explain the formation of Large
Scale Structure (LSS) \cite{Bond_80a,Thomas_10a}.
Alternative candidates were sterile neutrinos \cite{Dodelson_94a}
and axions, introduced to explain CP violation \cite{peccei_quinn_1977}.
Similarly, supersymmetry produces further candidates such as the axino,
the s-neutrino, the gravitino and the neutralino.
These particles need to be stable so they must be
the lightest supersymmetric particle. This leaves
a small number of candidates, basically the
neutralino and the gravitino \cite{Bertone_05a}.
The situation could be more complex if the DM
is not described by a single field but by a whole particle sector
with nontrivial structure. In string theory, the second piece of the symmetry
$E_8\otimes E_8$ could describe a sector that would interact with
baryonic matter only via gravity \cite{Gross_1985,Das_CR_2011}.
However, in spite of many candidates that
have been proposed and exhaustive searches that have been carried
out in the last decades, no concrete evidence of the particle nature
of the DM has emerged.

The nature of DE is even a more troubling question. When the theoretical
description of DE is made very general, models can be constructed using
a wide variety of choices at the expense of loosing predictability.
This great freedom indicates that the description of DE is more a
scenario than a physical theory, similarly to what happens with
inflationary models. The best guiding principles are
simplicity and the consistency of the theoretical foundation.
Let us assume that DE can be described in terms of quantum fields.
Its pressure should be negative to generate a period of accelerated
expansion ($\omega_d\le -1/3$, see eq.~(\ref{eq:friedmann1a})). Even in this
simplified approach, quantum field theory already imposes severe restrictions
if $\omega_d<-1$ \cite{Wang_2005Gong}. The difficulty of constructing
suitable quantum field models is illustrated by the fact that
several models correspond to non-renormalizable Lagrangians
\cite{Marulli_2012,Baldi_2012f,Amendola_2008b}. There are also models
with fermionic \cite{Ribas2008,Ribas2011,Saha2004} and vectorial DE
\cite{WeiCai_06e,Koivisto_08e,ArmendarizPicon_07f,Zuntz_2010f}. Although
gravity and other fields are purely classical and in spite of gravity being
itself non-renormalizable, the need to consider non-renormalizable models
is a clear indication that, at the moment, the description of DE must be
phenomenological.

\subsection{Field description and the DM/DE Interaction.}
\label{sec:qf_interaction}

The simplest DM description is in terms of fermions with
pressure vanishing at decreasing momenta (small energy). Let us
consider the following canonical fermionic field Lagrangian,
\begin{equation}
{\cal L}=\sqrt{-g}\bar\Upsilon\left(i\not\!{\cal D}-m\right)\Upsilon
+{\hbox{non derivative interactions}}.
\end{equation}
The energy-momentum tensor is defined as
$T_{\mu\nu}\equiv e^{-1}e_\mu^a(\delta{\cal L}/\delta e^{\nu a})$
where $e_{\mu a}$ is the vierbein and $e$ the corresponding determinant.
For a fermion field $\Upsilon$, it is given by
\begin{equation}
T_{\mu\nu}= \frac{i}{4}\left(\overline{\Upsilon}
\gamma_\mu\nabla_\nu\Upsilon+\overline{\Upsilon}
\gamma_\nu\nabla_\mu\Upsilon-\nabla_\mu\overline{\Upsilon}
\gamma_\nu\Upsilon-\nabla_\nu\overline{\Upsilon}\gamma_\mu\Upsilon\right).
\end{equation}
For a homogeneous Universe the spatial part of the energy-momentum tensor
vanishes and so does the pressure. This is not correct for relativistic fermions
since the average momentum does not vanish and originates a pressure that,
like in the case of massive neutrinos, would alter the formation of LSS. If we consider that DM and DE interact, then this constraint
can be evaded since the pressure of each component is not well defined.
An interaction gives the freedom to choose what fraction of the pressure
corresponds to the DM or to the DE. A natural choice is to take the
interaction term to be in the fermionic component, then the corresponding
background pressure vanishes and matter behaves as a pressureless fluid.
Therefore, hereafter we will describe the DM as a  non-relativistic fermion
with zero pressure, i.e., the DM is `cold'. A discussion on what models are
compatible with observational constraints is given in \cite{Pavan_2012a,bolotin_15f}.

Scalar fields are the quantum fields that provide the simplest description
of DE. If $K$ is the kinetic and $V$ the potential energy of the field $\varphi$,
the energy density and pressure associated to the field would be
$\rho_d\sim K+V$ and $p_d\sim K-V$, respectively.
If $|V|>K$, it is possible to find
configurations where the EoS is negative enough (i.e., $\omega_d<-1/3$)
to give rise to a cosmological period of accelerated expansion.
We shall see that in a theoretical field
formulation, the interaction is not only allowed but is actually inevitable.
In this section we will discuss scalar fields with renormalizable
lagrangians and defer to the next section the non-renormalizable case.

A fermionic DM and a renormalizable DE model can be described by
the Lagrangian
\begin{equation}
{\cal L}=\overline{\Upsilon}\left(i\not\partial\right)\Upsilon
+{\cal L}_s(\varphi)+F(\varphi)\overline\Upsilon\Upsilon .
\label{eq:lagrangian_general}
\end{equation}
where $F\equiv F(\varphi)$ is an effective interaction.
Any generic lagrangian would contain an interaction term
except if such term is forbidden by a given symmetry \cite{wetterich_88f}.
To continue further, let us assume that the DE can be described as
an uncharged scalar $\varphi$ obeying the lagrangian
\begin{equation}
\label{lagr}
\mathcal{L}_s(\varphi)=\ell\frac{1}{2}\partial^{\mu}\varphi\partial_{\mu}\varphi
-V(\varphi) ,
\label{eq:lagrangian_majorana}
\end{equation}
where $V(\varphi)$ is the scalar field potential (of arbitrary shape).
The sign $\ell=-1$ describes a phantom field. For simplicity,
we will restrict our study to $\ell=+1$ (see \cite{Pavan_2012a} for details)
and to the linear relation $F(\varphi)= M-\beta\varphi$.
Then $M$ is the usual fermion mass and $\beta$ a Yukawa coupling constant.

The interaction term in eq.~(\ref{eq:lagrangian_general}) couples DM and DE.
The Hubble function  (eq.~\ref{eq:friedmann1a}) for a FRW universe that also
includes baryons and radiation becomes \cite{kolb_90f}
\begin{equation}
H^2=\frac{8\pi G}{3}\left(\rho_r+\rho_b+\rho_c+\frac 12\dot{\varphi}^2+V(\varphi)\right) .
\end{equation}
In this simplified model, the different components evolve separately
and their energy densities are independently conserved except for DM
and DE. For these two components, the energy-momentum conservation
equations are
\begin{eqnarray}
&&\dot{\rho}_c+3H\rho_c=-\rho_c\dot{\varphi}\varrho/(1-\varrho\varphi), \label{eq:dm_coupled}\\
&&\ddot{\varphi}+3H\dot{\varphi}+V'(\varphi)=\rho_c \varrho/(1-\varrho\varphi),
\label{eq:de_coupled}
\end{eqnarray}
where ${\varrho}=\beta/M$; dots correspond to time derivatives and primes to derivatives
with respect to the scalar field $\varphi$. Eqs.~(\ref{eq:dm_coupled},
\ref{eq:de_coupled}) show that if DM and DE are members of a unified
quantum field description, they interact.

Although from the theoretical point of view, quantum field models constitute an
improvement over the more simple phenomenological interaction \cite{Costa_2014b},
the coupling is still undetermined. Several attempts have been tried,
including modifications of the space-time dimensions \cite{Yin_2007wa}.
Alternative exponential forms of $F(\varphi)$ have been extensively considered
in the literature giving different coupling kernels
\cite{Marulli_2012,Baldi_2012f,Amendola_2008b, Xue2015}.
The field description is a possible understanding on the interaction between dark sectors, however it brings another hidden fine tuning problem which needs to be carefully dealt with.

\subsection{Scalar fields as k-essence and Tachyons.} \label{sec:k-essence}

When renormalizability is not required, models become increasingly
more complex. For example, $k$-essence is a model of a scalar field
defined by a non-standard kinetic term
\begin{equation}
{\cal L}=p(\varphi,X), \qquad
X=\frac{1}{2}\left(D_\mu\varphi D^\mu\varphi \right) .
\label{eq:k-essence_lagrangian}
\end{equation}
If the kinetic term is separable in its variables $\varphi$ and $X$, then
the $k$-essence field can be
transformed from a tracking background into an effective cosmological
constant at the epoch of matter domination \cite{ArmendarizPicon_2001a}.
We will restrict our study to this particular Lagrangian because of its
simplicity. Our interest is driven by string theory and supergravity
where such non-standard kinetic terms appear quite often.
The lagrangian of eq.~(\ref{eq:k-essence_lagrangian}) generalizes the simplest
scalar field models. In the limit of small spatial derivatives
the Lagrangian is equivalent to that of a canonical field.

Another non-renormalizable class of models is related to tachyons in string theory.
The tachyon Lagrangian, derived from brane developments in this theory
is given by \cite{SenAshoke_2003a,SenAshoke_2003b,SenAshoke_2003c,
SenAshoke_2002a,SenAshoke_2002b,SenAshoke_2002c}
\begin{equation}
{\cal L}_{tach}=-V(\varphi)\sqrt{(1-\alpha\partial^\mu\varphi\partial_\mu\varphi)} .
\label{eq:tach_lagrangian}
\end{equation}
This Lagrangian has the form discussed by \cite{ArmendarizPicon_2001a} and
has been used to give general descriptions of the components of the dark sector
\cite{Bento_2002a,Bento_2004a}. It can be implemented in models with
interaction. One such interacting Lagrangian is
\begin{equation}
{\cal L}={\cal L}_{tach}+\frac{i}{2}\left[
\overline{\Upsilon}\gamma^\mu\nabla_\mu\Upsilon-
\overline{\Upsilon}
\overleftarrow{\nabla}_\mu
\gamma^\mu\Upsilon\right]- F(\varphi)\overline{\Upsilon}\Upsilon \quad ,
\label{eq:tach_lagrangian_coupled}
\end{equation}
where $\Upsilon$ is a fermionic field for DM and $\varphi$ a bosonic field for DE.
The linear (for canonical bosons renormalizable) model $F(\varphi)=M-\beta\phi$
has been studied in detail and shown to be compatible with the observational
constraints, although it is not  renormalizable because of the bosonic
non-linearities \cite{Micheletti_2009a}. The equations of motion can be derived
from eq.~(\ref{eq:tach_lagrangian_coupled}) and for the linear case they read
\begin{eqnarray}
&&i\gamma^\mu\nabla_\mu\Upsilon-(M-\beta\varphi)\Upsilon=0 , \\ \label{eq:dirac}
&&\alpha\nabla_\mu\partial^\mu\varphi
+\alpha^2\frac{\partial^\mu\varphi(\nabla_\mu\partial_\sigma\varphi)
\partial^\sigma\varphi}{1-\alpha\partial_\mu\varphi\partial^\mu\varphi}+
\frac{d\ln V(\varphi)}{d\varphi}=\frac{\beta\bar{\Upsilon}\Upsilon}{
V(\varphi)}\sqrt{1-\alpha\partial^\mu\varphi\partial_\mu\varphi} .\label{eq:mov_tachyons}
\end{eqnarray}
Neglecting spatial gradients, the motion of the scalar field becomes
\begin{equation}
\ddot{\varphi}=-(1-\alpha\dot{\varphi}^2)\bigg[\frac{1}{\alpha}
\frac{d\ln V(\varphi)}{d\varphi}
+3H\dot{\varphi}-\frac{\beta\bar{\Upsilon}\Upsilon}{\alpha V(\varphi)}
\sqrt{1-\alpha\dot{\varphi}^2}\bigg]\ ,
\label{homotaq}
\end{equation}
where $H=\dot{a}/a$ is the Hubble function. Fermionic current conservation implies
\begin{eqnarray}
\frac{d(a^3\bar{\Upsilon}\Upsilon)}{dt}=0\ ,
\label{eqpsi}\quad &\Rightarrow& \quad
\bar{\Upsilon}\Upsilon=\bar{\Upsilon}_0\Upsilon_0a^{-3} .
\label{eq:gravity_background}
\end{eqnarray}
Let us now show that the lagrangian of eq.~(\ref{eq:tach_lagrangian_coupled})
gives rise to a cosmological model with an interaction in the dark sector.
To that purpose, we compute the energy-momentum tensor (see \cite{Micheletti_2009a}
for details). The energy density and pressure of each component is given by
\begin{eqnarray}
\label{rofi}
&\rho_{\varphi}=\frac{V(\varphi)}{\sqrt{1-\alpha\dot{\varphi}^2}}\ ,\qquad
&p_{\varphi}=-V(\varphi)\sqrt{1-\alpha\dot{\varphi}^2} ,\\
\label{ropsi}
&\rho_{\Upsilon}= (M-\beta \varphi)\bar{\Upsilon}\Upsilon\ , &p_{\Upsilon}=0 .
\end{eqnarray}
An important consequence of eq.~(\ref{rofi}) is that the EoS
parameter of the fluid associated to the DE field is
$\omega_{\varphi}\equiv p_{\varphi}/\rho_{\varphi}=-(1-\alpha\dot{\varphi}^2)$.
If $\alpha\dot{\varphi}^2\ll 1$, then the DE acts as an effective cosmological
constant. In addition, from eqs.~(\ref{rofi},\ref{ropsi}) the time evolution of
the DM and DE energy densities are
\begin{eqnarray}
\dot{\rho}_{\varphi}+3H\rho_{\varphi}(\omega_{\varphi}+1)&=&
\beta\dot{\varphi}\bar{\Upsilon}_0\Upsilon_0a^{-3} ,
\label{conserrofi}\\
\dot{\rho_{\Upsilon}}+3H\rho_{\Upsilon}&=&-\beta\dot{\varphi}\bar{\Upsilon}_0\Upsilon_0a^{-3} .
\label{conserropsi}
\end{eqnarray}
and the Friedmann eq.~(\ref{eq:friedmann1a}) becomes
\begin{equation}
H^2=\frac{8\pi G}{3}\bigg[\rho_r+\rho_b
+(M-\beta\varphi)\bar{\Upsilon}_0\Upsilon_0 a^{-3}
+\frac{V(\varphi)}{\sqrt{1-\alpha\dot{\varphi}^2}}\bigg] .
\label{friedmann}
\end{equation}
Together with the equations of evolution of baryons and radiation,
eqs.~(\ref{conserrofi},\ref{conserropsi},\ref{friedmann}) fully describe
the background evolution of the Universe.
These equations are very similar to the ones used in
phenomenological models \cite{Feng_08f,He_08f,He_09af,Wang_07f}.
The  RHS of eqs.~(\ref{conserrofi},\ref{conserropsi})
does not contain the Hubble parameter $H$ explicitly,
but it does contain the time derivative of the scalar field, which
should behave as the inverse of the cosmological time,  thus replacing
the Hubble parameter in the phenomenological models.

Analytic solutions have been found in \cite{Padmanabhan_2002,
Feinstein_2002,Abramo_2003} in the pure bosonic case with the potential
$V(\varphi)=m^{4+n}\varphi^{-n}$ and $n>0$.
Choosing $n=2$, leads to a power law expansion of the universe.
This model has been shown  to be compatible with the observational
data \cite{Micheletti_2009a}.

\subsection{Holographic DE models.}

Another set of models are loosely based on heuristic arguments taken from
particle physics. The concept of holography
\cite{hooft,susskind_95d} has been used to fix the order
of magnitude of the DE \cite{li_04d}. To explain the origin
of these ideas, let us consider
the world as three-dimensional lattice of spin-like degrees of
freedom and let us assume that the distance between every two neighboring
sites is some small length $\ell$. Each spin can be in one of two
states. In a region of volume $L^{3}$ the number of quantum states
will be $N(L^{3}) = 2^{n}$, with $n= (L/\ell)^{3}$ the number of
sites in the volume, whence the entropy will be $S \propto
(L/\ell)^{3} \ln 2$. One would expect that if the energy density
does not diverge, the maximum entropy would vary as $L^{3}$,
i.e., $S \sim L^{3} \, \lambda_{UV}^{3}$, where $\lambda_{UV} \equiv
\ell^{-1}$ is to be identified with the ultraviolet cut-off.
Even in this case, the energy is large enough for
the system to collapse into a black hole larger than $L^{3}$.
Bekenstein suggested that the
maximum entropy of the system should be proportional to its area
rather than to its volume \cite{bekenstein_94d}. In the same vein `t Hooft
conjectured that it should be possible to describe all phenomena
within a volume using only the degrees of freedom residing on its
boundary. The number of degrees of freedom should not exceed that
of a two-dimensional lattice with about one binary degree of
freedom per Planck area.

Elaborating on these ideas, an effective field theory that saturates
the inequality $L^{3}\, \lambda^{3}_{UV} \leq S_{BH}$ necessarily
includes many states with $R_{s} > L$, where $R_{s}$ is the
Schwarzschild radius of the system under consideration \cite{cohen_99d}.
Therefore, it seems reasonable to propose a stronger constraint on the
infrared cutoff $L$ that
excludes all states lying within $R_{s}$, namely, $L^{3}\,%
\lambda^4_{UV} \leq m_{Pl}^{2}\, L$ (clearly, $\lambda^{4}_{UV}$ is the
zero--point energy density associated to the short-distance
cutoff) and we can conclude that $L\sim \lambda_{UV}^{-2}$ and
$S_{max} \simeq S_{BH}^{3/4}$. Saturating the inequality
and identifying $\lambda_{UV}^{4}$ with the
holographic DE density is given by \cite{li_04d}
\begin{equation}
\rho_d = \frac{3\wp }{8\pi G L^2} , \label{rhox}
\end{equation}
where $\wp$ is a positive, dimensionless parameter, either constant or
very slowly varying with the expansion.

Suggestive as they are, the above ideas provide no indication
about how to choose the infrared cutoff in a cosmological context.
Different possibilities have been tried with varying degrees of
success, namely, the particle horizon \cite{fischler_98d,cataldo_01d}, the future
event horizon \cite{li_04d,guberina_05d,huang_04d,gong_05d,
Wang_2005Gong,wang_06d} and the Hubble horizon. The first
choice fails to produced an accelerated
expansion. The second presents a circularity problem: for the
cosmological event horizon to exist the Universe must accelerate
(and this acceleration must not stop), i.e., it needs the
existence of DE. The third option is the most natural,
but $L=H^{-1}$ corresponds to an energy density with $\rho \propto
a^{-3}$, i.e., to dust and not to DE. Nevertheless, as we shall see below,
if the holographic DE interacts with pressureless matter then it can drive
a period of accelerated expansion and alleviate, or even solve,
the coincidence problem \cite{pavon_06d,zimdahl_07d}.

\subsubsection{Interacting holographic DE.}

An effective theory based on the holographic principle
that produces a period of accelerated expansion requires
the following assumptions: (a)
the DE density is given by Eq.~(\ref{rhox}), (b) $L =
H^{-1}$, and (c) DM and holographic DE interact
with each other obeying eqs.~(\ref{eq:cons_c},\ref{eq:cons_d}).
As an example, we will consider the kernel $Q=\xi_1\rho_{c}$ with $\xi_1> 0$.
In a spatially flat Universe, the EoS parameter of the DE for this kernel
can be expressed in terms of the interaction $\xi_1$ parameter and the
ratio $r=\rho_c/\rho_d$, namely, $\omega_d=
-(1+r)\xi_1/(3rH)$. As the DE decays into pressureless DM, it
gives rise to a negative $\omega_d$ and
the ratio of the energy densities is a constant, $r_{0} =
(1-\wp)/\wp$, irrespectively of the value of $\xi_1$ \cite{pavon_06d}.
When $\xi_1\propto H$ then $\rho_c,\rho_d$ $\propto a ^{-3m}$
with $m = (1+r_{0}+\omega_d)/(1+r_{0})$ and $a \propto t^{2/(3m)}$. Then,
the Universe will be accelerating if $\omega_d< -(1+r_{0})/3$ but if $\xi_1= 0$, the
choice $L =H^{-1}$ does not lead to acceleration.

In conclusion, the
interaction will simultaneously solve the coincidence problem
and produce a late period of accelerated expansion.
Prior to the current epoch the Universe
had to undergo a period of radiation and matter domination to preserve
the standard picture of the formation of cosmic structure. The usual way
to introduce these epochs
is to assume that the ratio $r$ has not been constant but was
(and possibly still is) decreasing. In the present context, a time
dependence of $r$ can only be achieved if $\wp$ varies slowly with
time, i.e., $0<\dot{\wp}/\wp\ll H $. This hypothesis is
not only admissible but it is also reasonable since it
is natural to expect that the holographic bound only gets fully saturated
in the very long run or even asymptotically \cite{radicella-10d}.
There is, however,  a different way to recover an early
matter dominated epoch. It is straightforward to show that
\begin{equation}
\dot{r} = 3Hr\left[\omega_d+\frac{1+r}{r}\frac{\xi_1}{3H}\right] . \label{dotr}
\end{equation}
Then, if $\xi_1/H \ll 1$ then $|\omega_d|\ll 1$ and the DE itself behaves as
pressureless matter, even if $r\simeq const$. If we
neglect the dynamical effect of curvature, baryons and radiation,
from eq.~(\ref{dotr}) and $\rho_d= 3H^2\wp/(8\pi G)$ we obtain $\wp(t)=1/(1+r(t))$.
At late times, $r\rightarrow r_{0}$ and $\wp\rightarrow \wp_0$. In
this scenario $w_d$ would depend on the fractional change of $\wp$ according to
\begin{equation}
\omega_d=-\left(1+\frac{1}{r}\right)\left[\frac{\xi_1}{3H}+\frac{\dot{\wp}}{3H\wp}\right] .
\label{wx}
\end{equation}
Holographic DE must satisfy the dominant energy condition and
it is not compatible with a phantom EoS \cite{bak_00d} and this additional
restriction $\omega_d \geq -1$ sets further constraints on $\xi_1$ and $\wp$ that need
to be fulfilled when confronting the model with observations \cite{duran_11f}.
The model is a simple and elegant option to
account for the present era of cosmic accelerated expansion
within the framework of standard gravity. Finally, its validity will be decided observationally.

\subsubsection{Transition to a new decelerated era?}

It has been speculated that the present phase of
accelerated expansion is just transitory and the Universe
will eventually revert to a fresh decelerated era. This can be
achieved by taking as DE a scalar field whose energy
density obeys a suitable ansatz. The EoS
parameter $\omega_d$ would evolve from values above but close to $-1$ to much
less negative values; the deceleration parameter increases
to positive values \cite{carvalho_06d} and the troublesome
event horizon that afflicts superstring theories disappears.
Interacting holographic models that provide a transition from
the deceleration to the acceleration can be shown to be
compatible with such a transition, reverting to a decelerating phase.
Inspection of eq.~(\ref{wx}) reveals that $w_d \le -1/3$ when
either any of the two terms in the square parenthesis (or both)
reach sufficiently small values or the first term is nearly constant and
the second becomes enough negative.
These possibilities are a bit contrived, especially the
second one since -contrary to intuition- the saturation parameter
would be decreasing instead of increasing. This counterintuitive
behavior is the result of requiring that a decelerated phase follows
the period of accelerated expansion for the
sole purpose of eliminating the event horizon.
But even if data does not suggest existence of a future
period of decelerated expansion, we cannot dismiss this possibility
offhand.  In any case, it should be noted  that  holographic dark
energy proposals that identify the infrared cutoff $L$ with the
event horizon radius are unable to produce  such transition.

\subsection{On the direction of the interaction.}\label{sec:sec2.direction}

An important open question in interacting DM/DE models
is in which direction is transferred the energy; does
DE decays into DM ($\xi > 0$) or is the other way around ($\xi< 0$)?
Although this question will be eventually settled observationally, at present we can
 explore different options based on
physical principles.

Thermodynamic considerations suggest that DE must decay into DM.
If the interaction is consistent with the principles of thermodynamics,
their temperatures will evolve according to $\dot{T}/T = - 3H
(\partial p/\partial \rho)_{n}$, where $n$ is the the number
density of particles. Then, the temperature of the DM and the DE
fluids will evolve differently due to the different time
evolution of their energy densities. When a system is
perturbed out of thermodynamic equilibrium it will react to
restore it or it will evolve to achieve a new equilibrium \cite{reif_65d}.
Then, if both DM and DE are amenable to a phenomenological
thermo-fluid description and follow the Le Ch\^atelier-Braun principle,
the transfer of energy-momentum
from DE to DM will increase their temperature difference more
slowly than if there were no interaction ($\xi=0$) or if
it is transferred in the opposite direction, the
temperature difference will increase faster \cite{Pavon_09f}.
Thus, both components, DM and DE, will stay closer to thermal
equilibrium if energy transfers from DE to DM than otherwise.

Even if the DE field is non-thermal, i.e., it
corresponds to a scalar field in a pure quantum state, a
transfer of energy from DM to DE involves an uncompensated
decrease of entropy. By contrast, a transfer in the opposite
direction creates entropy by producing DM particles. The former
process violates the second law of thermodynamics while
the latter does not.
This is also true if the DM particles are fermions and the
DE is described as a scalar field. Due to the conservation of
quantum numbers, DM decaying into DE would
violate the second law while the inverse process would not.
This latter process is similar to the production of particles in warm inflation
\cite{berera_95d} and the production of particles by the
gravitational field acting on the quantum vacuum \cite{parker_09d}.
In Chapter~\ref{sec:sec7}, we will discuss which is the direction of the
energy flow that is favored by the observations. We will show that the
data marginally favors a flow consistent with
the second law of thermodynamics and is such that alleviates the coincidence problem.

\subsection{The connection between modified gravity and interacting DM/DE}\label{sec:sec2_fR}

A DM/DE interaction is closely related to
modified theories of gravity. One example is $f(R)$ gravity.
In this theory matter is minimally coupled to
gravity in the Jordan frame, while after carrying out a conformal
transformation to the Einstein frame, the non-relativistic matter
is universally coupled to a scalar field that  can play the
role of DE \cite{Felice2010}.
Interestingly, it was found that a general $f(R)$ gravity in the
Jordan frame can be systematically and self-consistently constructed
through conformal transformation in terms of the mass dilation rate
function in the Einstein frame \cite{HeJH_2011f}.  The mass dilation
rate function marks the coupling strength between DE and DM (see
detailed discussions in \cite{HeJH_2011f}). The new $f(R)$ model
constructed in this way can generate  a  reasonable cosmic expansion.
For this $f(R)$ cosmology, the requirement to avoid the instability
in high curvature regime and to be consistent with CMB observations
is exactly equivalent to the requirement of
an energy flow from DE to DM in the interaction model to ensure
the alleviation of the coincidence problem in the Einstein frame
\cite{Felice2010,HeJH_2011f}. This result shows the conformal equivalence
between the $f(R)$ gravity in the Jordan frame and the interacting
DM/DE model in the Einstein frame. Furthermore, this
equivalence is also present at the linear perturbation level
\cite{he_2013}. The $f(R)$ model constructed from the mass dilation
rate has been shown that it can give rise to a matter dominated period
and an effective DE equation of state in consistent
with the cosmological observations \cite{heJH2013b,he_2013}.
The equivalence of the Einstein and Jordan frames
has also been discussed in \cite{Postma2014}.
In \cite{Chiba2013} it was argued that there exists a correspondence
between the variables in the Jordan frame and those in the Einstein
frame in scalar-tensor gravity and that the cosmological
observables/relations (redshift, luminosity distance,
temperature anisotropies) are frame-independent. Other discussions on the connection between modified gravity and interacting DM/DE can also be found, for example, in \cite{Kofinas2016}.

In addition to a conformal transformation, one can consider whether
there are more general transformations with similar properties.
These new transformations could provide more general
couplings between matter and gravity through a scalar field.
The question was first studied in \cite{Bekenstein1993}
where a new class of transformations, called disformal transformations,
were proposed. The idea behind such transformations is that
matter is coupled to a metric which is not just a rescaling of the
gravitational metric but it is stretched in a particular direction,
given by the gradient of a scalar field. Disformal transformations
can be motivated from brane world models and from massive gravity
theories \cite{Brax2012,Zumalacarregui2013}. Interactions between DM and DE allowing disformal couplings have also been studied in the background evolution, anisotropies in the cosmic microwave background and LSS \cite{Mota12,Bruck2015}. Recently the
idea of the disformal transformation has also been extended to study
more general theories of gravity such as the
Horndeski theory \cite{Deffayet2011,Kobayashi2011}. Similarly to the conformal transformation, in the
disformal transformation, physics must be invariant and such cosmological disformal invariance exists \cite{Domenech2015}.
All these results could provide further insight on how Cosmology can
test gravity at the largest scales and provide evidence of
generalized theories of gravity.

\section{Background Dynamics.}\label{sec:sec3}

In this section we will consider the evolution of a flat Universe whose
dynamics is influenced by the interaction between DE and DM.  The evolution
of the main cosmological parameters will differ from that of the concordance
model and their comparison with observations could, in principle, prove
the existence of interactions within the Dark Sector. To illustrate the background
evolution we will choose a particle field description of the dark sectors. For
the phenomenological fluid model, the discussions are more simplified
and the readers can refer to \cite{Chimento_2003a,Feng_08f,Feng_07f,Zimdahl_01f}.

\subsection{Attractor Solutions of Friedmann Models.}

The action describing the dynamics of a fermion DM field $\Upsilon$ coupled to a scalar
DE field $\varphi$ evolving within an expanding Universe is
\begin{eqnarray}
S&=&\int d^4x \sqrt{-g}\left(-\frac{R}{4}	
     +\frac{1}{2}\partial_\mu\varphi\partial^\mu\varphi-V(\varphi)\right.\nonumber\\
&+&\left.\frac{i}{2}\left[\bar{\Upsilon}\gamma^\mu\nabla_\mu\Upsilon
	-\bar{\Upsilon}\overleftarrow{\nabla}_\mu\gamma^\mu\Upsilon\right]
	-F(\varphi)\bar{\Upsilon}\Upsilon\right) .\label{eq:02}
\end{eqnarray}
The metric is the Friedmann-Robertson-Walker metric given
by eq.~(\ref{eq:frw-canonical}) with $K=0$, $R$ is the Ricci scalar,
$V(\varphi)$ is the scalar field potential and
$F(\varphi)$ is the interaction term. The lagrangian is slightly more general than
eq.~(\ref{eq:lagrangian_majorana}) since $F(\varphi)$ is an arbitrary function
to be specified. From the action of eq.~(\ref{eq:02})  we can derive
the equations that describe the background evolution of the Universe
\begin{eqnarray}
\label{eq:03}
\ddot{\varphi}&+&3H\dot{\varphi}+V^\prime=-F^{'}\bar{\Upsilon}\Upsilon\quad \ ,\\
\label{eq:05}
H^{2}&=&\frac{1}{3M_{p}^{2}}\left\{\frac{\dot{\varphi}^2}{2}+V(\varphi)+
  F(\varphi)\bar{\Upsilon}\Upsilon\right\}\ ,\\
\label{eq:06}
\dot{H}&=&-\frac{1}{2M_{p}^{2}}\left\{\dot{\varphi}^2+
  F(\varphi)\bar{\Upsilon}\Upsilon\right\}\ ,
\end{eqnarray}
where  $M_{p}^2=1/8\pi G$ is the reduced Planck mass. Primes represent derivatives with
respect to the scalar field $\varphi$. The fermion equation of motion can be
exactly solved to describe the DM sector in terms of the scale factor as  given by
eq.~(\ref{eq:gravity_background}).

To construct analytic solutions we define $W$ such that
$H(t)=W(\varphi(t))$. This definition
restricts the search of solutions to smooth and monotonic functions $\varphi(t)$
that are invertible; it does not solve the general case. Then,
$\dot{H}=W_{\varphi}\dot{\varphi}$, where
$W_{\varphi}\equiv\partial W /\partial \varphi$ and eq.~(\ref{eq:06})
can be rewritten as
\begin{equation}
\label{eq:08}
-W_{\varphi}\ \dot{\varphi}\ 2M_p^{2}=\dot{\varphi}^{2}+
F(\varphi)\frac{\bar{\Upsilon}_0\Upsilon_0}{a^3} .
\end{equation}
Further, we choose $a(t)^{-3}=\sigma \ \dot{\varphi}^{n}\ J(\varphi)$,
where $\sigma$ is a real constant, $n$ an integer and $J(\varphi)$ an
arbitrary function of the scalar field. This expression is general
enough to allow us to obtain a large class of exact solutions with
interacting DM/DE; by choosing conveniently
$n$ and $J(\varphi)$ we can reduce the order of the equations of motion.
Introducing this notation in eq.~(\ref{eq:08}) we obtain
\begin{equation}
\dot{\varphi}^{n-1}+\frac{[\dot{\varphi}+2M_p^2W_{\varphi}]}{
    F(\varphi)\bar{\Upsilon}_0\Upsilon_0\sigma\ J(\varphi)}=0\;,
\label{eq:10}
\end{equation}
which can be solved as an algebraic equation for $\dot{\varphi}$ for each value of $n$.
Let us consider two examples:

\subsubsection{Example I.}
If we take $F(\varphi)=M-\beta\varphi$ and choose the de-Sitter solution
($\dot a/a=const=H_{0}$) then eq.~(\ref{eq:gravity_background}) allows us
to write eq.~(\ref{eq:03}) as
\begin{equation}
\ddot\varphi+3H_{0}\dot\varphi+V'=\frac{\beta\bar{\Upsilon}_0\Upsilon_0}{a^3}\;.
\end{equation}
that has the following solution
\begin{equation}
\varphi(t)=K_1+K_{2} e^{-3H_0t}+K_{3} e^{-\frac{3}{2}H_0t}\; ,
\end{equation}
where $K_1$, $K_2$ and $K_3$ are constants.

For a power-law scale factor $a=K t^{p}$, with $K$ and $p$ positive
constants, we have for $\varphi(t)$
\begin{equation}
\varphi(t)=Y_1+Y_2\left[\frac{(\ln t)^2}{2}+Y_3 \ln t\right]\; ,
\end{equation}
where $Y_1$, $Y_2$ and $Y_3$ are constants. This solution is clearly
non-invertible and, therefore, outside the subset of solutions we are considering.
Several solutions have been obtained, though most of them
 were unphysical \cite{Pavan_2012a}. Moreover,
the construction assumes the relation between bosonic field and time
is invertible, what is not the case here but other solutions can
exist.

\subsubsection{Example II.}
If we choose $n=3$, $\sigma =1$, $W(\varphi )={\mu^4}/({\varphi M_P^2})$ and
$J(\varphi)=-{\varphi^2}/({4\bar{\Upsilon}_0\Upsilon_0\mu^4 F(\varphi)})$, where
$\mu$ is a parameter with dimensions of mass, then
\begin{eqnarray}
\label{varphi_6}
F(\varphi)&=&-C_1\frac{\mathrm{e}^{({3\varphi^{2}}/{8M_p^2})}}{\varphi^4} ,
\quad V_3(\varphi)=\frac{3\mu^8}{4M_p^2\varphi^2} , \\
\varphi(t)&=&\left(6\mu^4t\right)^{1/3} , \quad a(t)=
\left(\frac{\bar{\Upsilon}_0\Upsilon_0C_1}{2\mu^8}\right)^{1/3}
e^{\left(6\mu^4t\right)^{2/3}/{8M_p^2}} .
\end{eqnarray}
This solution corresponds to a massless fermionic DM interacting with DE.
The interaction kernel $F(\varphi$) is the product of an exponential and an
inverse power-law; the coefficient $C_1$ measures the strength of the coupling.
Notice that if $t>({2\sqrt 2M_p^3}/{6\mu^4})$ the expansion is accelerated.

\begin{figure}[!htb]
\centering
\includegraphics[scale=0.80]{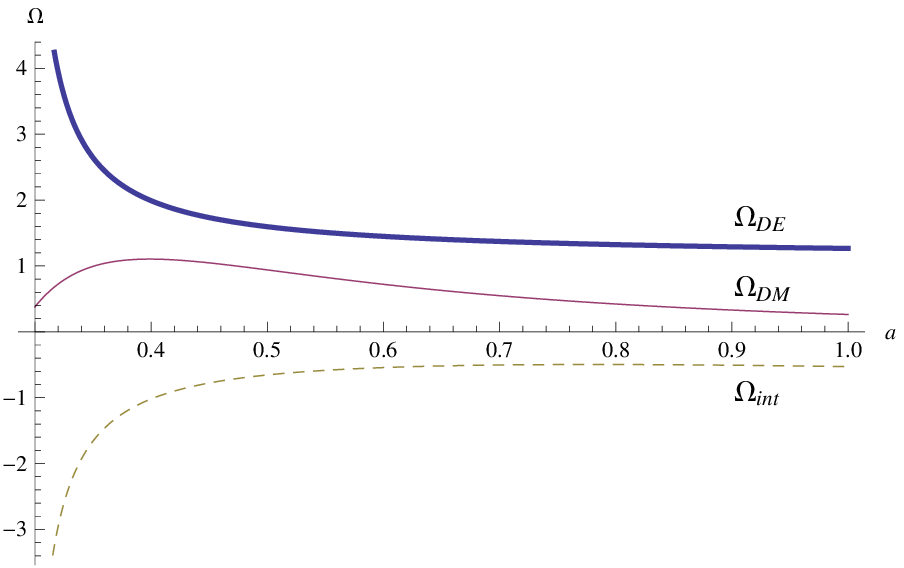} \hspace{0.3cm} %
\includegraphics[scale=0.80]{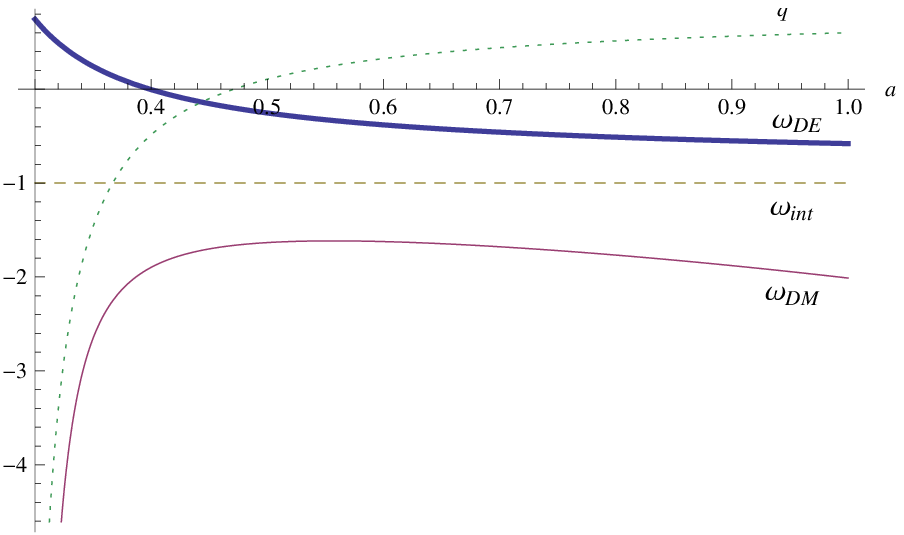}
\caption{Density parameter $\Omega$ (left panel) and equation of
state parameter $w$ and deceleration parameter (right panel) for the
model given by eqs.~(\ref{varphi_6}-\ref{rho_6_dm}). The interaction term
has been  explicitly separated. }
\label{solution6}
\end{figure}

In this model, eqs.~(\ref{eq:03}-\ref{eq:06}) can be solved analytically.
The energy density, pressure and EoS parameter for DE are given by
\cite{Pavan_2012a}
\begin{eqnarray}
\rho_d(a)&=&\frac{\mu^8}{(32M_p^4)}\frac{1}{\ln^2(\gamma a)}\left(1+3\ln(\gamma a)\right),\\
p_d(a)&=&\frac{\mu^8}{(32M_p^4)}\frac{1}{(\ln(\gamma a))^2}\left( 1-3\ln(\gamma a)\right),\\
w_d(a)&=&\frac{1-3\ln(\gamma a)}{1+3\ln(\gamma a)},
\label{rho_6_dm}
\end{eqnarray}
where $\gamma=\left({2\mu^8}/{C_{1}\bar{\Upsilon}_0\Upsilon_0}\right)^{1/3}$.
For illustration, in Fig.~\ref{solution6} we plot the solution
of eqs.~(\ref{varphi_6}-\ref{rho_6_dm})  describing the evolution of
the Universe in the limit that baryons and radiation  are not dynamically
important. In the left panel we represent
the fractional energy densities and in the right panel the deceleration and the EoS parameters.
We also plotted the interaction term and the DM
density and equation of state, respectively.
Notice that DM and DE densities have similar amplitude today, at $a=1$
when the acceleration parameter changes sign. This solution presents a transition
 from a decelerated to an accelerated expansion in agreement with observations.

The measured values of DM and DE energy densities from
Table~\ref{table:parameters} indicate that $P\sim O(10^{-7})$
and $\gamma \approx 2.06$. This gives the coupling constant $|C_1|\sim 10^{-17}$,
i.e., the interaction is very weak \cite{Pavan_2012a}.

This example shows that even with very simplifying assumptions, exact solutions
can be found that display cosmologically viable DM and DE evolutions. The only
requirement is that the coupling constant must be very small, an indication
that, observationally, the model does not differ significantly from the
concordance model while it retains all the conceptual advantages of
a  field description. Other studies on the dynamics of coupled quintessence can be referred to, for example \cite{Shahalam2015, Landim2016}

\subsection{Challenges for Scaling Cosmologies.} \label{sec:challenges}

The purpose of the interacting models is to generate cosmological solutions
where the radiation epoch is followed by a period of matter domination and
a subsequent accelerated expansion, as in the concordance model.
To solve or alleviate the coincidence problem, an almost
constant DM to DE ratio is also required.
For the idea to be of interest, the final accelerating phase must be an
attractor otherwise we would have  a new coincidence problem. Such a sequence of
cosmological eras: radiation, matter and DE dominated periods, poses a
fundamental restriction to viable models. The canonical
scalar-tensor model with an exponential scalar
potential is ruled out since it does not lead to a matter dominated period
\cite{Amendola_2000a} but even more general $k$-essence
models have difficulties to  generate viable cosmologies.
As described in Sec.~\ref{sec:k-essence}, in these models,
the Lagrangian density is $\mathcal{L}=p(X,\varphi)$, with
$X=-\frac 12 g^{\mu\nu}\partial_\mu\varphi\partial_\nu\varphi$.
To obtain scaling solutions, it is necessary that
$p(X,\varphi)=Xf(Y)$ where $Y=Xe^{\lambda\varphi}$
and $f(Y)$ is a generic function \cite{Piazza_2004a,Tsujikawa_2004a}.

For the above Lagrangian the equations of motion are
\begin{eqnarray}
H^2&=&\frac{8\pi G}{3}\left\lbrack X (f+2f_1)+\rho_c+\rho_r\right\rbrack\quad ,\\
\dot H &=& -4\pi G\left\lbrack 2X(f+f_1)+\rho_c+\frac 43 \rho_{r}\right\rbrack\quad ,\\
\ddot\varphi&=& -3AH(f+f_1)\dot\varphi -\lambda X\left(1-A(f+2f_1)\right)-AQ\rho_c\quad ,
\end{eqnarray}
where $A=(f+5f_1+2f_2)^{-1}$, $f_n=Y^n\frac{\partial^n f}{\partial Y^n}$
and $Q$ is the interaction kernel  given by
\begin{equation}
Q=-\frac{1}{\rho_c\sqrt{-g}}\frac{\partial {\cal L}_m}{\partial\varphi} .
\end{equation}
 where $g$ is the determinant of the metric $g_{\mu\nu}$.
The  equations above
can be simplified by introducing the dimensionless variables
\begin{eqnarray}
x&=&\frac{\sqrt{4\pi G}\dot\varphi}{3H}, \quad
y=\frac{\sqrt{8\pi G}e^{-\lambda\varphi/2}}{3H},\quad
z=\frac{\sqrt{8\pi G\rho_r}}{3H} ,\\
\Omega_c&=&\frac{8\pi G\rho_c}{9H^2}=1-\Omega_\varphi-z^2,
\quad \Omega_\varphi=x^2(f+2f_1) .
\end{eqnarray}
 The corresponding equations have been analyzed in \cite{Amendola_2006a} where it was
shown that a large class of coupled scalar field Lagrangians with
scaling solutions do not give rise
to a sufficiently long matter-dominated epoch  before acceleration
in order to provide enough time to give rise to galaxies and large scale structures.
As a result, DM/DE interacting models based on a scalar field description
are strongly constrained at the background level. This reflects our lack of
a solid physical foundation of the nature of DE. Particular examples
of scalar field models that are not limited by the background evolution
exist and are discussed in Sec.~\ref{sec:sec7}, but are not generic.
Therefore, in the next section we will particularize the
study of perturbation theory to the phenomenological fluid models.

\section{Perturbation theory.}\label{sec:sec4}

Models with non-minimally coupled  DM
and DE can successfully describe the accelerated
expansion of the Universe. Currently DE
and DM have only been detected via their
gravitational effects and any change in the DE
density is conventionally attributed to its
equation of state $\omega_d$. This leads to an
inevitable degeneracy between the
signature of the  interaction within the dark
sector and other cosmological parameters.
Since the coupling modifies the evolution of matter and
radiation perturbations and the clustering properties
of galaxies, to gain further insight
we need to examine the evolution of
density perturbations and test model predictions using
the most recent data on CMB temperature anisotropies
and large scale structure. Our purpose
is to identify the unique signature of the
interaction on the evolution of density perturbations
in the linear and non-linear phases.

In this Section we discuss linear
perturbation theory. We present a systematic
review on the first order perturbation equations,
discuss the stability of their solutions and
examine the signature of the interaction
between dark sectors in the CMB temperature
anisotropies. Finally, we study the growth of
the matter density perturbations. Details can be found in
\cite{He_09af,He_09bf,He_09cf,Xu_11f,He_11f}.
Alternative formulations are described in
\cite{Amendola_03af,Koivisto_05f,
Olivares_06f,Mainini_07f,Bean_08af,Bean_08bf,
Corasaniti_08f,Vacca_08f,Pettorino_08f,Schaefer_08f,
Caldera_09f,Chongchitnan_09f,
Gavela_09f,Jackson_09f,Koyama_09f,Kristiansen_09f,Vacca_09f,
Vergani_09f,Majerotto_09f, Aviles2011}.

\subsection{First order perturbation equations.}\label{sec:sec4.1}

In this subsection we will discuss the linear
perturbation theory in DM/DE interacting models.
The space-time element of eq.~(\ref{eq:frw-conformal})
perturbed at first order reads
\begin{eqnarray}
ds^2 &=& a^2(\tau)[-(1+2\psi)d\tau^2+2\partial_iBd\tau
dx^i
+(1+2\phi)\delta_{ij}dx^idx^j+D_{ij}Edx^idx^j],\nonumber\\
\label{perturbedspacetime}
\end{eqnarray}
where $\tau$ is the conformal time defined by $d\tau=dt/a$, $\psi, B, \phi, E$ represent the scalar
metric perturbations, $a$ is the cosmic scale
factor and
$
D_{ij}=(\partial_i\partial_j-\frac{1}{3}\delta_{ij}\nabla^2)
$.

\subsubsection{Energy-momentum balance.}

We work with the energy-momentum tensor $T^{\mu\nu}=\rho
u^{\mu}u^{\nu}+p(g^{\mu\nu}+u^{\mu}u^{\nu})$,
for  a two-component system consisting of DE and
DM. The covariant description of the
energy-momentum transfer between DE and DM is
given by $\nabla_{\mu}{T_{(\lambda)}}^{\mu\nu}
={Q_{(\lambda)}}^{\nu}\label{energy-stress}$
where ${Q_{(\lambda)}}^{\nu}$ is a four vector
governing the energy-momentum transfer between the
different components \cite{Sasaki_84f}. The subindex $\lambda$
refers to DM and DE respectively. For the whole
system, DM plus DE, the energy and momentum are conserved,
and the transfer vector satisfies
$\sum_{\lambda}{Q_{(\lambda)}}^{\nu}=0$.

The perturbed energy-momentum tensor reads,
\begin{eqnarray}
\delta \nabla_{\mu}T^{\mu0}_{(\lambda)} &=&
\frac{1}{a^2}\{-2[\rho_{\lambda}'+3\mathcal
{H}(p_{\lambda}+\rho_{\lambda})]\psi+\delta
\rho_{\lambda}'+(p_{\lambda}+\rho_{\lambda})\theta_{\lambda}\nonumber\\
&+&3\mathcal{H}(\delta p_{\lambda}+\delta
\rho_{\lambda})
+3(p_{\lambda}+\rho_{\lambda})\phi'\}=\delta Q^0_{\lambda},\nonumber
\label{perturbation}\\
\partial_i \delta \nabla_{\mu}T^{\mu i}_{(\lambda)} &=&
\frac{1}{a^2}\{[p'_{\lambda}+\mathcal
{H}(p_{\lambda}+\rho_{\lambda})]\nabla^2B+
[(p'_{\lambda}+\rho'_{\lambda})+4\mathcal{H}(p_{\lambda}+\rho_{\lambda})]
\theta_{\lambda}\nonumber\\
&+&(p_{\lambda}+\rho_{\lambda})\nabla^2B'+\nabla^2\delta
p_{\lambda}+(p_{\lambda}+\rho_{\lambda})\theta_{\lambda}'
+(p_{\lambda}+\rho_{\lambda})\nabla^2\psi\}\\
&=& \partial_i \delta Q^i_{(\lambda)}\quad ,\nonumber
\end{eqnarray}
where $\theta = \nabla^2 v$, $v$ is the potential
of the three velocity and primes denote
derivatives with respect to the conformal time
$\tau$. At first order, the perturbed Einstein equations are
\begin{eqnarray}
&&-4\pi Ga^2\delta \rho=\nabla^2\phi+3\mathcal{H}\left(\mathcal{H}\psi-\phi'\right)
+\mathcal{H}\nabla^2B-\frac{1}{6}[\nabla^2]^2E\; ,\nonumber \\
&&-4\pi G a^2(\rho+p)\theta=\mathcal{H}\nabla^2\psi-\nabla^2\phi'+2\mathcal{H}^2\nabla^2B
-\frac{a^{''}}{a}\nabla^2B+\frac{1}{6}[\nabla^2]^2E'\; ,\nonumber\\
&&8\pi Ga^2\Pi^{i}_{j}=-\partial^{i}\partial_{j}\psi-\partial^{i}\partial_{j}\phi
+\frac{1}{2}\partial^{i}\partial_{j}E^{''}+\mathcal{H}\partial^{i}\partial_{j}E'
+\frac{1}{6}\partial^{i}\partial_{j}\nabla^2E\nonumber\\
&&-2\mathcal{H}\partial^{i}\partial_{j}B-\partial^{i}\partial_{j}B; ,\label{PerEinstein}
\end{eqnarray}
where $\delta \rho = \sum_{\lambda} \delta \rho_{\lambda}$ and $(p+\rho)\theta=
\sum_{\lambda}(p_{\lambda}+\rho_{\lambda})\theta_{\lambda}$
is the total energy density perturbation.

\subsubsection{The general perturbation equations.}

Considering an infinitesimal transformation of the coordinates\cite{Sasaki_84f}
\begin{eqnarray}
\tilde{x}^{\mu}= x^{\mu}+\delta x^{\mu},\quad
\delta x^0 = \xi^0(x^{\mu}),\quad
\delta x^{i} = \partial^{i}\beta
(x^{\mu})+v_*^{i}(x^{\mu})
\end{eqnarray}
where $\partial_iv_*^{i}=0$. The perturbed quantities behave as
\begin{eqnarray}
\tilde{\psi} &=& \psi -\xi^{0'}-\frac{a'}{a}\xi^0\; ,\qquad
\tilde{B} = B +\xi^{0}-\beta'\; ,\nonumber\\
\tilde{\phi} &=& \phi
-\frac{1}{3}\nabla^2\beta-\frac{a'}{a}\xi^0\; ,\quad
\tilde{E}=E-2\beta\nonumber\\
\tilde{v}&=&v+\beta'\; ,\qquad \qquad \qquad
\tilde{\theta}=\theta+\nabla^2\beta'\; .\label{quantities}
\end{eqnarray}
Inserting eq.~(\ref{quantities}) in eq.~(\ref{perturbation}), we obtain
\begin{eqnarray}
\tilde{\delta Q}^0=\delta Q^0-Q^{0'}\xi^0+Q^0\xi^{0'}\; ,\quad
\tilde{\delta Q}_p=\delta Q_p+Q^0\beta',\label{Qtransform}
\end{eqnarray}
where $\delta Q_p$ denotes the potential of three vector $\delta Q^i$ and verifies
\begin{equation}
\delta Q^i= \partial^i\delta Q_p+ \delta Q_*^i ,\nonumber
\end{equation}
with $\partial_i\delta Q_*^i=0$. This is consistent with the results
obtained using Lie derivatives,
\begin{eqnarray}
\mathcal{L}_{\delta x} Q^{\nu}=\delta
x^{\sigma}Q^{\nu}_{,\sigma}-Q^{\sigma}\delta
x^{\nu}_{,\sigma}\label{Lie},\quad
\delta \tilde{Q}^{\nu}=\delta
Q^{\nu}-\mathcal{L}_{\delta x} Q^{\nu}\; ,
\end{eqnarray}
which shows that $Q^{\nu}$ is covariant.

We expand metric perturbations in Fourier
space by using scalar harmonics,
\begin{eqnarray}
\tilde{\psi}Y^{(s)} &=& (\psi -\xi^{0'}-\frac{a'}{a}\xi^0)Y^{(s)}\quad ,\qquad
\tilde{B}Y_i^{(s)} = (B -k\xi^{0}-\beta')Y_i^{(s)}\quad ,\nonumber\\
\tilde{\phi}Y^{(s)} &=& (\phi
-\frac{1}{3}k\beta-\frac{a'}{a}\xi^0)Y^{(s)}\quad ,\quad
\tilde{E}Y_{ij}^{(s)}=(E+2k\beta)Y_{ij}^{(s)}\quad ,\nonumber\\
\tilde{\theta}Y^{(s)}&=&(\theta+k\beta')Y^{(s)}\quad ,\label{gaugeInfourier}
\end{eqnarray}
and the perturbed conservation equations of eq.~(\ref{perturbation}) read
\begin{eqnarray}
&&\delta_{\lambda}'+3\mathcal{H}(\frac{\delta
p_{\lambda}}{\delta
\rho_{\lambda}}-\omega_{\lambda})\delta_{\lambda} =
-(1+\omega_{\lambda})kv_{\lambda}
-3(1+\omega_{\lambda})\phi'  \nonumber\\
&&+(2\psi-\delta_{\lambda})\frac{a^2Q^0_{\lambda}}{\rho_{\lambda}}+\frac{a^2
\delta
Q^0_{\lambda}}{\rho_{\lambda}}\quad ,\label{perturbed}\\
&&(v_{\lambda}+B)'+\mathcal{H}(1-3\omega_{\lambda})(v_{\lambda}+B)
= \frac{k}{1+\omega_{\lambda}}\frac{\delta
p_{\lambda}}{\delta
\rho_{\lambda}}\delta_{\lambda}
-\frac{\omega_{\lambda}'}{1+\omega_{\lambda}}(v_{\lambda}+B)\nonumber\\
&&+k\psi
-\frac{a^2Q^0_{\lambda}}{\rho_{\lambda}}v_{\lambda}
-\frac{\omega_{\lambda}a^2Q^0_{\lambda}}{(1+\omega_{\lambda})\rho_{\lambda}}B
+\frac{a^2\delta Q_{p\lambda}}{(1+\omega_{\lambda})\rho_{\lambda}}\nonumber\,.
\end{eqnarray}
Introducing the gauge invariant quantities \cite{Sasaki_84f}
\begin{eqnarray}
&&\Psi=\psi -
\frac{1}{k}\mathcal{H}(B+\frac{E'}{2k})-\frac{1}{k}(B'+\frac{E^{''}}{2k})\; ,\quad
\Phi=\phi+\frac{1}{6}E-\frac{1}{k}\mathcal{H}(B+\frac{E'}{2k})\; ,\nonumber\\
&&\delta Q^{0I}_{\lambda }=\delta
Q^0_{\lambda}-\frac{Q^{0'}_{\lambda}}{\mathcal{H}}(\phi+\frac{E}{6})
+Q^{0}_{\lambda}\left[\frac{1}{\mathcal{H}}(\phi+\frac{E}{6})\right]'\;
,\quad V_{\lambda} = v_{\lambda}-\frac{E'}{2k}\; ,\nonumber\\
&&\delta Q_{p\lambda}^{I}=\delta Q_{p\lambda}-Q^0_{\lambda}\frac{E'}{2k}\; , \qquad
D_{\lambda}=\delta_{\lambda}-\frac{\rho_{\lambda}'}{\rho_{\lambda}\mathcal{H}}
\left(\phi+\frac{E}{6}\right)\label{invariantQ}\;,
\end{eqnarray}
we obtain the gauge invariant linear perturbation
equations for the dark sector.  $D_{\lambda}$ is the gauge invariant density perturbation of DM or DE, and $V_{\lambda}$ is the gauge invariant velocity perturbation for DM and DE respectively.  For the DM they are
\begin{eqnarray}
&&D'_{c}+\left \{\left(\frac{a^2Q_c^0}{\rho_c\mathcal{H}}\right)'
+\frac{\rho_c'}{\rho_c\mathcal{H}}\frac{a^2Q_c^0}{\rho_c}\right \}\Phi
+\frac{a^2 Q_c^0}{\rho_c}D_{c}+\frac{a^2Q_c^0}{\rho_c\mathcal{H}}\Phi'
= -kV_c
\nonumber\\
&&+2\Psi\frac{a^2Q_c^0}{\rho_c}+\frac{a^2\delta Q_c^{0I}}{\rho_c}
+\frac{a^2Q_c^{0'}}{\rho_c\mathcal{H}}\Phi
-\frac{a^2Q_c^0}{\rho_c}\left(\frac{\Phi}{\mathcal{H}}\right)'\;, \nonumber\\
&&V_c'+\mathcal{H}V_c=k\Psi-\frac{a^2Q_c^0}{\rho_c}V_c+
\frac{a^2\delta Q_{pc}^{I}}{\rho_c}\;,\label{DMV}
\end{eqnarray}
while for the DE we have
\begin{eqnarray}
&&D'_{d}+\left\{\left(\frac{a^2Q_d^0}{\rho_d\mathcal{H}}\right)'
-3\omega_d'+3(C_e^2-\omega_d)\frac{\rho_d'}{\rho_d}
+\frac{\rho_d'}{\rho_d\mathcal{H}}\frac{a^2Q_d^0}{\rho_d}\right\}\Phi\nonumber\\
&&+\left\{3\mathcal{H}(C_e^2-\omega_d)+\frac{a^2Q_d^0}{\rho_d}\right\}D_{d}
+\frac{a^2Q_d^0}{\rho_d\mathcal{H}}\Phi'\nonumber\\
=&&-(1+\omega_d)kV_d+3\mathcal{H}(C_e^2-C_a^2)\frac{\rho_d'}{\rho_d}
\frac{V_d}{k}+2\Psi\frac{a^2Q_d^0}{\rho_d}+\frac{a^2\delta
Q_d^{0I}}{\rho_d}\nonumber\\
&&+\frac{a^2Q_d^{0'}}{\rho_d\mathcal{H}}\Phi-\frac{a^2Q_d^0}{\rho_d}
\left(\frac{\Phi}{\mathcal{H}}\right)',\label{DEV}\\
&&V_d'+\mathcal{H}(1-3\omega_d)V_d
=\frac{kC_e^2}{1+\omega_d}D_{d}+\frac{kC_e^2}{1+\omega_d}
\frac{\rho_d'}{\rho_d\mathcal{H}}\Phi-\frac{a^2Q_d^0}{\rho_d}V_d\nonumber\\
&&-\left(C_e^2-C_a^2\right)\frac{V_d}{1+\omega_d}\frac{\rho_d'}{\rho_d}
-\frac{\omega_d'}{1+\omega_d}V_d+k\Psi+\frac{a^2\delta Q_{pd}^{I}}
{(1+\omega_d)\rho_d}\; ,\nonumber
\end{eqnarray}
in these expressions we have introduced
\begin{equation}
\frac{\delta p_d}{\rho_d} =
C_e^2\delta_d-(C_e^2-C_a^2)\frac{\rho_d'}{\rho_d}\frac{v_d+B}{k}\label{Dpp}
\end{equation}
with $C_e^2$ is the effective sound speed of DE at the rest frame and
$C_a^2$ is the adiabatic sound speed \cite{Valiviita_08f}.

To alleviate the singular behavior caused by $\omega_d$
crossing $-1$, we substitute $V_{\lambda}$ into
$U_{\lambda}$ in the above equations where
\begin{equation}
U_{\lambda}=(1+\omega_d)V_{\lambda}\; .
\end{equation}
Thus we can rewrite eqs.~(\ref{DMV},\ref{DEV}) as
\begin{eqnarray}
&&D'_{c}+\left \{\left(\frac{a^2Q_c^0}{\rho_c\mathcal{H}}\right)'
   +\frac{\rho_c'}{\rho_c\mathcal{H}}\frac{a^2Q_c^0}{\rho_c}\right \}\Phi
   +\frac{a^2 Q_c^0}{\rho_c}D_{c}+\frac{a^2Q_c^0}{\rho_c\mathcal{H}}\Phi'=\nonumber\\
&&-kU_c +2\Psi \frac{a^2Q_c^0}{\rho_c}+\frac{a^2\delta Q_c^{0I}}{\rho_c}
   +\frac{a^2Q_c^{0'}}{\rho_c\mathcal{H}}\Phi
   -\frac{a^2Q_c^0}{\rho_c}\left(\frac{\Phi}{\mathcal{H}}\right)',\nonumber\\
&&U_c'+\mathcal{H}U_c = k\Psi-\frac{a^2Q_c^0}{\rho_c}U_c
   +\frac{a^2\delta Q_{pc}^{I}}{\rho_c}, \label{DMU}\\
&&D'_{d}+\left\{\left(\frac{a^2Q_d^0}{\rho_d\mathcal{H}}\right)'
   -3\omega_d'+3(C_e^2-\omega_d)\frac{\rho_d'}{\rho_d}
   +\frac{\rho_d'}{\rho_d\mathcal{H}}\frac{a^2Q_d^0}{\rho_d}\right\}\Phi\nonumber\\
&&+\left\{3\mathcal{H}(C_e^2-\omega_d)+\frac{a^2Q_d^0}{\rho_d}\right\}D_{d}
   +\frac{a^2Q_d^0}{\rho_d\mathcal{H}}\Phi'
   =3\mathcal{H}(C_e^2-C_a^2)\frac{\rho_d'}{\rho_d}
    \frac{U_d}{(1+\omega_d)k}\nonumber\\
&&-kU_d+2\Psi\frac{a^2Q_d^0}{\rho_d}+\frac{a^2\delta Q_d^{0I}}{\rho_d}
   +\frac{a^2Q_d^{0'}}{\rho_d\mathcal{H}}\Phi
   -\frac{a^2Q_d^0}{\rho_d}\left(\frac{\Phi}{\mathcal{H}}\right)', \label{DEU}\\
&&U_d'+\mathcal{H}(1-3\omega_d)U_d=kC_e^2D_{d}
   +kC_e^2\frac{\rho_d'}{\rho_d\mathcal{H}}\Phi-\left(C_e^2-C_a^2\right)
   \frac{U_d}{1+\omega_d}\frac{\rho_d'}{\rho_d}\nonumber\\
&&+(1+\omega_d)k\Psi-\frac{a^2Q_d^0}{\rho_d}U_d+\frac{a^2\delta Q_{pd}^{I}}{\rho_d}.
\end{eqnarray}
The quantity $\Phi$ is given by
\begin{equation}
\Phi=\frac{4\pi Ga^2\sum \rho_i\{D^i+3\mathcal{H}U^i/k\}}{k^2
    -4\pi Ga^2\sum \rho_i'/\mathcal{H}}\label{quad1}.
\end{equation}
Eqs.~(\ref{DMU},\ref{DEU}) are the most generic form of
the linear perturbation equations for the DM and the DE,
regardless of the specific form of the interaction ${Q_{(\lambda)}}^{\nu}$.
The transfer of energy-momentum between DM and DE has to be specified in a
covariant form. In the next subsection we will specify ${Q_{(\lambda)}}^{\nu}$
for each particular coupling.

\subsubsection{Covariant couplings.} \label{sec:cov_couplings}

The four vector ${Q_{(\lambda)}}^{\nu}$ can be
phenomenologically decomposed into two parts with
respect to a given observer $\eta$ with four
velocity ${U_{(\eta)}}^{\mu}$,
\begin{equation}
{Q_{(\lambda)}}^{\mu}=Q{_{(\lambda
\eta)}}{U_{(\eta)}}^{\mu}+{F_{(\lambda
\eta)}}^{\mu} ,
\label{eq:Q_lambda}
\end{equation}
where $Q{_{(\lambda \eta)}}=-{U_{(\eta)}}_{\nu}Q{_{(\lambda)}}^{\nu}$
is the energy transfer rate of the $\lambda$ component observed by $\lambda'$ observer;
${F_{(\lambda \eta)}}^{\mu}={{h_{(\eta)}}^{\mu}}_{\nu}{Q_{(\lambda)}}^{\nu}$
is the corresponding momentum transfer and $h^{\mu}_{\nu}$ is the projection operator.
In \cite{He_11f} it has been  probed  that such a decomposition of
${Q_{(\lambda)}}^{\nu}$ and its perturbed form are covariant.
As discussed in \cite{He_11f}, we can specify the coupling vector
${Q_{(\lambda)}}^{\nu}$ in the co-moving frame as
\begin{eqnarray}
{Q_{(\lambda)}}^{\nu}&=&\left[\frac{Q_{(\lambda)}}{a},0,0,0\right]^{T}\label{couplev}
\end{eqnarray}
where $Q_{(\lambda)}$ is the module of the four vector
${Q_{(\lambda)}}^{\nu}$. The perturbed form
$\delta Q_{(\lambda)}^0$  can be uniquely
determined from the background energy-momentum
transfer $Q_{(\lambda)}^{\mu}$. From
\begin{equation}
{Q_{(\lambda)}}^{\mu}Q_{(\lambda)\mu}=g_{00}({Q_{(\lambda)}}^0)^2=-{Q_{(\lambda)}}^2\quad ,
\end{equation}
where $Q_{(\lambda)}=a{Q_{(\lambda)}}^{0}$ is a
scalar in the FRW space  and the minus sign
indicates that ${Q_{(\lambda)}}^{\mu}$ is
time-like, we can obtain the form of the perturbed
part of the energy-momentum transfer
\begin{equation}
\delta Q_{(\lambda)}^0=-\frac{\psi}{a}Q_{(\lambda)}+\frac{1}{a}\delta Q_{(\lambda)} .
\end{equation}
The coupling vector defined by eq.~(\ref{eq:Q_lambda}) is independent
of the choice of observers. Although it is decomposed in two parts
that depend on the observer $\eta$ and its four velocity
$U_{(\eta)}^{\mu}$, the decomposition cannot bring substantial
physics since eq.~(\ref{eq:Q_lambda})  is an identity. The perturbed
forms, including the zero-th component and the spatial component are
also identities. The zero-th component of the perturbed form can be
uniquely determined by the background $Q_{(\lambda)}^{\mu}$.
The spatial component of the perturbed
energy-momentum transfer $\delta Q^i_{(\lambda)}$
is independent of the zeroth component.  It
refers to the non-gravitational force and is
composed of two parts,
\begin{eqnarray}
\delta Q_{p\lambda}=\delta
Q_{p\lambda}^I\mid_{t}+Q_{(\lambda)}^0v_{t}\quad ,\label{deltaQ}
\end{eqnarray}
where $\delta Q_{p\lambda}$ is the potential of
the perturbed energy-momentum transfer $\delta
Q^i_{(\lambda)}$, $\delta Q_{p\lambda}^I\mid_{t}$
is the external non-gravitational force density
and $v_{t}$ is the average velocity of the energy
transfer. $v_{t}$ is a free quantity which needs
to be specified according to physics. In
\cite{Valiviita_08f}, $v_{t}$ was allowed to
follow the peculiar velocity of DM or DE
respectively. In fact, $v_{t}=v_c$ or
$v_{t}=v_d$ reproduce the results of
\cite{Valiviita_08f}. In our analysis we will
consider that there are no other interactions
than gravitation acting on the coupled DM/DE system;
only the inertial drag effect due to the stationary
energy transfer between DE and DM appears
\cite{Simpson_11f}. Thus, we set $v_{t}=0$ and
$\delta Q_{p\lambda}^I\mid_{t}=0$, which leads to
a vanishing perturbation, $\delta Q^i=0$.

When constructing the four vector defined by
Eq.~(\ref{couplev}), the module $Q_{\lambda}$ can
be chosen to be any combination of scalars in the
FRW space, such as the energy density
${\rho_{(\lambda)}}={T_{(\lambda)}}^{\mu
\nu}U_{(\lambda)\mu}U_{(\lambda)} \nu$, expansion
$H_{(\lambda)}=\frac{1}{3}\nabla_{\mu}{U_{(\lambda)}}^{\mu}$,
or any other scalar function. Considering that
$Q_{\lambda}$ is independent of the observer so is
the energy density as well as its perturbed form, we
require $H$ to be a global quantity invariant under change of observers.
In a general phenomenological description, we can assume the DM/DE coupling is
\begin{eqnarray}
Q_{c}=-Q_d=3H(\xi_1\rho_c+\xi_2\rho_d) .\label{eq:gauge}
\end{eqnarray}
The perturbed forms read
\begin{eqnarray}
\delta Q_{c}&=&-\delta Q_d= 3H(\xi_1\delta\rho_c+\xi_2\delta\rho_d) ,\label{eq:gauge-a}\\
\delta Q_{c}^0 &=&-\delta Q_{d}^0= -3H(\xi_1\rho_c+\xi_2\rho_d)\frac{\psi}{a}
+3H(\xi_1\delta \rho_c + \xi_2 \delta \rho_d)\frac{1}{a} .\label{eq:gauge-b}
\end{eqnarray}
The gauge invariant quantities $\delta Q_c^{0I}$
and $\delta Q_d^{0I}$ in eqs.~(\ref{DMU},\ref{DEU}) are given by
\cite{He_09bf}
\begin{eqnarray}
\frac{a^2\delta Q_c^{0I}}{\rho_c}&=&-3\mathcal{H}(\xi_1+\xi_2/r)\Psi
    +3\mathcal{H}\{\xi_1D_{c}+\xi_2D_{d}/r\}+3(\xi_1\frac{\rho_c'}{\rho_c}
    +\frac{\xi_2}{r}\frac{\rho_d'}{\rho_d})\Phi \nonumber\\
&&-\frac{a^2}{\rho_c}\frac{Q_c^{0'}}{\mathcal{H}}\Phi
    +\frac{a^2Q_c^0}{\rho_c}\left[\frac{\Phi}{\mathcal{H}}\right]' ,\label{eq:gauge-c}\\
\frac{a^2\delta Q_d^{0I}}{\rho_d}&=&3\mathcal{H}(\xi_1r+\xi_2)\Psi
    -3\mathcal{H}\{\xi_1D_{c}r+\xi_2D_{d}\}-3(\xi_1r\frac{\rho_c'}{\rho_c}
    +\frac{\rho_d'}{\rho_d}\xi_2)\Phi\nonumber\\
&&-\frac{a^2Q_d^{0'}}{\rho_d}\Phi+\frac{a^2Q_d^0}{\rho_d}
    \left[\frac{\Phi}{\mathcal{H}}\right]' ,\label{eq:gauge-d}
\end{eqnarray}
where as before $r=\rho_c/\rho_d$ is the DM to DE ratio.

\subsubsection{Phenomenological gauge-invariant perturbation equations.}

Inserting the phenomenological interaction eqs.~(\ref{eq:gauge}--\ref{eq:gauge-d})
into eqs.~(\ref{DMU},\ref{DEU}) and neglecting the
spatial perturbations $\delta Q_{p\lambda}^{I}=0$, we obtain the
phenomenological general gauge-invariant perturbation equations for
DM and DE, respectively,
\begin{eqnarray}
D_{c}'&=&-kU_c+3\mathcal{H}\Psi(\xi_1+\xi_2/r)-3(\xi_1+
\xi_2/r)\Phi'+3\mathcal{H}\xi_2(D_{d}-D_{c})/r, \nonumber\\
U_c'&=&-\mathcal{H}U_c+k\Psi-3\mathcal{H}(\xi_1+\xi_2/r)U_c, \label{cvelocity}\\
D_{d}'&=&-3\mathcal{H}(C_e^2-\omega_d)D_{d}
   +\left\{3\omega_d'-9\mathcal{H}(\omega_d-C_e^2)
   \left(\xi_1r+\xi_2+1+\omega_d\right )\right\}\Phi \nonumber\\
&-&9\mathcal{H}^2(C_e^2-C_a^2)\frac{U_d}{k}+3(\xi_1r+\xi_2)\Phi'
   -3\Psi\mathcal{H}(\xi_1 r+\xi_2)\nonumber \\
&-&9\mathcal{H}^2(C_e^2-C_a^2)(\xi_1r+\xi_2)\frac{U_d}{(1+\omega_d)k}-kU_d
+3\mathcal{H}\xi_1r(D_{d}-D_{c}), \nonumber\\
U_d'& = &-\mathcal{H}(1-3\omega_d)U_d-3kC_e^2\left(\xi_1
r+\xi_2+1+\omega_d\right )\Phi \nonumber\\
&&+3\mathcal{H}(C_e^2-C_a^2)(\xi_1r+\xi_2)\frac{U_d}{(1+\omega_d)}
+3(C_e^2-C_a^2)\mathcal{H}U_d\nonumber\\
&&+kC_e^2D_{d}+(1+\omega_d)k\Psi+3\mathcal{H}
(\xi_1r+\xi_2)U_d . \label{dvelocity}
\end{eqnarray}
The general gauge invariant formalism fully
removes the ambiguity of gauge choice.  However, numerical
solutions can be obtained after choosing a gauge without loosing
generality (see chapter III of \cite{Sasaki_84f} for
details). The results will be the same for
different gauges if the gauge is fully fixed
\cite{He_11f}. Following \cite{He_09af}, in our subsequent
discussion we will choose  the conformal
Newtonian gauge with adiabatic
initial conditions.

\subsection{Stability analysis.}\label{sec:sec4.stability}

Models with a background evolution characterized by
adiabatic initial conditions were studied in perturbation theory
and found to have  unstable growing modes when the interaction
couplings were much larger than the gravitational strength
\cite{Bean_08af}. In parallel, \cite{Valiviita_08f} the authors
have considered models with an interacting DE
component and a constant equation of state; they
found that perturbations were unstable for
couplings proportional to the DM energy density;
these models exhibited extremely rapid growth of DE fluctuations
on superhorizon scales in the early Universe.
While this result would appear to rule out all couplings
of the above form and with constant $\omega_d$, the
explicit examples in \cite{Valiviita_08f} did not
include models where the interaction rate was
proportional to the density of DE or the DE EoS
varied with time that have been shown to have
stable solutions \cite{He_09af,Xu_11f,Corasaniti_08f,Jackson_09f}.
Also, the results of \cite{Valiviita_08f} do
not rule out models with dynamical DE or DM or  models
where the coupling depends on elementary fields
\cite{Costa_2014b,Micheletti_2009a}, so the stability of
the solutions have to be considered on the remaining cases.

In the above phenomenological gauge-invariant
linear perturbation equations, $C_a^2 =\omega_d <0$.
However, it is not clear what expression should
we have for $C_e^2$. In \cite{Valiviita_08f} it
has been argued in favor of $C_e^2=1$. This is
correct for scalar fields, but it is not
obvious for other cases, especially for a fluid
with a constant equation of state. From the stability
point of view, the most dangerous possibility
is $C_e^2=1\not = C_a^2=\omega_d<0$ since
the term in eq.~(\ref{dvelocity}) can lead to a run away
solution when the constant DE EoS is $\omega_d\simeq -1$.
Hereafter, in eq.~(\ref{dvelocity})  we will assume  $C_e^2=1$,
$C_a^2=\omega_d$.  Using the gauge-invariant quantity
$\zeta=\phi-\mathcal{H}\delta \tau$ and letting
$\zeta_c =\zeta_d=\zeta$, we obtain the adiabatic
initial condition
\begin{equation}
\frac{D_c}{1-\xi_1-\xi_2/r}=\frac{D_d}{1+\omega_d+\xi_1r+\xi_2} .
\end{equation}
The curvature perturbation given in eq.~(\ref{quad1}) can be computed
using the CMBFAST code \cite{Seljak_99f}. First, let us consider the interaction
proportional to the DM energy density, $\xi_2=0$, with
an DE EoS verifying $\omega_d\neq -1$. If
$\omega_d>-1$ is constant, we observe that $\xi_1r$
exhibits a scaling behavior, which remains constant
both at early and late times.
This behavior is not changed when $\xi_2\ne 0$.

The scaling behavior of $\xi_1r$ influences the
curvature perturbation $\Phi$. When $\omega_d>-1$ and
$\xi_1\neq 0$,  $\Phi$ blows up, what agrees
with the result obtained in \cite{Valiviita_08f}.
The instability starts at an earlier time
when $\omega_d$ approaches $-1$ from above and it
happens regardless of the value of $\xi_2$.
Let us now demonstrate that this instability disappears
when the constant EoS is $\omega_d<-1$. The study can be
made analytic if in eq.~(\ref{dvelocity}) we neglect
all contributions except those terms that give rise to
the instability. The approximate equations are
\begin{eqnarray}
&&D_d'\approx(-1+\omega_d+\xi_1r)3\mathcal{H}D_d-9\mathcal{H}^2
(1-\omega_d)(1+\frac{\xi_1r+\xi_2}{1+\omega_d})\frac{U_d}{k} \, , \nonumber\\
&&U_d'\approx2\left[1+\frac{3}{1+\omega_d}(\xi_1r+\xi_2)\right]
\mathcal{H}U_d+kD_d.\label{approx}
\end{eqnarray}
If $\xi_1\neq0$ and $\xi_2=0$, assuming that $\xi_1r\approx-\omega_d$,
we can simplify the above equations to obtain
\begin{eqnarray}
D_d'\approx
-3\mathcal{H}D_d-9\mathcal{H}^2\frac{1-\omega_d}
{1+\omega_d}\frac{U_d}{k},\hspace{1cm} U_d'\approx
2\frac{1-2\omega_d}{1+\omega_d}\mathcal{H}U_d+kD_d\quad .
\end{eqnarray}
A second order differential equation for $D_d$  is
\begin{equation}
D_d''\approx \left(2\frac{\mathcal{H}'}{\mathcal{H}}-\frac{1+7\omega_d}
{1+\omega_d}\mathcal{H}\right)D_d'+3(\mathcal{H}'-\mathcal{H}^2)D_d .\label{secondorder}
\end{equation}
In the radiation dominated period, we have
$\mathcal{H}\sim\tau^{-1},\mathcal{H}'\sim-\tau^{-2},
(\mathcal{H}'/\mathcal{H})\sim-\tau^{-1}$
and eq.~(\ref{secondorder}) can be approximated as
$ D_d''\approx-3\frac{1+3\omega_d}{1+\omega_d}\frac{D_d'}{\tau}-\frac{6}{\tau^2}D_d,$
whose solution is
$ D_d\approx C_1\tau^{r_1}+C_2\tau^{r_2}, $
where
$ r_1=-\frac{1+4\omega_d-\sqrt{-5-4\omega_d+10\omega_d^2}}{1+\omega_d},
r_2=-\frac{1+4\omega_d+\sqrt{-5-4\omega_d+10\omega_d^2}}{1+\omega_d}.  $
It is easy to check that when
$\omega_d<-1$, both $r_1$ and $r_2$ are negative; this
results in the decay of the perturbation of
$D_d$. The solution is stable, regardless of the
value of $\xi_1$.

To conclude, we have demonstrated that when
the DE EoS is constant, $\omega_d>-1$ and the coupling is
proportional to $\rho_c$ ($\xi_1 \neq 0$) the curvature
perturbation diverges; however, the divergence does not
exist when $ \omega_d<-1$. When the interaction is
proportional to $\rho_d$ ($\xi_2\ne 0$), the solutions
of the perturbation equations are stable in both
cases, $\omega_d>-1$ and $\omega_d<-1$.
Those terms in  eq.~(\ref{approx}) that give rise
to the unstable growth discussed above now reduce to
\begin{eqnarray}
D_d'&\approx&(-1+\omega_d)3\mathcal{H}D_d-9\mathcal{H}^2(1-\omega_d)
\left(1+\frac{\xi_2}{1+\omega_d}\right)\frac{U_d}{k} , \nonumber\\
U_d'&\approx&2 \left(1+\frac{3\xi_2}{1+\omega_d}\right)\mathcal{H}U_d+kD_d .
\end{eqnarray}
We can rewrite the second order differential
equation for $D_d$ in the form
\begin{eqnarray}
&&D_d''=\left[\left(-1+3\omega_d+\frac{6\xi_2}{1+\omega_d}\right)\mathcal{H}+
2\frac{\mathcal{H}'}{\mathcal{H}}\right]D_d'\nonumber\\
&&+3(1-\omega_d)\left[\mathcal{H'}
+\mathcal{H}^2\left(-1+\frac{3\xi_2}{1+\omega_d}\right)\right]D_d ,
\end{eqnarray}
which, in the radiation dominated era, reduces to
\begin{equation}
D_d''=\left(-3+3\omega_d+\frac{6\xi_2}{1+\omega_d}\right)\frac{D_d'}{\tau}+
3(1-\omega_d)\left(-2+\frac{3\xi_2}{1+\omega_d}\right)\frac{D_d}{\tau^2} .
\end{equation}
Introducing the auxiliary quantities
$\Gamma=3\omega_d^2+\omega_d+6\xi_2-2$ and
$\Delta = 9\omega_d^4+30\omega_d^3+13\omega_d^2+(-28+12\xi_2)\omega_d+36\xi_2^2+
12\xi_2-20$ \cite{He_09af}, when $\Delta>0$ we find
\begin{eqnarray}
D_d&\sim& C_1\tau^{r_1}+C_2\tau^{r_2}\quad ,\label{solution}\\
r_1&=&\frac{1}{2}\frac{\Gamma}{1+\omega_d}+\frac{1}{2}
\frac{\sqrt{\Delta}}{1+\omega_d} ,\qquad
r_2=\frac{1}{2}\frac{\Gamma}{1+\omega_d}-\frac{1}{2}\frac{\sqrt{\Delta}}
{1+\omega_d} ,
\end{eqnarray}
while, for $\Delta<0$, it becomes
\begin{equation}
D_d\sim C_1\tau^{\frac{1}{2}\frac{\Gamma}{1+\omega_d}}
\cos\frac{1}{2}\frac{\sqrt{\mid\Delta\mid}}{1+\omega_d}\ln\tau+C_2
\tau^{\frac{1}{2}\frac{\Gamma}{1+\omega_d}} \sin
\frac{1}{2}\frac{\sqrt{\mid\Delta\mid}}{1+\omega_d}\ln\tau .\label{solution2}
\end{equation}

It has been shown in \cite{He_09af} that $\Delta$ can be
positive only in the vicinity of $\omega_d=-1$. When
$\xi_2\ll 1$ then $\Delta\ll 1$. The singularity occurs when
$\omega_d=-1$ since it will lead to the divergence in $r_1$ that translates into a divergence in the density perturbation solution
eq.~(\ref{solution}). When $\omega_d>-1$ and $\Delta>0$,
the blow-up in the density perturbation can also
occur since $\Gamma/2(1+\omega_d)$ is also positive.
But when $\omega_d$ grows further above $-1$, $\Delta$
will become negative and so does $\Gamma/2(1+\omega_d)$,
which will lead to the convergent result  of
eq.~(\ref{solution2}). When $\omega_d<-1$, $\Gamma/2(1+\omega_d)$
is always negative, the density perturbation will
decay even when $\omega_d$ is close to $-1$ from below
and $\Delta$ is small and positive.

In summary, when the interaction
$\xi_2 \neq 0$, the system is stable
for any constant $\omega_d<-1$. For $\omega_d>-1$,
when the coupling is $\xi_2\ll 1$, in the range of values
of $\omega_d$ compatible with observations, the instability is
also avoided.  However, the system could become unstable
in the observationally allowed range $\omega_d>-1$
when the interaction parameter is $\xi_2\sim 1$.

In \cite{He_09af},  the case when
the interaction kernel is $Q=\xi_1\rho_{m}+\xi_2\rho_d$ has been studied. It has been found that when
$\omega_d>-1$ is constant and $\xi_1\ne 0$ the
instability occurs in agreement with the results
of \cite{Valiviita_08f}. For phantom DE,
constant $\omega_d<-1$, the perturbation is
stable regardless of the value of the coupling.
These conclusions were confirmed by \cite{Jimenez_2003a}.

When the time dependence of the DE EoS is
of the Chevallier-Polarski-Linder type
\cite{Chevallier_01f,Linder_03f},  the stability of
the linear perturbation have also been studied \cite{Xu_11f}
who found that the evolution of density perturbations
at linear order is stable. Similar stability analysis
for interacting scalar fields have been described by \cite{Corasaniti_08f}.
For field theory models (such as canonical bosons
and fermions as DE and DM, respectively)
the perturbations are also well defined, at least for
a small range of parameters \cite{Costa_2014b}.

\subsection{Cosmic microwave background temperature anisotropies.}
\label{sec:cmb_perturbation_theory}

The formalism developed in Sec.~\ref{sec:sec4.1}
can be used to study the evolution of matter and radiation perturbations
that can then be tested against observations \cite{Olivares_08bf}.
CMB temperature anisotropies provide a wealth of information
that overshadows observables of the Hubble expansion.
SNIa data are rather insensitive
to the coupling between dark sectors
\cite{Olivares_06f,Olivares_08af}, while
the integrated Sachs-Wolfe (ISW) component is
a more sensitive probe \cite{Zimdahl_05f,Olivares_08bf}.
CMB observation are expected to break the degeneracy between the
coupling and other cosmological parameters, such
as the DE EoS parameter $\omega_d$ and DM
abundance, providing tighter constraints on the
interaction within the dark sector.
Many interacting models have been studied in the literature.
See for example \cite{Amendola_03bf,
Wang_07f,Guo_07f,Feng_07f,Feng_08f,Xia_09f,
Martinelli_10f,Valiviita_10f,Fabris_10f,Honorez_10f,
Xu_11f,Xu_12f,Salvatelli_13f,Xia_13f,
Costa_14f,Salvatelli_14f} among others.
In this section we will discuss the effect
of the interaction in the pattern of CMB temperature
anisotropies. The formalism reviewed here is mainly based
on \cite{He_09bf} and \cite{He_11f}.

\begin{figure*}
\centering
\epsfxsize=0.7\textwidth\epsfbox{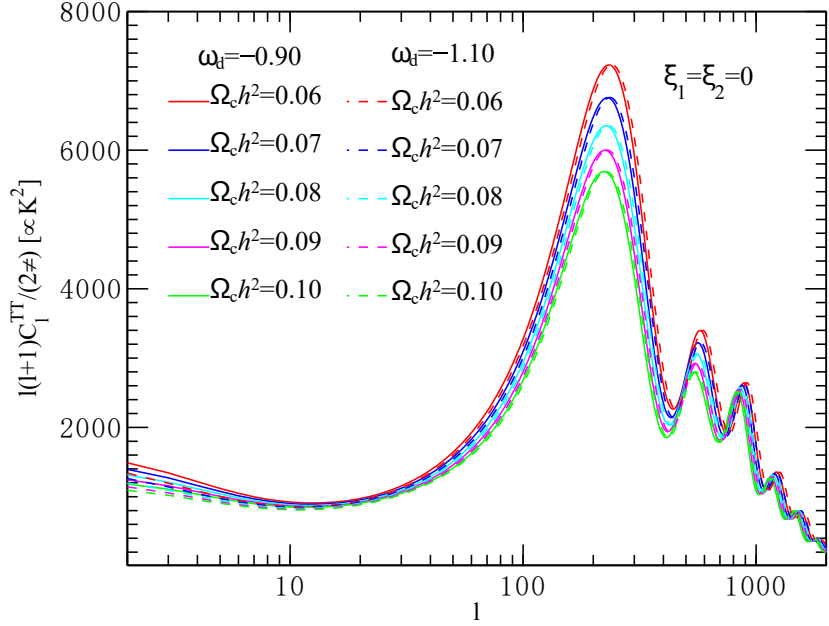}
\caption{Variation of the CMB radiation power spectrum with
cosmological parameters.}
\label{omegac}
\end{figure*}

The temperature anisotropy power spectrum can be
calculated by \cite{sachs_67f}
\begin{equation}
C_\ell=4\pi \int
\frac{dk}{k}\mathcal{P}_{\chi}(k)\mid \Delta_\ell(k,
\tau_0) \mid^2,
\end{equation}
where $\Delta_\ell$ gives the transfer function for
each $\ell$, $\mathcal{P}_{\chi}$ is the primordial
power spectrum and $\tau_0$ is the conformal time
at present. On large scales the transfer functions are
\begin{equation}
\Delta_\ell(k,\tau_0)=\Delta_\ell^{SW}(k)+\Delta_\ell^{ISW}(k),
\end{equation}
where $\Delta_\ell^{SW}(k)$ is the contribution from
the last scattering surface given by the ordinary
Sachs-Wolfe (SW) effect and $\Delta_\ell^{ISW}(k)$
is the contribution due to the change of the
gravitational potential when photons passing
through the Universe on their way to the observer
\cite{sachs_67f}. The ISW contribution can be written as
\begin{equation}
\Delta_\ell^{ISW}= \int_{\tau_i}^{\tau_0} d\tau
j_\ell(k[\tau_0-\tau])e^{\kappa(\tau_0)-\kappa(\tau)}[\Psi'-\Phi'],
\end{equation}
where $j_\ell$ is the spherical Bessel function and
$\kappa$ is the optical depth due to Thompson
scattering. From Einstein equations, we obtain,
\begin{eqnarray}
\Psi' - \Phi' &=& 2\mathcal{H}\left[ \Phi + 4\pi
Ga^2 \sum_{i} U^{i}\rho^i/(\mathcal{H}k) +
\mathcal{T} \right ]- \mathcal{T}'
\end{eqnarray}
where
\begin{eqnarray}
\Phi' = -\mathcal{H}\Phi - \mathcal{H}\mathcal{T}
- 4\pi Ga^2 \sum_{i} U^{i}\rho^i/k, \quad
\mathcal{T} = \frac{8 \pi G a^2}{k^2} \left \{
p^{\gamma}\Pi^{\gamma}+ p^{\nu} \Pi ^{\nu}\right
\}\nonumber
\end{eqnarray}
and $\Pi$  is the anisotropic stress of
relativistic components which can be neglected in
the following discussion.

\begin{figure*}
\centering
\includegraphics[width=3.0in,height=2in]{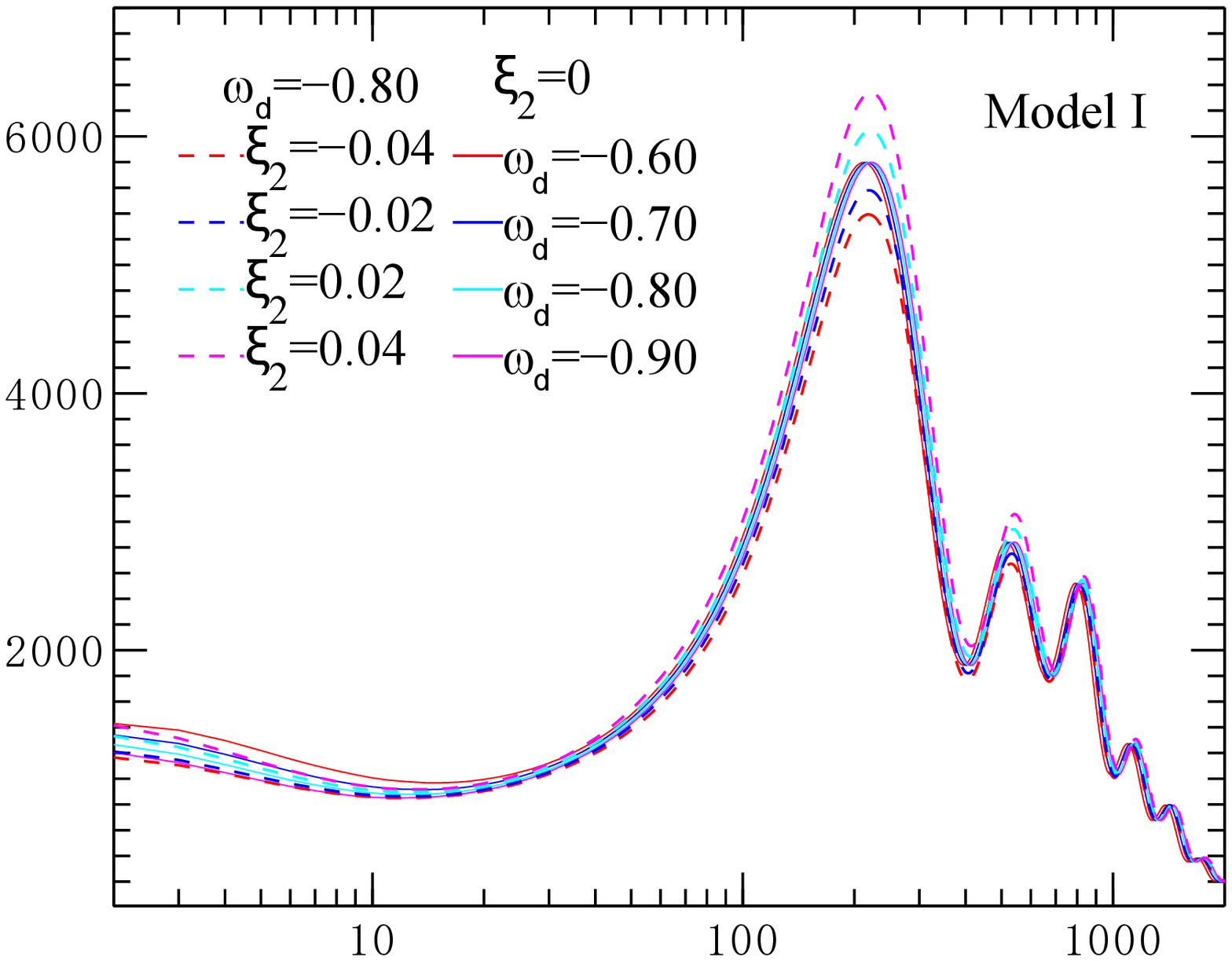}
\includegraphics[width=3.0in,height=2in]{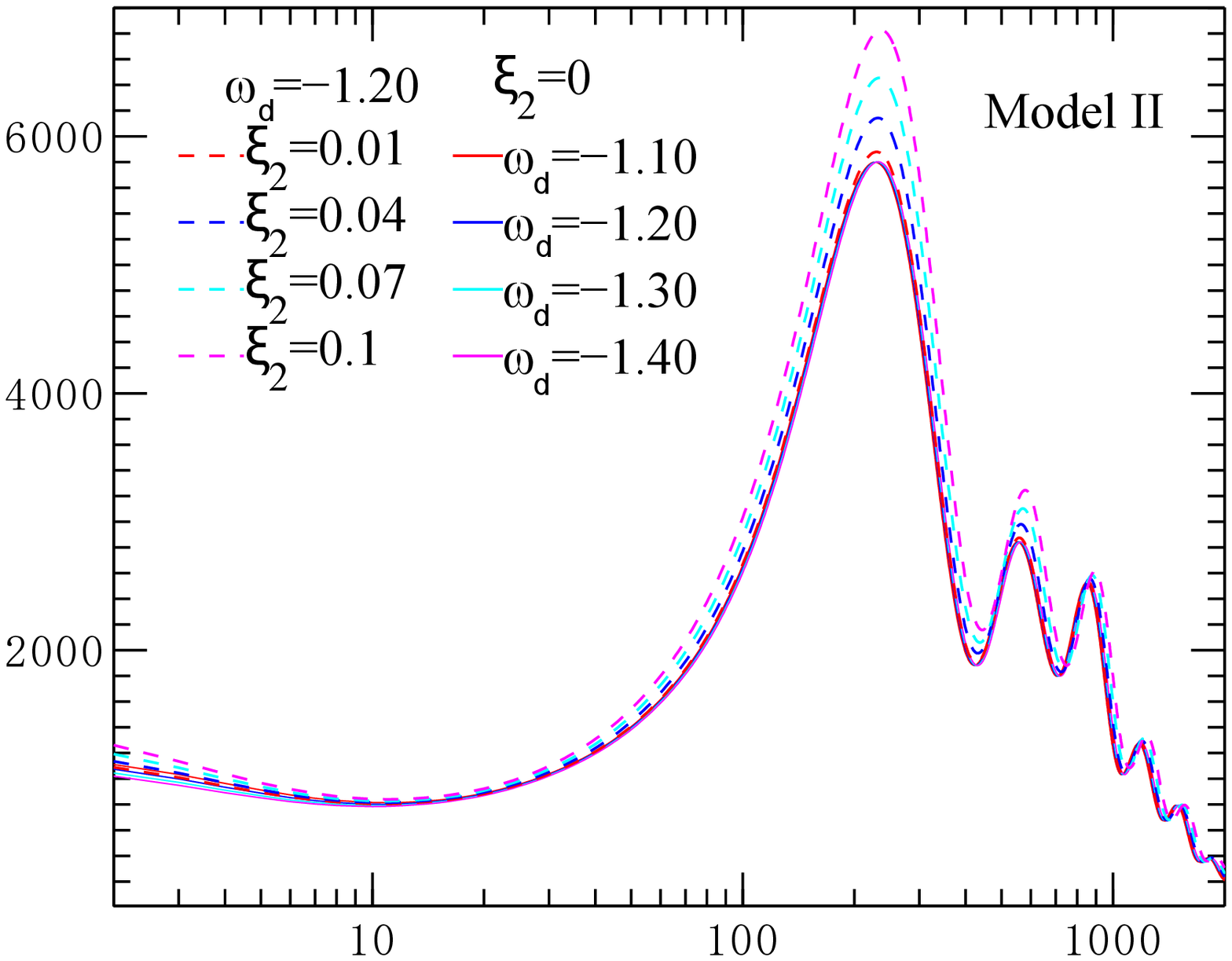}
\includegraphics[width=3.0in,height=2in]{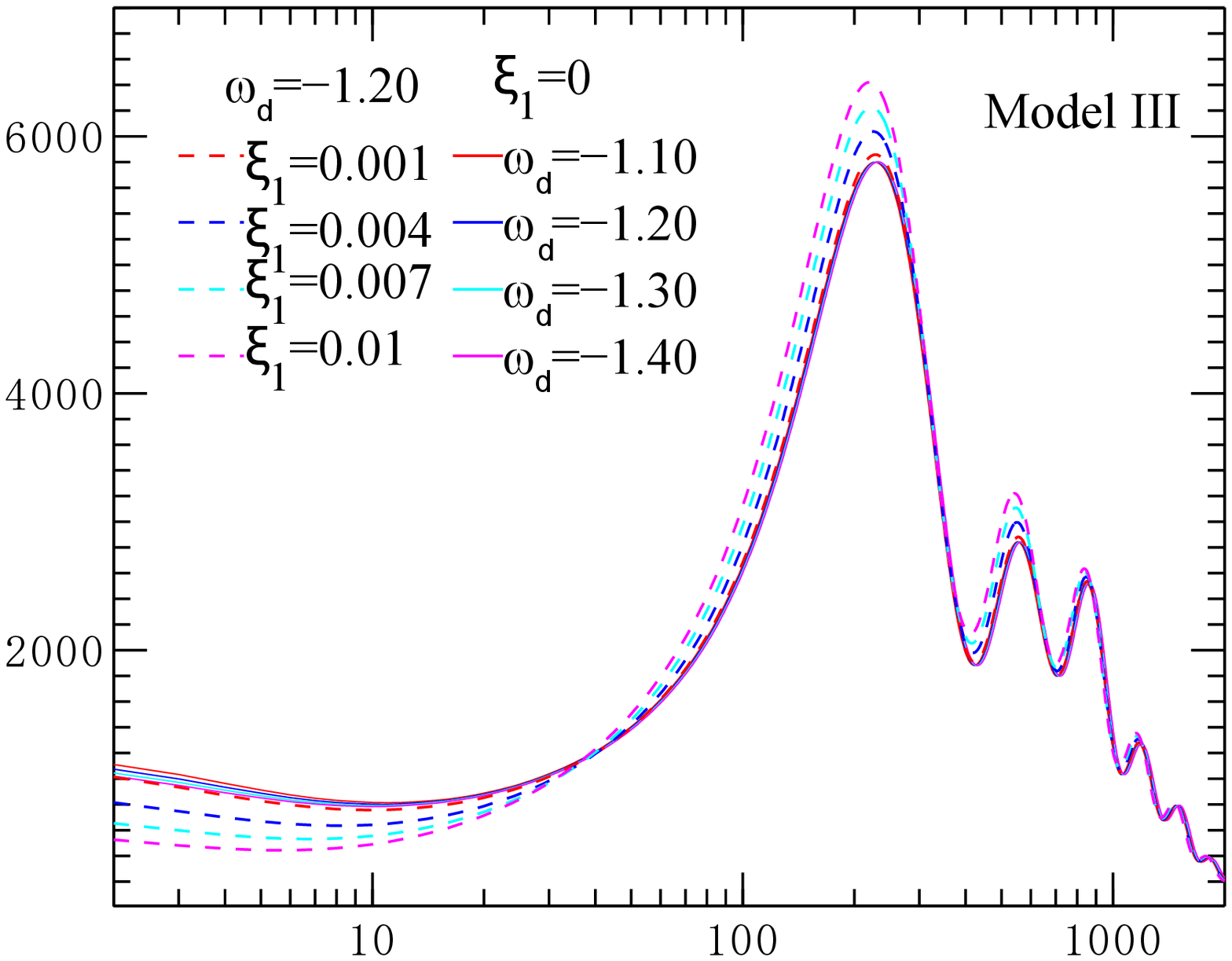}
\includegraphics[width=3.0in,height=2in]{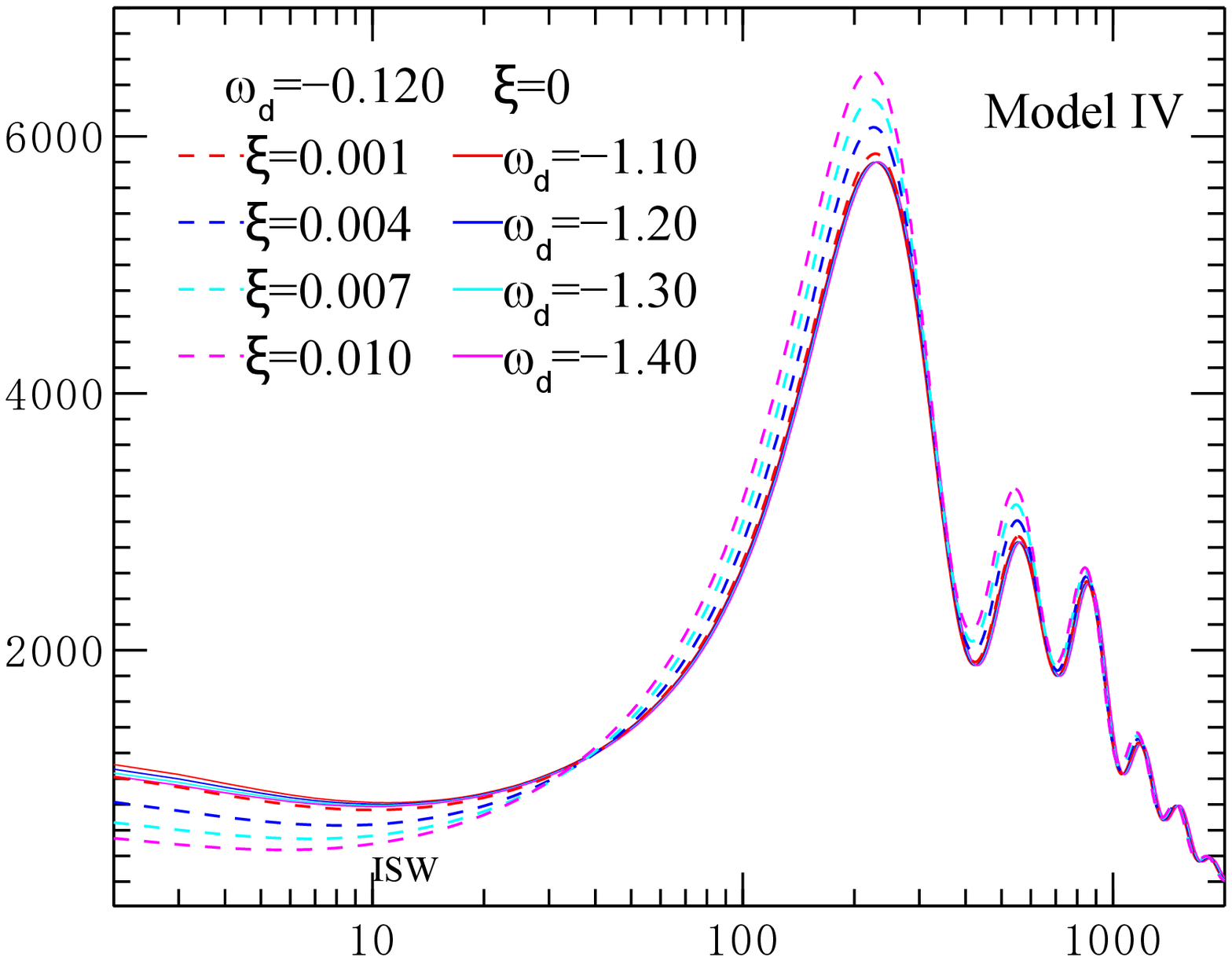}
\caption{The CMB TT power spectrum for the
phenomenological kernels given in Table~\ref{table:models}. Solid lines are for models without interactions between dark sectors. Dashed lines are for different strengths of interacting models. }\label{cmb}
\end{figure*}

In Fig.~\ref{omegac} we illustrate the effect of
the DE EoS parameter $\omega_d$ and the fraction
of DM, parametrized as the commonly used variable
$\omega_{cdm}=\Omega_ch^2$, on the CMB radiation
power spectrum for a cosmological model without
interaction. We shall see that the interaction
also alters the height and location of the
acoustic peaks and draughts, and this effect
needs to be distinguished from that of other parameters.
For comparison, in Fig.~\ref{cmb} we plot the CMB
power spectrum of temperature anisotropies for
different interaction models. Let us first
concentrate on models with constant
$\omega_d$ and a constant speed of sound.
For simplicity, we limit our study to the three
commonly studied phenomenological interaction
kernels of Table~\ref{table:models}). In Fig~\ref{cmb}
we fixed the energy densities to the values
given in Table~\ref{table:parameters}. Solid lines
correspond to variations in the EoS parameter.
Fig.~\ref{cmb}a shows that the power at low-$\ell$
increases with increasing value of $\omega_d$, but
the effect on the acoustic peaks is negligible.
In Fig.~\ref{cmb}b,c,d the variation is in the same
direction but the effect of changing $\omega_d$ is
smaller than when $\omega_d>-1$.

The effect of the interaction, represented in
Fig~\ref{cmb} by dashed lines, is more evident.
The interaction changes the spectrum at low multipoles
through its effect on the gravitational potentials
and the ISW effect. When the coupling increases, the
low-$\ell$ spectrum is further suppressed. When the
interaction between dark sectors is proportional
to the DM or total dark sector energy density,
the low-$\ell$ spectrum is more sensitive to the
change of the coupling than that of the DE EoS.
Both effects, the change on the low multipoles due
to the ISW effect and  on the acoustic peaks
are important since
they could help to break the degeneracy between
the interaction between dark sectors and the DE
EoS and other cosmological parameters as illustrated
by Fig~\ref{omegac}.

The ISW effect has two components: early and
late time effects. The early ISW effect occurs
when the gravitational potentials evolve in time
since matter--radiation equality to the moment
when the radiation is no longer dynamically
significant. Its contribution is largest around
the first acoustic peak and below \cite{hu_95f}.
The late time ISW effect arises when
the DE becomes important and the gravitational
potentials decay. When a photon
passes through a decaying potential well, it will
have a net gain in energy. Consequently, the ISW effect
can be used to probe the dynamical effect of the DE.
This component has a significant contribution to the
large scale CMB radiation power spectrum. Since
galaxies trace the large scale gravitational field,
cross-correlating matter templates constructed from
galaxy catalogs with CMB temperature anisotropy maps
can be used to isolate the ISW contribution
\cite{Crittenden_96f,Cooray_02f} and test the effect of the interaction
\cite{Olivares_08bf,Schaefer_08f,Xia_09f,mainini_12f}.

In the absence of interaction, the late ISW effect depends
on the EoS parameter $\omega_d$ and sound speed $C_e^2$.
For constant $C_e^2\lsim 1$ and $\omega_d>-1$, the spectrum
of CMB temperature anisotropies on large scales are larger than
in the concordance model \cite{Weller_03f,Bean_04f}.
Increasing $C_e^2$ leads to further
suppression of DM perturbations, increasing the contribution to the ISW
effect \cite{Bean_04f}. However, when $\omega_d<-1$ the
effect is the contrary, the contribution
to the ISW effect increases as the sound speed of
DE decreases \cite{Weller_03f}. The interplay
between perturbations in the DE and DM and the
ISW effect is very subtle and can not be disentangled
easily from the radiation power spectrum. A more direct
probe is to cross-correlate the late ISW effect to its
source term, the time variation of the gravitational
potential \cite{Weller_03f}.

Both the power spectrum due to the late ISW effect and the
cross-correlation with a matter template constructed from a galaxy
catalog can be expressed in terms of quadratures \cite{Crittenden_96f}.
The auto- and cross-correlation power spectra are given by
\begin{eqnarray}
\label{eq:isw_auto}
C_l^{gg}  =  4\pi \int \frac{dk}{k}
\mathcal{P}_{\chi}(k)I_l^{g}(k)I_l^g(k)\\
\label{eq:isw_cross}
C_l^{gI}  =  4\pi \int \frac{dk}{k}
\mathcal{P}_{\chi}(k)I_l^{g}(k)\Delta_l^{ISW}(k),
\end{eqnarray}
where the projected density of galaxies is
given by $ I_l^{g}(k)= \int dz
b_g(z) n(z)(D_{c}+D_{b})j_l[k\chi(z)], $ where $n(z)$ is the number density of objects at a given redshift and $D_c(z), D_b(z)$ are the growth function of CDM and baryons respectively.  Here
$b_g(z)$ is the galaxy bias and $\chi(z)$ is the
conformal distance, or equivalently the look-back
time from the observer at redshift $z=0$, $
\chi(z) = \int_0^z \frac{dz'}{H(z)} =
\int_{\tau_i(z)}^{\tau_0}d\tau = \tau_0 -
\tau_i(z)$ (see Sec.~\ref{sec:look-back}).
We assume $b(z)\sim 1$ for simplicity
and adopt the redshift distribution of the
form \cite{Lesgourgues_08f}
$n(z) = \frac{3}{2} \frac{z^2}{z_0^3}{\rm
exp}\left[ -(\frac{z}{z_0})^{3/2}\right], $
the normalization constant is fixed by setting $\int n(z) dz=1.$ This expression has a maximum near the median redshift $z_m=1.4z_0$.
For illustrative purpose, we choose $z_m=0.1$ and $z_m=0.4$.
The first value would correspond to a shallow survey like 2MASS
\cite{Bilicki_14f} while the second would correspond to the SDSS
photo-$z$ galaxy sample \cite{Way_09f} (see Sec.~\ref{sec:sec6} for details).

\begin{figure}
\begin{center}
  \begin{tabular}{cc}
\includegraphics[width=3.6in,height=3.2in]{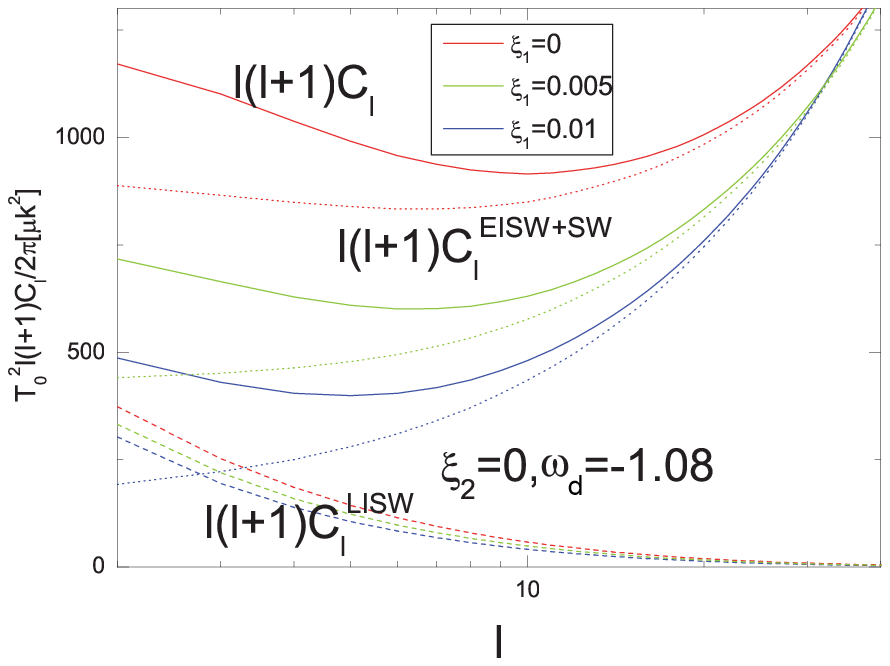}& \hspace*{-1.2cm}
\includegraphics[width=3.6in,height=3.2in]{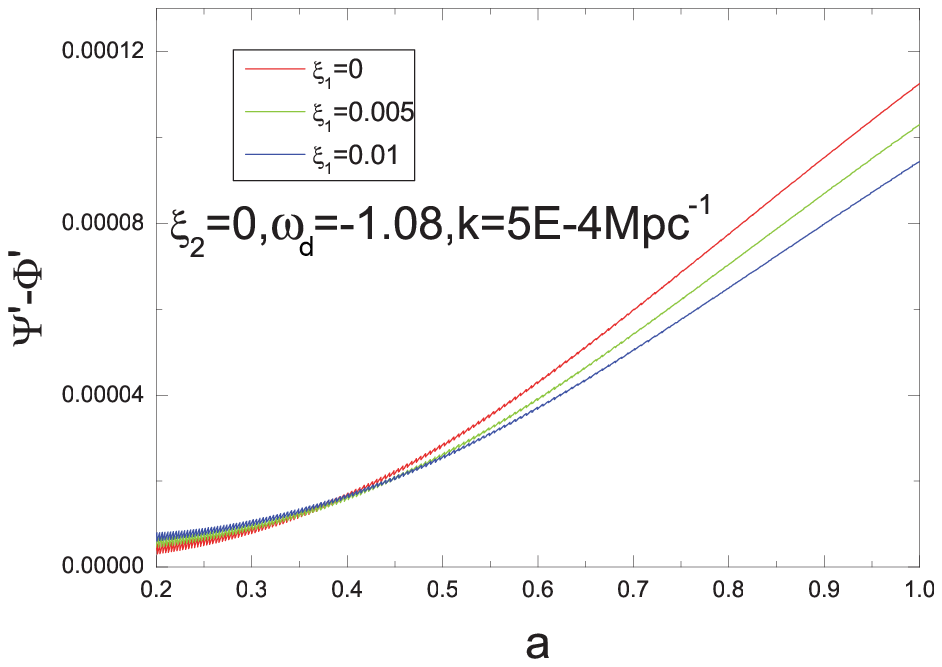}\nonumber \\
    (a)&(b)\nonumber \\
\includegraphics[width=3.6in,height=3.2in]{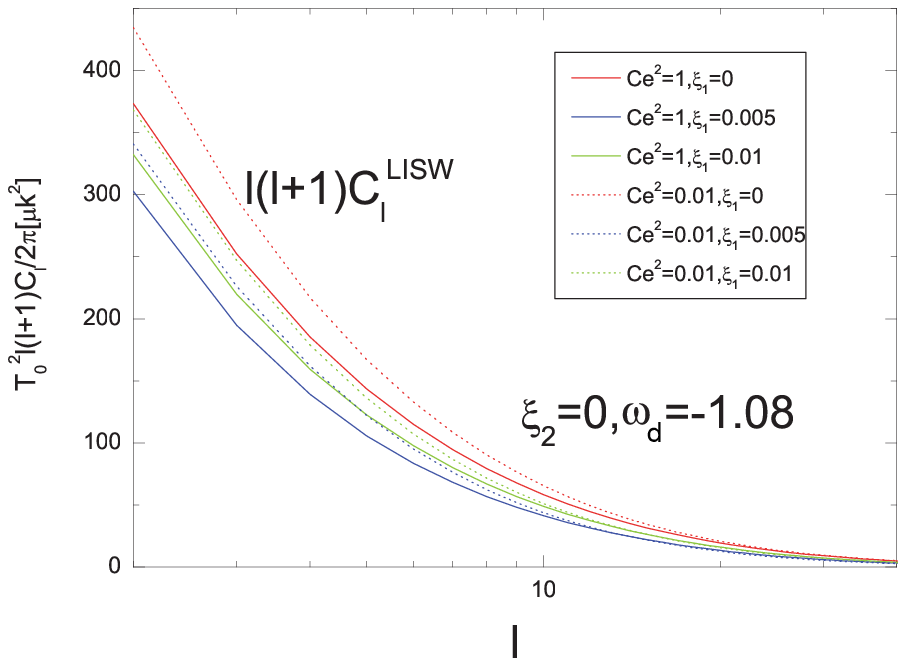}& \hspace*{-1.2cm}
\includegraphics[width=3.6in,height=3.2in]{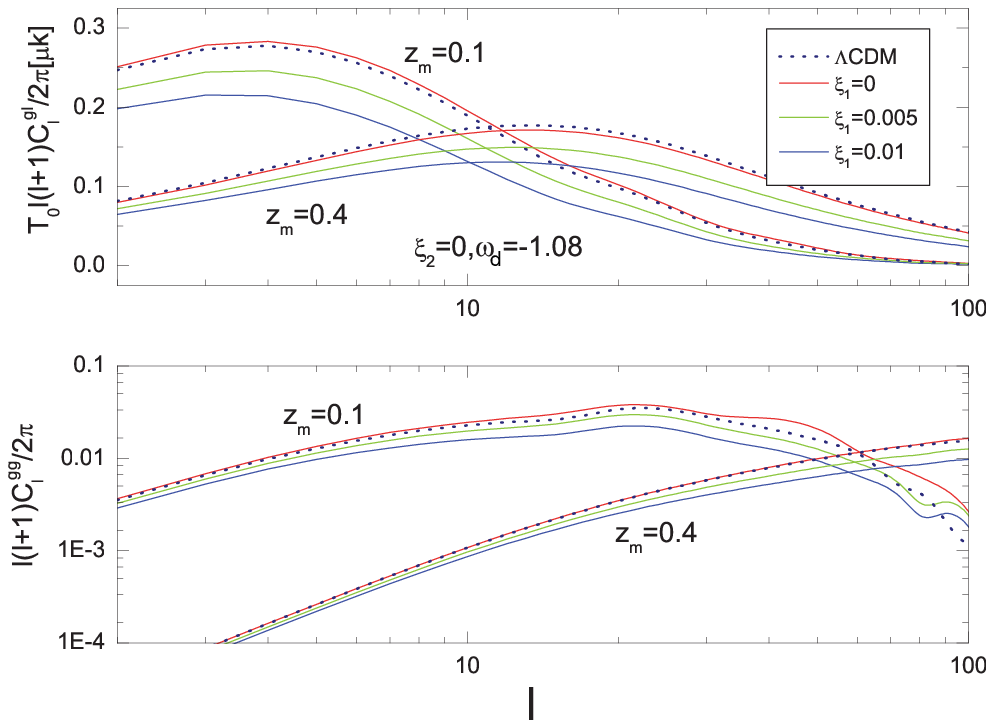}
\nonumber \\
    (c)&(d)\nonumber
  \end{tabular}
\end{center}
\caption{Different magnitudes for phenomenological fluid Model III.
(a) CMB radiation power spectra for small $\ell$.
(b) Evolution of the gravitational potentials. (c) Evolution
of the Integrated Sachs-Wolfe component. (d) Upper panel
ISW and galaxy cross-spectra as given by eq.~(\ref{eq:isw_cross}); the
lower panel shows the galaxy power spectra of eq.~(\ref{eq:isw_auto}).}
\label{figthree}
\end{figure}

\begin{figure}
\begin{center}
\begin{tabular}{cc}
\includegraphics[width=3.6in,height=3.2in]{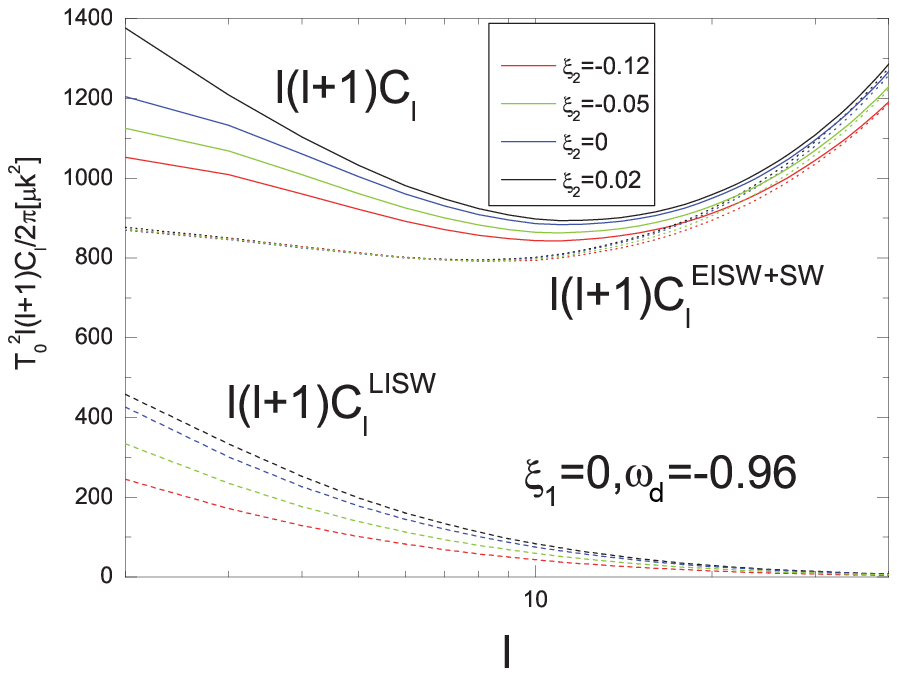}& \hspace*{-1.2cm}
\includegraphics[width=3.6in,height=3.2in]{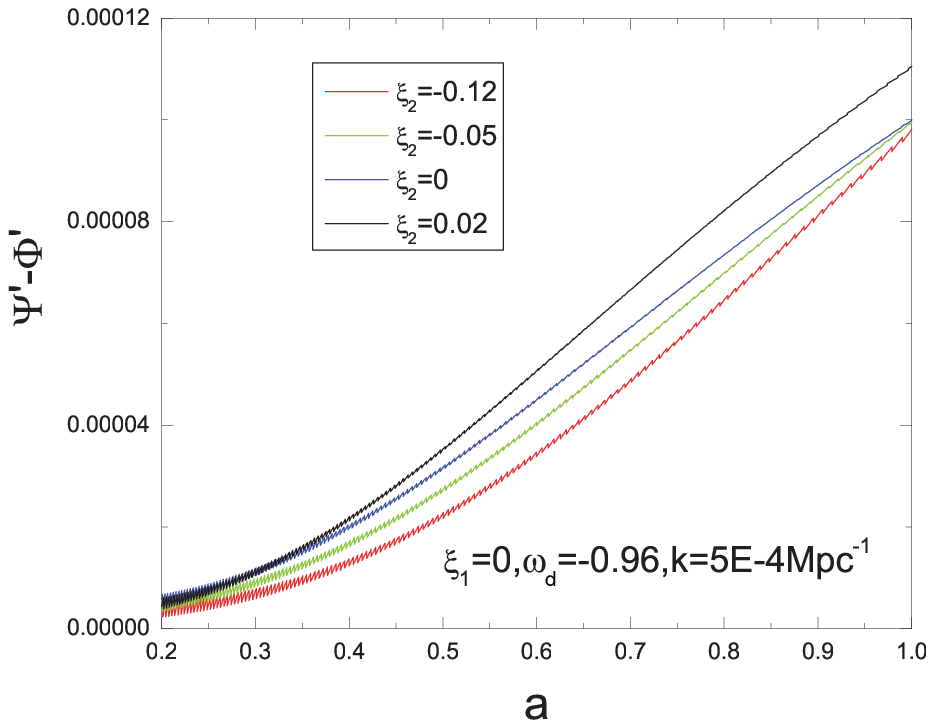}\nonumber \\
    (a)&(b)\nonumber \\
\includegraphics[width=3.6in,height=3.2in]{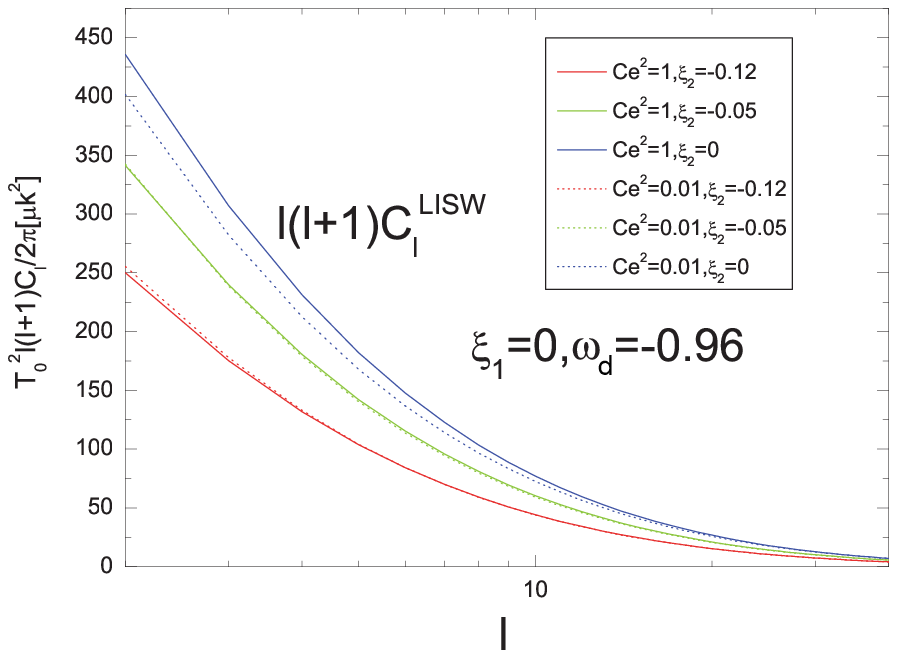}& \hspace*{-1.2cm}
\includegraphics[width=3.6in,height=3.2in]{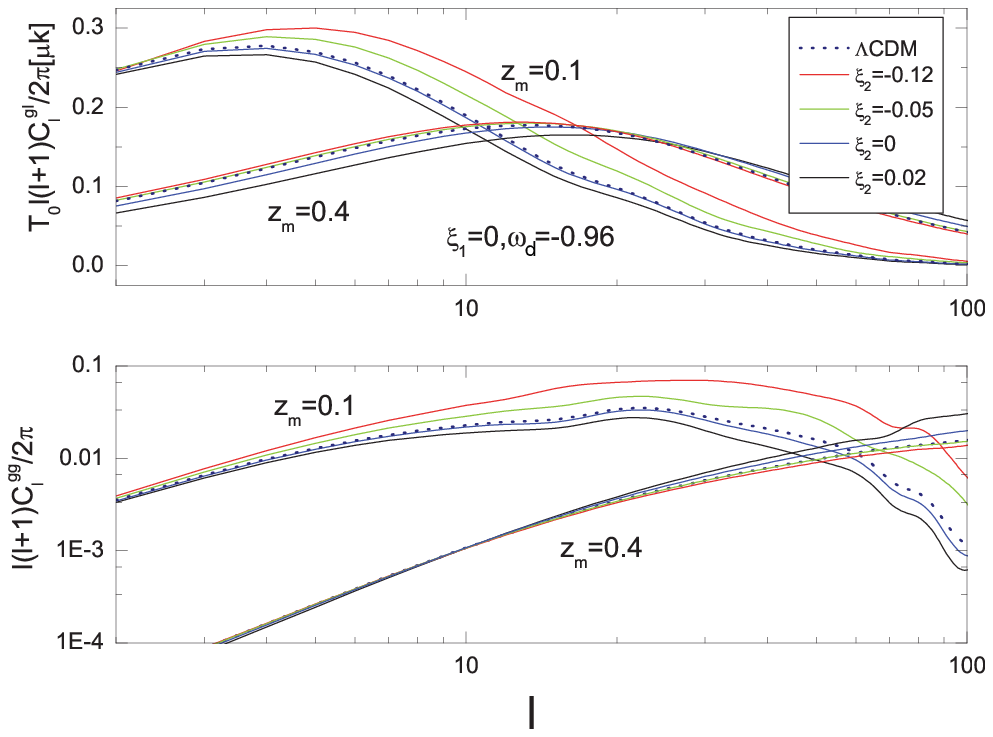}
\nonumber \\
    (c)&(d)\nonumber
  \end{tabular}
\end{center}
\caption{Same as in Fig~\ref{figthree} but for Model I.}\label{figone5}
\end{figure}

The late ISW effect is a promising tool to
measure the EoS and sound speed of DE. Let us now
analyze if it can provide useful evidence of the interaction.
First, we will analyze the case of $\xi_1\ne 0,\,\xi_2=0$
with constant EoS parameter $\omega_d<-1$
(Model III in Table~\ref{table:models})
and sound speed $C_e^2=1$. The radiation power spectrum is
given in Fig~\ref{figthree}$a$. In the figure, the lower set
of dashed lines correspond to the late ISW (LISW) and the
upper set of dotted lines  correspond to the early ISW (EISW)
plus SW effect. Solid lines correspond to the total anisotropy.
For this coupling the SW+EISW effect shows a larger
variation than the LISW. The results show that larger the coupling bigger
suppression of CMB anisotropies. This behavior can
be expected in the light of the time evolution of
the potentials given in Fig~\ref{figthree}$b$ that demonstrates
that the potentials evolve slower when the interaction is larger. In
Fig~\ref{figthree}$c$ we illustrate the variation of the
radiation power spectrum with sound speed. The figure shows
that a smaller DE sound speed increases the LISW effect.
The effect of the interaction in Models IV is similar to model III
and is not explicitly shown.

The results for Model I (see Table~\ref{table:models}) are presented
in Fig~\ref{figone5}. Lines in (a) and (c) follow the same
conventions than Fig.~\ref{figthree}a,c.
For this model the interaction does not modify the SW+EISW
effect but it changes the LISW significantly.
In this case, a positive coupling increases the
amplitude of the LISW and a negative coupling decreases
it compared with the uncoupled case. The behavior follows
the evolution of the ISW source term shown in
Fig~\ref{figone5}b that demonstrates how the change of the gravitational
potential increases with increasing value of the coupling.
In Fig~\ref{figone5}$c$ we present the effect of
the sound speed. When DM and DE
interact, a variation on the sound speed from
$C_e^2=1$ to $C_e^2=0.01$ does not lead to significant
variations on the LISW. Finally, in Model II with $C_e^2=1$
the effect of the interaction produces the same behavior
as in Model I.

Since the influence of the interaction in the LISW effect
is relatively strong, it is interesting to discuss if the
cross-correlation with templates constructed from the
large scale matter distribution can measure how the potentials evolve with time.
Progress in CMB and LSS surveys have enabled
detections of the ISW-LSS cross correlation at
$\sim 3\sigma$ level \cite{Boughn_04f,Nolta_04f,
Afshordi_04f,Cabre_07f}. The signal-to-noise ratio (S/N)
can be further improved by a factor of a
few for future all sky LSS surveys (See Secs.~\ref{sec:sec6}
and \ref{sec:sec8}).  Cross-correlation of data
has several advantages. The main advantage is that the primary CMB does
not correlate with the LSS and does not bias the ISW
measurement, it only contributes to the error
bar. Then, using several traces of the matter distribution
at different redshifts,  it is possible
to reconstruct the redshift evolution of the
gravitational potential.

Although the ISW-LSS cross correlation is  potentially powerful probe of the interaction between dark sectors, due to the low S/N of the current ISW-LSS
measurements and the complexities of the theoretical
interpretation (e.g. galaxy bias), we will not confront the
model predictions against the existing ISW-LSS cross-correlation
measurements.  Instead, we will just calculate the expected
cross-correlation signal between the ISW effect and galaxies
given by eq.~(\ref{eq:isw_cross}) for some representative cases to show how the cross-correlation is modified due to the interaction.
The results for Model III are shown in Fig.~\ref{figthree}d and
Fig.~\ref{figone5}d. The top panels show the cross-correlation
with two galactic templates of different depth and the
low panels show the auto correlation power of the
LISW traced by those templates. In both cases,
a positive coupling decreases the auto-correlation and
the cross-correlation spectra compared with the $\Lambda$CDM
model while negative couplings increase them.
The effect is more important for the shallow galaxy survey
($z_m=0.1$) than for the deep one ($z_m=0.4$).


Finally, the study of models with a variable DE EoS parameter
have been discussed in \cite{Xu_11f}.
DE perturbation are not longer negligible and
contribute to the ISW effect. The effect on models
based on a field theory description have also been
discussed in \cite{Pu_14f}.

\subsection{Matter density perturbations.}

In addition to CMB temperature anisotropies, the
interaction modifies the evolution of matter density
perturbations.  If DE is not a cosmological constant
and couples to DM, it must be dynamical and
fluctuate in space and in time. In these models
structure formation would be different than
in the concordance model \cite{Amendola_06f,Olivares_06f,
Tsujikawa_07f,Gong_08f,Ballesteros_08f,Tsagas_08f},
changing the growth index
\cite{Manera_06f,He_09cf,Caldera_09f}.
Furthermore, the collapse and subsequent
dynamical equilibrium of the resulting objects
such as clusters is modified compared to the
concordance model
\cite{Bertolami_07f,Abdalla_09f,Abdalla_10f}.
Comparing the naive virial masses of a large
sample of clusters with their masses estimated from
X-ray and weak lensing data, a small positive
coupling has been tightly constrained
\cite{Abdalla_09f,Abdalla_10f}, which agrees
with the results given in
\cite{He_09bf,He_11f} from CMB.

In this subsection we will analyze the effect of the
interaction on the growth of DM density perturbations.
First we restrict our attention to models with constant
sound speed and constant DE EoS parameter. We will
show that the effect of the interaction is larger than
that of the existence of DE perturbations, providing
another test of interacting models.

Let us first derive the equations of evolution of DM density
perturbations. From eq.~(\ref{invariantQ}) and defining
\begin{eqnarray}
\delta\rho^I_{\lambda}&=&\delta \rho_{\lambda}-\rho'_{\lambda}\frac{v_{\lambda}+B}{k},\quad
\delta p^I_{\lambda}=\delta p_{\lambda}-p'_{\lambda}\frac{v_{\lambda}+B}{k},\nonumber\\
\Delta_{\lambda}&=&\delta_{\lambda}-\frac{\rho'_{\lambda}}{\rho_{\lambda}}
\frac{v_{\lambda}+B}{k},\quad V_{\lambda}=v_{\lambda}-\frac{E'}{2k},\label{gaugeinvariant}
\end{eqnarray}
the perturbed Einstein equations eq.~(\ref{PerEinstein}) become
\begin{eqnarray}
&&\Phi=4\pi G\frac{a^2}{k^2}\sum_{\lambda}\left(\Delta_{\lambda}
      +\frac{a^2Q_{\lambda}^0}{\rho_{\lambda}}
       \frac{V_{\lambda}}{k}\right)\rho_{\lambda},\nonumber\\
&&k\left(\mathcal{H}\Psi-\Phi'\right)=4\pi Ga^2
   \sum_{\lambda}\left(\rho_{\lambda}+p_{\lambda}\right)V_{\lambda},\quad
   \Psi=-\Phi, \label{Einstein}
\end{eqnarray}
where we have neglected the pressure perturbation of DE,
$\Pi^{i}_j=0$. The gravitational potential is consistent
with eq.~(\ref{quad1}) when we combine eq.~(\ref{gaugeinvariant})
and eq.~(\ref{invariantQ}) with eqs.~(\ref{Einstein},\ref{quad1}).

Using the gauge invariant quantities of eq.~(\ref{gaugeinvariant}),
we can obtain the linear perturbation equations for the DM,
\begin{eqnarray}
&&\Delta_c'+\left[\frac{\rho_c'}{\rho_c}\frac{V_c}{k}\right]'=
  -kV_c-3\Phi'+2\Psi\frac{a^2Q_c^0}{\rho_c}-\Delta_c\frac{a^2Q_c^0}{\rho_c}
  -\frac{\rho_c'}{\rho_c}\frac{V_c}{k}\frac{a^2Q_c^0}{\rho_c}\nonumber\\
&&+\frac{a^2\delta Q^{0I}_c}{\rho_c}+\frac{a^2Q_c^{0'}}{\rho_c\mathcal{H}}\Phi
  -\frac{a^2Q_c^0}{\rho_c}\left[\frac{\Phi}{\mathcal{H}}\right]',
  \label{linearDM1perturbation}\\
&&V_c'=-\mathcal{H}V_c+k\Psi-\frac{a^2Q_c^0}{\rho_c}V_c+
  \frac{a^2\delta Q_{pc}^{I}}{\rho_c}\label{DM2}.
\end{eqnarray}
Considering the pressure perturbation of DE
eq.~(\ref{Dpp}), we obtain the gauge invariant form of
DE perturbation equations \cite{Valiviita_08f}
\begin{eqnarray}
\Delta_d'&+&\left[\frac{\rho_d'}{\rho_d}\frac{V_d}{k}\right]'+3\mathcal{H}C_e^2(\Delta_d
    +\frac{\rho_d'}{\rho_d}\frac{V_d}{k})
    -3\mathcal{H}(C_e^2-C_a^2)\frac{\rho_d'}{\rho_d}\frac{V_d}{k} \nonumber \\
&&-3\omega_d\mathcal{H}(\Delta_d+\frac{\rho_d'}{\rho_d}\frac{V_d}{k})=
    -k(1+\omega_d)V_d-3(1+\omega_d)\Phi'+2\Psi\frac{a^2Q_d^0}{\rho_d} \nonumber\\
&&-(\Delta_d+\frac{\rho_d'}{\rho_d}\frac{V_d}{k})\frac{a^2Q_d^0}{\rho_d}
    +\frac{a^2\delta Q^{0I}_d}{\rho_d}+\frac{a^2Q_d^{0'}}{\rho_d\mathcal{H}}\Phi
    -\frac{a^2Q_d^0}{\rho_d}\left[\frac{\Phi}{\mathcal{H}}\right]',\\
V_d'&+&\mathcal{H}(1-3\omega_d)V_d=
    \frac{k}{1+\omega_d}\left[C_e^2(\Delta_d+\frac{\rho_d'}{\rho_d}\frac{V_d}{k})
    -(C_e^2-C_a^2)\frac{\rho_d'}{\rho_d}\frac{V_d}{k}\right] \nonumber \\
&&  -\frac{\omega_d'}{1+\omega_d}V_d+k\Psi-\frac{a^2Q_d^0}{\rho_d}V_d
    +\frac{a^2\delta Q_{pd}^{I}}{(1+\omega_d)\rho_d}.
\end{eqnarray}
In the subhorizon approximation $k>>aH$, the above perturbation equations reduce to
\begin{eqnarray}
\Delta_c'&=& -kV_c-\Delta_c\frac{a^2Q_c^0}{\rho_c}
   +\frac{a^2\delta Q^{0I}_c}{\rho_c},\nonumber\\
V_c'&=&-\mathcal{H}V_c+k\Psi-\frac{a^2Q_c^0}{\rho_c}V_c
   +\frac{a^2\delta Q_{pc}^I}{\rho_c};\label{DMsub}\\
\Delta_d'&=& 3\mathcal{H}(\omega_d-C_e^2)\Delta_d-k(1+\omega_d)V_d
   -\Delta_d\frac{a^2Q^{0}_d}{\rho_d}+\frac{a^2\delta Q^{0I}_d}{\rho_d},\nonumber\\
V_d'&+&\mathcal{H}(1-3\omega_d)V_d=\frac{kC_e^2}{1+\omega_d}
   \Delta_d+\frac{C_a^2}{1+\omega_d}\frac{\rho_d'}{\rho_d}V_d+k
   \Psi-\frac{\omega_d'}{1+\omega_d}V_d\nonumber\\
&&-\frac{a^2Q_d^0}{\rho_d}V_d+\frac{a^2\delta Q_{pd}^I}{(1+\omega_d)\rho_d}.\label{DEsub}
\end{eqnarray}
Eliminating $V_c$, in the subhorizon approximation $k>>aH$, we obtain the second order
equation for the DM perturbation
\begin{eqnarray}
\Delta_c^{''}&=&-(\mathcal{H}+\frac{2a^2Q_c^0}{\rho_c})\Delta_c'
+(-\Delta_c\frac{a^2Q_c^0}{\rho_c}+\frac{a^2\delta
Q^{0I}_c}{\rho_c})(\mathcal{H}+\frac{a^2Q_c^0}{\rho_c})\nonumber\\
&&-\Delta_c(\frac{a^2Q_c^0}{\rho_c})'+(\frac{a^2\delta
Q^{0I}_c}{\rho_c})'-\frac{a^2k\delta Q_{pc}^{I}}{\rho_c}-k^2\Psi.
\end{eqnarray}
Similarly for the DE perturbation we have
\begin{eqnarray}
\Delta_d^{''}&=&-3\mathcal{H}'C_e^2\Delta_d
    -(\frac{a^2Q_d^0}{\rho_d})'\Delta_d+\left\{\mathcal{H}(1-3\omega_d)
    -\frac{\omega_d}{1+\omega_d}\frac{\rho_d'}{\rho_d}+\frac{a^2Q_d^0}{\rho_d}\right \}
    \nonumber\\
&&\times\left\{-3\mathcal{H}C_e^2+3\omega_d\mathcal{H}
    -\frac{a^2Q_d^0}{\rho_d}\right\}\Delta_d\nonumber\\
&&-[\mathcal{H}+3\mathcal{H}C_e^2-6\omega_d\mathcal{H}+\frac{2a^2Q_d^0}{\rho_d}
    -\frac{\omega_d}{1+\omega_d}\frac{\rho_d'}{\rho_d}]\Delta_d'\nonumber\\
&&-k(\frac{a^2\delta Q_{pd}^{I}}{\rho_d})-k^2C_e^2\Delta_d-k^2(1+\omega_d)\Psi
+3(\omega_d'\mathcal{H}+\omega_d\mathcal{H}')\Delta_d\nonumber\\
&&+\frac{a^2\delta Q_{d}^{0I}}{\rho_d}[\mathcal{H}(1-3\omega_d)
    -\frac{\omega_d}{1+\omega_d}\frac{\rho_d'}{\rho_d}
    +\frac{a^2Q_d^0}{\rho_d}]+(\frac{a^2\delta Q_{d}^{0I}}{\rho_d})'.
\end{eqnarray}
From the perturbed Einstein equations eq.~(\ref{Einstein}) we obtain
the ``Poisson equation" in the subhorizon approximation,
\begin{equation}
-\frac{k^2}{a^2}\Psi=\frac{3}{2}H^2\left\{\Omega_c\Delta_c+(1-\Omega_c)\Delta_d\right\},
\end{equation}
which relates the matter inhomogeneities to the metric perturbations.
Finally, using the results of Sec.~\ref{sec:cov_couplings}, we can obtain
the perturbation equations for the dark sector with a constant DE EoS.

\begin{figure}
\begin{center}
  \begin{tabular}{cc}
\includegraphics[width=3.2in,height=3.2in]{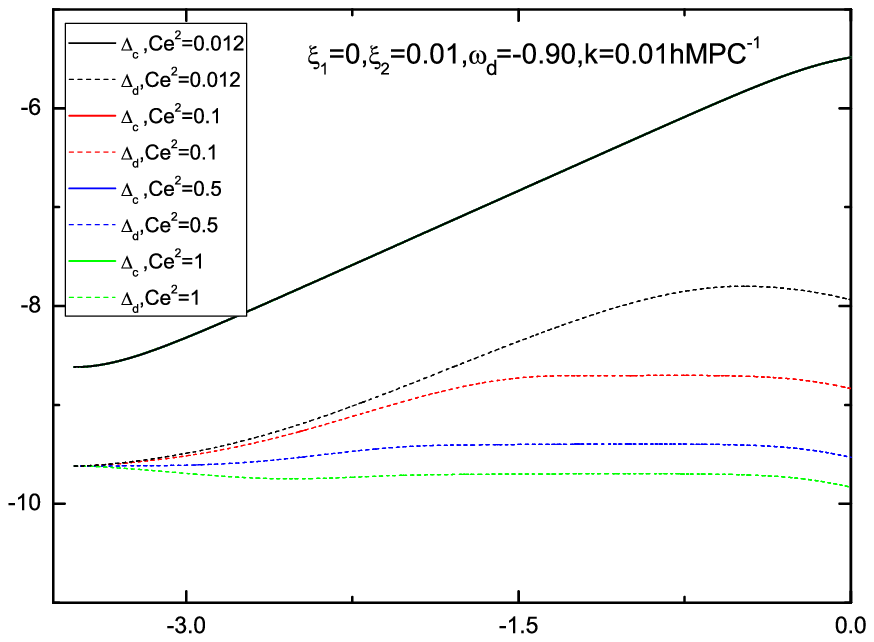}& \hspace*{-1.2cm}
\includegraphics[width=3.2in,height=3.2in]{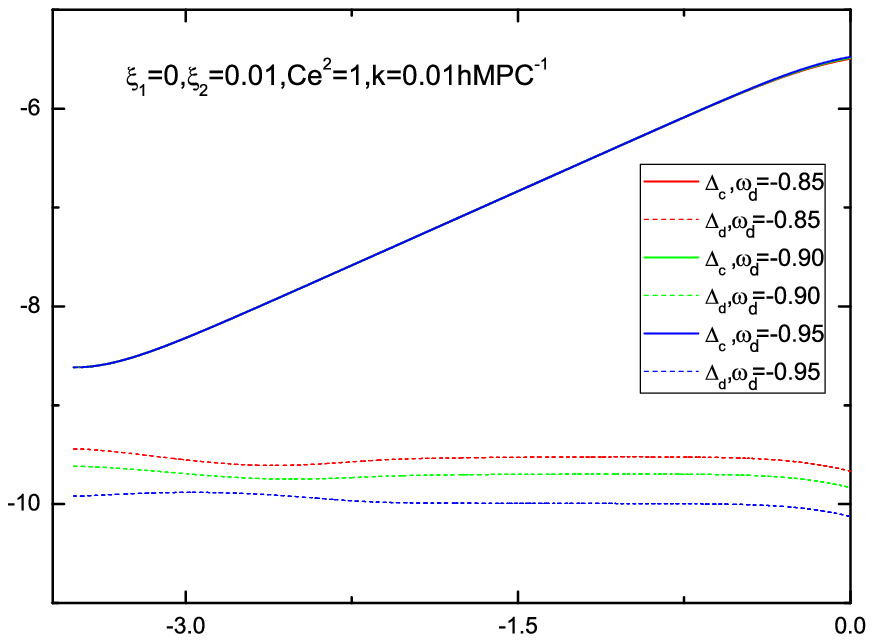}\nonumber \\
    (a)&(b)\nonumber \\
\includegraphics[width=3.2in,height=3.2in]{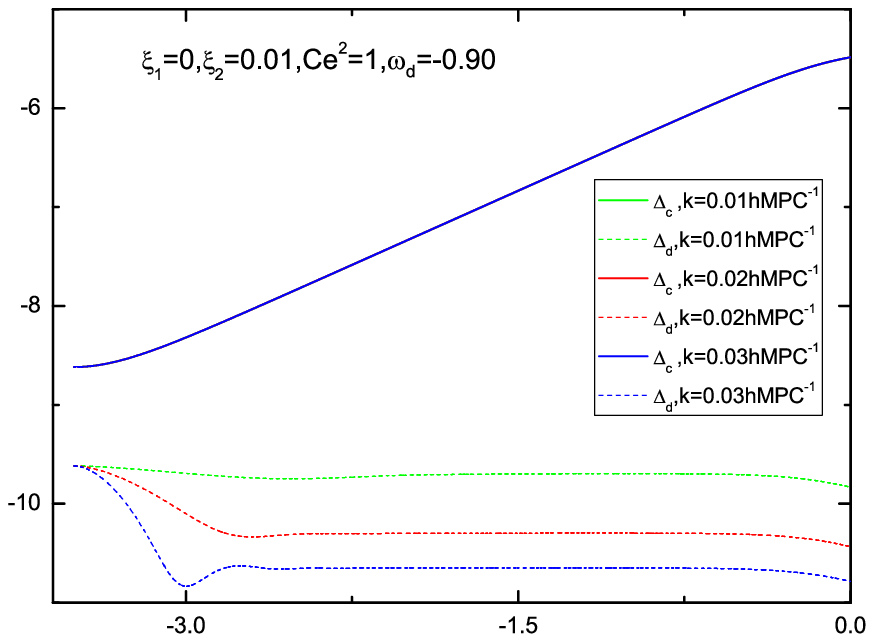}& \hspace*{-1.2cm}
\includegraphics[width=3.2in,height=3.2in]{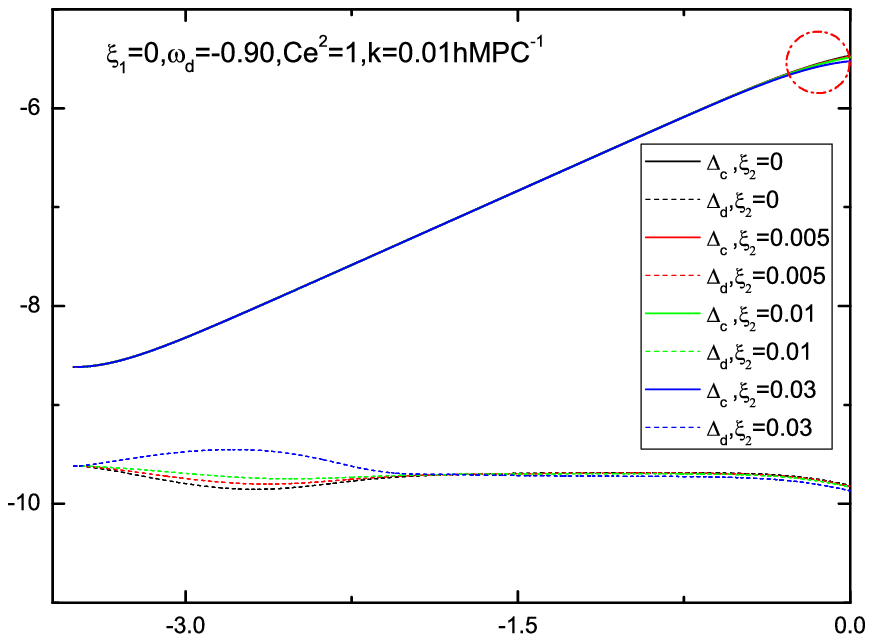}\nonumber \\
    (c)&(d)\nonumber
  \end{tabular}
\end{center}
\caption{Behavior of DE and DE perturbations for different
effective sound speed $C_e$, dark energy EoS parameter
$\omega_d$, wave number $k$ and coupling. Solid lines represent
DM perturbation and dotted lines DE perturbation. }
\label{fluctuation}
\end{figure}
\begin{figure}
\includegraphics[width=3.2in,height=2.8in]{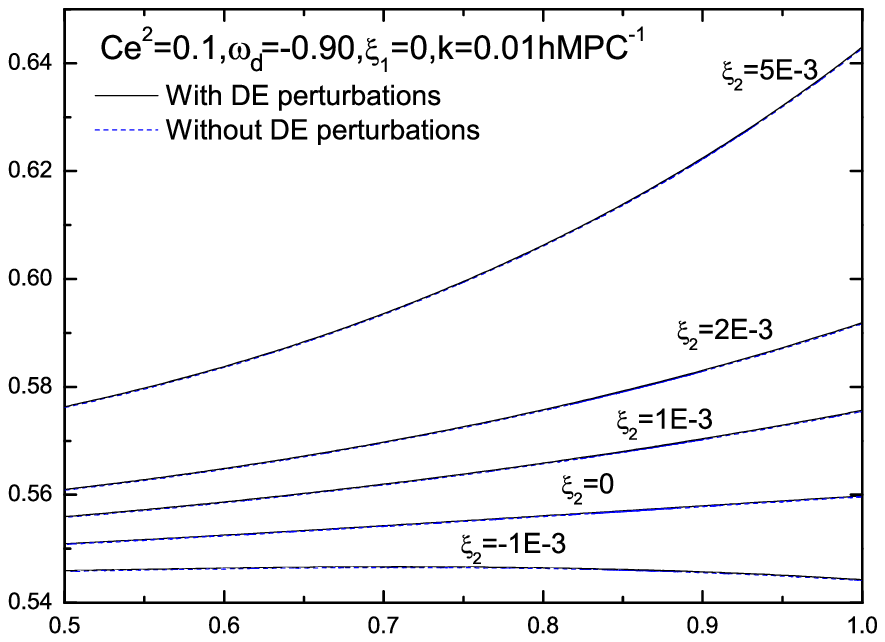}\hspace*{-0.7cm}
\includegraphics[width=3.2in,height=2.8in]{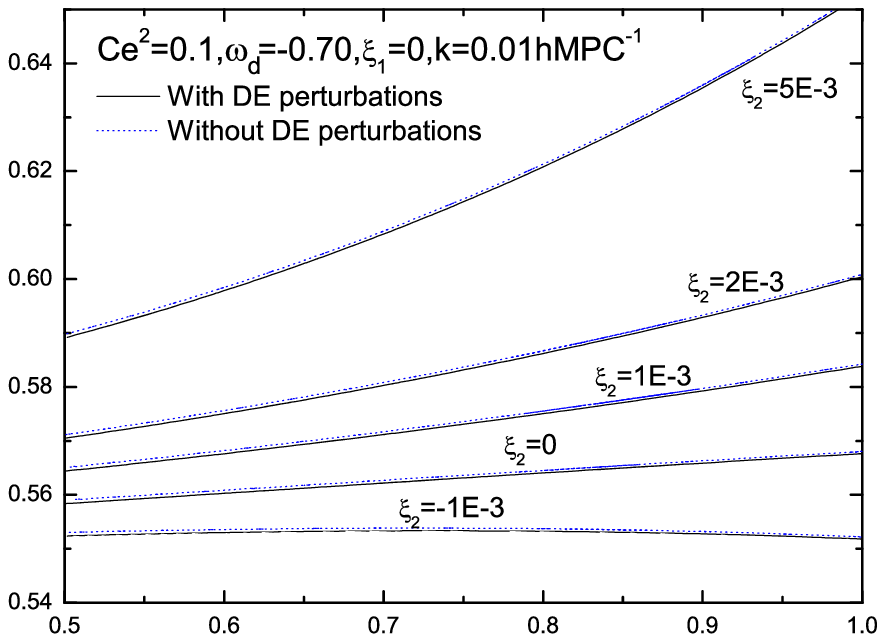}
\caption{Growth index for interacting models. Solid
lines are for the result with DE perturbation,
while dotted lines are for the result without DE
perturbation.}\label{couple_growth_index}
\end{figure}

The  stability of the growth function depends on the type
of coupling and on whether $\omega_d$ is constant in time or not.
When the interaction is proportional to the DM energy
density ($\xi_2=0$) the growth of matter density perturbations is
unstable.  The unstable solution can disappear
when $\omega_d$ varies with time. The growth is stable
when the interaction is proportional to the energy density
of DE ($\xi_1=0$). These results are consistent with the conclusions
of the evolution of curvature perturbations
on superhorizon scales in the early Universe \cite{He_09af}.
Hereafter we will restrict our analysis to the case $\xi_1=0$
since in this case all the solution are stable.
We will consider adiabatic initial conditions  at matter-radiation equality ($z=3200$). We assume that the time derivatives
of DM and DE perturbations are zero. To simplify the discussion,
we restrict the analysis to a constant EoS parameter with $\omega_d<-1$ and to those scales of the matter power spectrum that have been measured, i.e.,$k>0.01h\rm Mpc^{-1}$
\cite{Ballesteros_08f}. Also, the DE sound speed,  $C_e^2$,
will be positive and smaller than unity.
In Fig~\ref{fluctuation} we plot the
behavior of DE and DE perturbations for different
effective sound speed, dark energy EoS parameter, wave number and coupling. The adiabatic initial conditions are $\delta_c/(1-\xi_1-\xi_2/r)=\delta_d/(1+\omega_d+\xi_1 r+\xi_2)]$. Solid lines represent
DM perturbations and dotted lines DE perturbations. The solutions
largely overlap, except in the encircled region in panel (d).

In models without interaction, DE perturbations
have an effect on the evolution of DM density pertubations,
more important when the sound speed is $C_e^2\sim 0$ and
$\omega_d$ is significantly different from $-1$. Fig.~\ref{fluctuation}a illustrates
this behavior: a smaller the value of $C_e^2$ corresponds to a larger
 growth of DE perturbation; larger the difference
of $\omega_d$ from $-1$ larger the growth
of DE perturbations. The effect is even  more important on the
DM perturbations. In Fig.~\ref{fluctuation}b we illustrate
the behavior for a constant sound speed.
In this case, the DE perturbation grows larger when the difference
of $\omega_d+1$ from zero increases,  consistent with
\cite{Ballesteros_08f}. However, when $C_e^2\lsim 1$
and $\omega_d\simeq -1$ the effect of DE is suppressed.
Fig.~\ref{fluctuation}c illustrates that the effect of DE perturbations
is smaller at large than at small scales.
The previous results do not change qualitatively in the
presence of an interaction as illustrated by  Fig.~\ref{fluctuation}d. Only  at
very recent epochs, indicated  in the figure by a
red circle, there is a deviation due to the coupling.

An alternative probe of the evolution of DM perturbations is
the growth index, defined as $\gamma_c=(\ln\Omega_c)^{-1}
\ln\left[(a/\Delta_c)(d\Delta_c/da)\right]$, that is rather sensitive to
cosmological parameters \cite{Polarski_08f} and the underline theory
of gravity \cite{Gannouji_08f,Gannouji_09f}. It also sensitive to the
effect of a DM/DE coupling \cite{Ballesteros_08f}.
In Fig~\ref{couple_growth_index} we plot  this magnitude
for Model I with different  values of the interaction parameter.
Solid lines correspond to  models with DE perturbations, while
dotted lines correspond to  models with a DE density  homogeneously
distributed.  When there is no interaction and $\omega_d=const$
the matter growth index deviates from the value
of the concordance model more when there are DE
density perturbations than when the DE is  distributed homogeneously.
The deviation is more important when
$\omega_d$ differs significantly from $-1$
and when $C_e^2$ decreases.
The difference in $\gamma$  with and without DE perturbations
can reach $\Delta\gamma=0.03$. The differences between
large and small scales in the subhorizon approximation,
is not as large as the effect of $\omega_d$ and $C_e^2$,
 in agreement with \cite{Ballesteros_08f}.
Detailed discussions on the matter growth index in
the case of no interaction can be found in \cite{He_09cf}.
The effect of the interaction on the growth index is
more important than the presence or absence of DE perturbations.
 For instance, if $\xi_2\lsim 10^{-2}$,  value compatible
with the expansion history of the Universe
\cite{He_08f,Feng_08f}, the interaction dominates
the variation on the growth index over the effect
of DE perturbations.  This result opens the possibility that future
measurements of the growth factor could prove
the existence of the interaction irrespectively
of the DE distribution.

\section{The DM/DE Interaction Beyond Perturbation Theory.}\label{sec:sec5}

\subsection{Layzer-Irvine equation.}

In an expanding Universe, the Layzer-Irvine equation
describes how a collapsing system reaches a state of dynamical
equilibrium \cite{Layzer_63f,Peebles_93f}. The final state will be altered if there exists a
DM/DE interaction \cite{Bertolami_07f,Abdalla_09f,Abdalla_10f,He_10f,
Bertolami_11f,Delliou_14f,Pellicer_2012g}.
In this section we will derive the Layzer-Irvine equation
in the presence of such an interaction.  We start by redefining the
perturbed gauge invariant couplings of eqs.~(\ref{eq:gauge-c},\ref{eq:gauge-d})
as \cite{He_09af,He_09bf}
\begin{equation}
\frac{a^2\delta Q^{0I}_c}{\rho_c}\approx 3\mathcal{H}(\xi_1\Delta_c+\xi_2\Delta_d/r),\qquad
\frac{a^2\delta Q^{0I}_d}{\rho_d}\approx -3\mathcal{H}(\xi_1\Delta_cr+\xi_2\Delta_d),
\end{equation}
where $\Delta_c\approx \delta \rho_c/\rho_c=\delta_c$,
$\Delta_d\approx \delta \rho_d/\rho_d=\delta_d$ and $r=\rho_c/\rho_d$.
It is useful to rewrite eqs.~(\ref{DMsub},\ref{DEsub}) in real space
\begin{eqnarray}
&&\Delta_c'+\nabla_{\bar{x}}\cdot V_c= 3\mathcal{H}\xi_2(\Delta_d-\Delta_c)/r ,\nonumber \\
&&V_c'+\mathcal{H}V_c=-\nabla_{\bar{x}}\Psi-3\mathcal{H}(\xi_1+\xi_2/r)V_c ; \label{DM1}\\
&&\Delta_d'+(1+\omega_d)\nabla_{\bar{x}}\cdot V_d= 3\mathcal{H}(\omega_d-C_e^2)\Delta_d
+3\mathcal{H}\xi_1r(\Delta_d-\Delta_c) ,\nonumber \\
&&V_d'+\mathcal{H}V_d=-\nabla_{\bar{x}}\Psi-\frac{C_e^2}{1+\omega_d}
\nabla_{\bar{x}}\Delta_d-\frac{\omega_d'}{1+\omega_d}V_d ,\nonumber\\
&&+3\mathcal{H}\left\{(\omega_d-C_a^2)
+\frac{1+\omega_d-C_a^2}{1+\omega_d}(\xi_1r+\xi_2)\right\}V_d ;\label{DE1}
\end{eqnarray}
where ${\bar{x}}=a^{-1}x$ and $\nabla_{x}=a^{-1}\nabla_{\bar{x}}$.
Defining $\sigma_c=\delta \rho_c$,
$\sigma_d=\delta \rho_d$, and assuming that the
EOS of DE is constant, we can rewrite Eqs.~(\ref{DM1},\ref{DE1}) as
\begin{eqnarray}
&&\dot{\sigma}_c+3H\sigma_c+\nabla_{x}(\rho_cV_c)=3H(\xi_1\sigma_c+\xi_2\sigma_d) ,\nonumber \\
&&\frac{\partial}{\partial t}(a V_c)=-\nabla_x(a \Psi)-3H(\xi_1+\xi_2/r)(aV_c) ;\label{dmn}\\
&&\dot{\sigma}_d+3H(1+C_e^2)\sigma_d+(1+\omega_d)\nabla_{x}(\rho_dV_d)=
	-3H(\xi_1\sigma_c+\xi_2 \sigma_d) ,\nonumber \\
&&\frac{\partial}{\partial t}(aV_d)=
	-\nabla_x(a\Psi)-\frac{C_e^2}{1+\omega_d}\nabla_x\cdot (a\Delta_d) ,\nonumber\\
&&+3H\left[(\omega_d-c_a^2)+\frac{1+\omega_d-c_a^2}{1+\omega_d}(\xi_1r+\xi_2)\right](aV_d) .
\label{den}
\end{eqnarray}
The dot denotes the derivative with respect to
the coordinate time. The gravitational potential $\Psi$ can be decomposed as
$\Psi=\psi_c+\psi_d$, each component satisfying the following Poisson
equation \cite{Mota_04f},
\begin{equation}
\nabla^2\psi_\lambda=4\pi G (1+3\omega_\lambda)\sigma_\lambda .
\end{equation}
The subscript $\lambda=(c,d)$ denotes DM and DE, respectively
and $\sigma_{\lambda}$ represents the inhomogeneous fluctuation field. In a homogeneous
and isotropic background $<\psi_\lambda>=0$, since $<\sigma_\lambda>=0$. The corresponding gravitational potentials are
\cite{Mota_04f}
\begin{equation}
\psi_{c}=-4\pi G\int dV'\frac{\sigma_{c}}{\mid x-x' \mid}, \qquad
\psi_{d}=-4\pi G\int dV' \frac{(1+3\omega_d)\sigma_{d}}{\mid x-x' \mid}.
\end{equation}
For the DM, the rate of change of the peculiar velocity (eq.~\ref{dmn}) can
be recast as
\begin{equation}
\frac{\partial}{\partial t}(a V_c)=-\nabla_x(a\psi_c+a\psi_d)-3H(\xi_1+\xi_2/r)(aV_c).
\label{dmn1}
\end{equation}
Neglecting the influence of DE and the couplings, eq.~(\ref{dmn1})
represents the rate of change of the peculiar velocity of the DM
particle in an expanding universe described by the Newton's law;
this was the starting point of \cite{Layzer_63f}. To derive the
energy equation for the local inhomogeneities, we multiply
eq.~(\ref{dmn1}) by $aV_c\rho_c\hat{\varepsilon}$ and integrate
over the volume \cite{Layzer_63f}. Here
$\hat{\varepsilon}=a^3d\bar{x}\wedge d\bar{y}\wedge d\bar{z}$
is the volume element which satisfies $\frac{\partial}{\partial
t}\hat{\varepsilon}=3H\hat{\varepsilon}$.
Considering the continuity equation, the LHS of
eq.~(\ref{dmn1}) can be multiplied by $a V_m$ and
integrated to yield
\begin{equation}
\frac{\partial}{\partial t}\left(a^2T_c\right)-a^23H(\xi_1+\xi_2/r)T_c .
\label{Tm}
\end{equation}
where $ T_c=\frac{1}{2}\int V_c^2 \rho_c
\hat{\varepsilon}$ is the kinetic energy of DM
associated with peculiar motions of DM particles.

The RHS of eq.~(\ref{dmn1}) can be transformed in a similar manner.
Using partial integration, the potential part can be changed to
\begin{equation}
-\int aV_c \nabla_{x}(a \psi_c+a\psi_d) \rho_c
\hat{\varepsilon}=a^2 \int \nabla_{x}(\rho_c
V_c)\psi_c \hat{\varepsilon}+a^2 \int
\nabla_{x}(\rho_c V_c)\psi_d
\hat{\varepsilon} .\nonumber
\end{equation}
Taking into account the first equation in (\ref{dmn}), it can transformed to
\begin{eqnarray}
-\int aV_c \nabla_{x}(a \psi_c&+&a\psi_d) \rho_c \hat{\varepsilon}=-a^2(\dot{U}_{cc}
+HU_{cc})-a^2\int \psi_d \frac{\partial}{\partial t} (\sigma_c \hat{\varepsilon})\nonumber\\
&+&3a^2H\left\{\xi_1U_{cd}+\xi_2U_{dc}+2\xi_1U_{cc}+2\xi_2U_{dd}\right\}\label{firstdm}
\end{eqnarray}
where $U_{cc}=\frac{1}{2}\int \sigma_c \psi_c
\hat{\varepsilon} $, $U_{dc}=\int \sigma_d \psi_c
\hat{\varepsilon} $ , $U_{cd}=\int \sigma_c
\psi_d \hat{\varepsilon} $, and
$U_{dd}=\frac{1}{2}\int \sigma_d \psi_d
\hat{\varepsilon}$.

The second term in the RHS of eq.~(\ref{dmn1}) can be rewritten as
\begin{equation}
-\int(aV_c)^23H(\xi_1+\xi_2/r)\rho_c\hat{\varepsilon}=-a^26H(\xi_1+\xi_2/r)T_c \quad .
\label{seconddm}
\end{equation}
Combining Eqs.~(\ref{Tm},\ref{firstdm},\ref{seconddm}), one obtains
\begin{eqnarray}
\dot{T}_c+\dot{U}_{cc}&+&H(2T_c+U_{cc})=-\int\psi_d\frac{\partial}{\partial t}
(\sigma_c \hat{\varepsilon})-3H(\xi_1+\xi_2/r)T_c , \nonumber \\
&&+3H\left\{\xi_1U_{cd}+\xi_2U_{dc}+2\xi_1U_{cc}+2\xi_2U_{dd}\right\} .\label{DMv}
\end{eqnarray}
This equation describes how the DM reaches a dynamical equilibrium in a
collapsing system within an expanding universe. If the DE is distributed
homogeneously then $\sigma_d=0$ and eq.~(\ref{DMv}) reduces to
\begin{eqnarray}
\dot{T}_c+\dot{U}_{cc}&+&H(2T_c+U_{cc})=-3H(\xi_1+\xi_2/r)T_c
+6H\xi_1U_{cc}\quad .
\end{eqnarray}
For a system in equilibrium, $\dot{T}_c=\dot{U}_{cc}=0$ and we obtain the virial
condition $(2/3+\xi_1+\xi_2/r)T_c+(1/3-2\xi_1)U_{cc}=0$
\cite{Abdalla_09f}. Neglecting the interaction
$\bar{\xi_1}=\bar{\xi_2}=0$, we reach the
virial condition first derived by \cite{Layzer_63f}.
Let us remark that even when the DM is homogeneously distributed
the coupling changes both the time required by the system to reach
equilibrium and the equilibrium configuration itself \cite{Abdalla_09f}.

Let us now consider the case when DE is not homogeneous. The rate of
change of the peculiar velocity of DE is
described by the second equation of eq.~(\ref{den}).
Multiplying both sides of this equation by
$aV_d\rho_d\hat{\varepsilon}$ and integrating
over the volume, on the LHS we have
\begin{equation}
\frac{\partial}{\partial t}\left(a^2T_d\right)+3a^2H(\omega_d+\xi_1r+\xi_2)T_d .\label{de1}
\end{equation}
On the RHS the first term reads,
\begin{eqnarray}
-&&\int aV_d \nabla_{x}(a \psi_c+a\psi_d) \rho_d \hat{\varepsilon}=
 	-\frac{a^2}{1+\omega_d}(\dot{U}_{dd}+HU_{dd})\nonumber \\
&&-\frac{a^2}{1+\omega_d}3H\left\{2(C_e^2+\xi_2)U_{dd}+
	2\xi_1U_{cc}+\xi_1U_{cd}+(C_e^2+\xi_2)U_{dc}\right\}\nonumber \\
&&-\frac{a^2}{1+\omega_d}\int \psi_c\frac{\partial}{\partial t}
(\sigma_d \hat{\varepsilon}).\label{de2}
\end{eqnarray}
For the remaining terms, we have
\begin{eqnarray}
&&3H\left[(w-c_a^2)+\frac{1+w-C_a^2}{1+\omega_d}(\xi_1r+\xi_2)\right]
\int(aV_d)^2 \rho_d \hat{\varepsilon}\nonumber\\
&&-\frac{c_e^2}{1+\omega_d}\int aV_d\nabla_x(a \Delta_d)\rho_d
\hat{\varepsilon}\nonumber\\
&&=6a^2H\left[(\omega_d-C_a^2)+\frac{1+\omega_d-c_a^2}{1+\omega_d}
(\xi_1r+\xi_2)\right]T_d\nonumber\\
&&-\frac{c_e^2}{1+\omega_d}a^2\int V_d\nabla_x(\sigma_d)\hat{\varepsilon}.\label{de3}
\end{eqnarray}
Combining Eqs.~(\ref{de1},\ref{de2},\ref{de3}), we arrive at
\begin{eqnarray}
(1&+&\omega_d)\dot{T}_d+\dot{U}_{dd}+H[2(1+\omega_d)T_d+U_{dd}]
=-3H\left\{2(C_e^2+\xi_2)U_{dd}+2\xi_1U_{cc}\right.\nonumber \\
&&\left.+\xi_1U_{cd}+(C_e^2+\xi_2)U_{dc} \right\}
-\int\psi_c\frac{\partial}{\partial t}(\sigma_d
\hat{\varepsilon})-c_e^2\int
V_d\nabla_x(\sigma_d)\hat{\varepsilon}\nonumber \\
&&+3H\left[(1+\omega_d)(\omega_d-2C_a^2)+(1+\omega_d-2C_a^2)(\xi_1r+\xi_2)\right]T_d,
\label{DEvv}
\end{eqnarray}
which describes how in an expanding Universe
collapsing DE perturbations reach dynamical equilibrium.

In the non-interacting case
($\xi_1=\xi_2=0$) when $C_e^2=0$ the DE would cluster just like cold DM (compare (134) and (139)).
Examples of DE models with this property were investigated in
\cite{Creminelli_10f}. In the interacting case, the time and
dynamics required by the DE and the DM to reach
equilibrium in a collapsing system are different. This result was derived in linear theory.  It would be more
interesting to examine what occurs with the DE at the non-linear
perturbation level to obtain a clearer picture of the dynamical evolution of the DE during the formalism of structure galactic halos and large scale structure.

\subsection{Spherical collapse model.}

Let us now consider how the interaction
between DE and DM affects the evolution of collapsing systems.
The spherical collapse model is the simplest analytical model of
structure formation.
At the background level, the Universe expands
with the Hubble rate $H=\dot{a}/a$ and DE and DM
satisfy the continuity equations (eq.~\ref{eq:cons_c},\ref{eq:cons_d}).
A spherically symmetric region of radius $R$ with energy density
$\rho_{\lambda}^{cl}=\rho_{\lambda}+\sigma_{\lambda}$
(as before, $\lambda=(c,d)$ and $cl$ indicates  clustering),
will eventually collapse due to its self-gravity provided that
$\sigma_{\lambda}>0$. The equation of motion for
the collapsing system is governed by Raychaudhuri equation.
\begin{equation}
\dot\theta=-\frac{1}{3}\theta^2-4\pi G\sum_{\lambda}(\rho_{\lambda}+3 p_{\lambda})
\label{eq:Raychaudhuri}
\end{equation}
where $\theta=3(\dot{R}/R)$. Eq~(\ref{eq:Raychaudhuri}) can be written as,
\begin{equation}
\frac{\ddot{R}}{R}=-\frac{4\pi G}{3}
\sum_{\lambda}( \rho_{\lambda} + 3 p_{\lambda})
\label{newtonian}
\end{equation}
where $R$ is the local expansion scale factor. Its evolution is determined
by the matter inside the spherical volume and is not affected by the matter
outside.

Assuming that the DE is homogeneously distributed ($\sigma_d=0$),
the evolution of its energy density in an spherical volume is
\begin{equation}
\dot{\rho}_{c}^{cl}+3\frac{\dot R}{R}\rho_{c}^{cl}=
3H(\xi_1\rho_{c}^{cl}+\xi_2\rho_d) ,
\end{equation}
The Raychaudhuri equation applied to a spherical region has the form
\begin{equation}
\ddot{R}=-\frac{4\pi G}{3}[\rho_{c}^{cl}+
(1+3\omega_d)\rho_d]R\quad ,\label{newton}
\end{equation}
where $\rho_d$ is the background DE energy
density. Changing variables from time $t$ to the scale factor $a$,
we can write $\ddot{R}=(\dot{a})^2(d^2R/da^2)+\ddot{a}(dR/da)$
and change the Raychaudhuri equation to
\begin{equation}
2a^2(1+\frac{1}{r})\frac{d^2R}{da^2}-\frac{dR}{da}\left[1+(3\omega_d+1)/r\right]a=
-R\left[(3\omega_d+1)/r+\zeta\right] ,
\label{eq:dynam}
\end{equation}
where $\zeta=\rho_{c}^{cl}/\rho_c$. Therefore, we have
\begin{equation}
\frac{d\zeta}{da}=\frac{3}{a}(1-\xi_2/r)\zeta-3\frac{1}{R}\frac{dR}{da}\zeta+
\frac{3}{a}\xi_2/r .
\label{eq:zetam}
\end{equation}

In order to solve eqs.~(\ref{eq:dynam},\ref{eq:zetam}) we set the
initial conditions $R\sim a$ and $dR/da=1$ at matter-radiation equality, when we can consider that the spherical region is comoving with the background expansion.
In the subhorizon limit,  by
neglecting DE inhomogeneities we can
combine  eqs.~(\ref{eq:dynam},\ref{eq:zetam}) and integrate the resulting equation
from the initial moment $\zeta_i=\Delta_{ci}+1$ till  the spherical region
collapses, at $R(a_{coll})\approx 0$ \cite{He_09cf} to obtain the critical over density $\delta_c$ that determines which halos collapse. This linearly extrapolated
density threshold will fix the abundances of
collapsed halos at each redshift and mass scale.

If the DE distribution is not homogeneous, it will not fully trace the DM.
Their four velocities will be different, $u_{(d)}^a\neq u_{(c)}^a$, and
\begin{equation}
u_{(d)}^a=\gamma(u_{(c)}^a +
v_d^a)\label{Lorentz_boost}
\end{equation}
where $\gamma=(1-v_d^2)^{-1/2}$ is Lorentz-boost
factor and $v_d^a$ is the relative velocity of DE
fluid with respect to the DM rest frame. If the DE follows the
DM distribution, $v_d^a=0$. For non-comoving perfect fluids
we have \cite{Tsagas_08f}
\begin{eqnarray}
T_{(c)}^{ab}&=&\rho_{c}u_{(c)}^au_{(c)}^b\quad ,\nonumber \\
T_{(d)}^{ab}&=&\rho_du_{(d)}^au_{(d)}^b+p_dh_{(d)}^{ab}\quad ,
\label{non_comoving}
\end{eqnarray}
where $h^{ab}=g^{ab}+u^au^b$ is the projection operator. Inserting
eq.~(\ref{Lorentz_boost}) into the second eq.~(\ref{non_comoving})
and  denoting $u_{(c)}^a$ by $u^a$, the energy momentum tensor
for DE reads,
\begin{equation}
T_{(d)}^{ab}=\rho_du^au^b+p_dh^{ab}+2u^{a}q^{b}_{(d)}\quad ,
\end{equation}
where $q^a=(\rho_d + p_d)v_d^a$ is the
energy-flux of DE seen in the
DM rest frame. In the above equation, we
neglect the second order terms in
$v_d^a$ and assume that the energy-flux velocity
is much smaller than the speed of light
$v_d^a\ll 1,\gamma\sim 1$. In the spherical model,
we define the top-hat radius as the radius of the
boundary of the DM halo. When the DE does not trace the DM it
will not be bounded inside the top-hat radius and
will be stretched outside the spherical region. If we
assume that the DE leakage is still
spherically symmetric, the Birkhoff
theorem guarantees that in the spherical region the
Raychaudhuri equation takes the same form as in
(\ref{newtonian}). For the energy density
conservation law we have
\begin{equation}
\nabla_aT^{ab}_{(\lambda)}=Q^b_{(\lambda)} ,
\end{equation}
where $Q^b$ is the coupling vector and
``$\lambda$" denotes DE and DM respectively. The
timelike part of the above equation,
$u_b\nabla_aT^{ab}_{(\lambda)}=u_bQ^b_{(\lambda)}$,
gives
\begin{eqnarray}
&&\dot{\rho}_{c}^{cl}+3h\rho_{c}^{cl}=
3H(\xi_1\rho_{c}^{cl}+\xi_2\rho_d^{cl}),\nonumber \\
&&\dot{\rho}_{d}^{cl}+3h(1+\omega_d)\rho_{d}^{cl}=-\vartheta(1+\omega_d)\rho_d^{cl}
-3H(\xi_1\rho_{c}^{cl}+\xi_2\rho_d^{cl}) ,\label{decontin}
\end{eqnarray}
where $\vartheta=\nabla_xv_d$. The external term
incorporating $\vartheta$ in the DE density
evolution corresponds to the energy loss caused by the
leakage of DE out of the spherical region.

For the space-like part, $h^{a\phantom{b}}_b\nabla_aT^{ab}_{(\lambda)}=
h^{a\phantom{b}}_b Q^b_{(\lambda)}$, only the DE has a non-zero spatial
component and
\begin{equation}
\dot{q}^a_{(d)}+4hq^a_{(d)}=0\quad ,
\end{equation}
where $q^a_{(d)}$ is the DE flux.
Assuming that the energy and pressure are
distributed homogeneously, we obtain
\begin{equation}
\dot\vartheta+h(1-3\omega_d)\vartheta=3H(\xi_1\Gamma+\xi_2)\vartheta\label{vartheta} .
\end{equation}
In this expression, $\Gamma=\rho^{cl}_c/\rho^{cl}_d$ is the ratio of the DM and DE density in the collapsed region. We kept
only the linear terms of $\vartheta$ in eq.~(\ref{decontin}).
In eq.~(\ref{vartheta}), if $\vartheta$ vanishes
initially, it will remain so during the
subsequent evolution and the DE will fully trace the DM.
However, in most cases, even at the linear level
there is a small difference between $v_d$
and $v_c$, so that the initial condition for
$\vartheta$ is not zero. For illustration,
we took the initial condition $\vartheta\sim
k(v_d-v_c)\sim-5\times 10^{-3}\delta_{ci}$, which is
obtained from the prediction of linear equation
with $k=1{\rm Mpc}^{-1}$ at $z_i=3200$. We have
chosen $|\vartheta|<<1$; the negative sign
indicates that at the initial moment DM
expanded faster than that of DE. Taking
$\vartheta\sim -10^{-3}\delta_{ci}$ did not
change the results.

Defining $\zeta_c = \rho_{c}^{cl}/\rho_c$,
$\zeta_d=\rho_{d}^{cl}/\rho_d$ and
converting the time derivative from
$\frac{d}{dt}$ to $\frac{d}{da}$ we have the
evolution of DM and DE in the spherical region
described by
\begin{eqnarray}
&&\vartheta'+\frac{R'}{R}(1-3\omega_d)\vartheta=
\frac{3}{a}(\xi_1\Gamma + \xi_2)\vartheta,\nonumber\\
&&\zeta_c'=\frac{3}{a}\left\{1-\xi_2/r\right\}\zeta_c
-3\frac{R'}{R}\zeta_c+\frac{3}{a}\xi_2\zeta_d/r,\nonumber \\
&&\zeta_d'=\frac{3}{a}(1+\omega_d+\xi_1r)\zeta_d-3(1+\omega_d)\frac{R'}{R}\zeta_d
-\frac{3}{a}\xi_1\zeta_cr-\vartheta(1+\omega_d)\zeta_d.\label{zetamd}
\end{eqnarray}
When DE fully traces along DM, $\vartheta=0$,
only the last two equations  are needed. The
Raychaudhuri equation now becomes
\begin{equation}
2a^2(1+\frac{1}{r})R^{''}-R^{'}\left[1+(3\omega_d+1)/r\right]a=
-R\left[(3\omega_d+1)\zeta_d/r+\zeta_c\right].\label{dymd}
\end{equation}
Taking for the spherical region of radius $R$
the same initial conditions as above and
adopting the adiabatic initial
conditions: $\Delta_{di}=(1+\omega_d)\Delta_{ci}$,
$\zeta_{ci}=\Delta_{ci}+1$ and
$\zeta_{di}=\Delta_{di}+1$, which lead to
$\zeta_{di}=(1+\omega_d)\zeta_{ci}-\omega_d$, we can study the
spherical collapse of both the DM and the DE. However, it is
important to note that due to the coupling,
$\zeta_d$ may be negative at some moment during the
collapse which would be unphysical. To avoid reaching this point, we
remove the coupling when $\zeta_d$ becomes
negative. In this way, we guarantee $\zeta_d\geq 0$.
In the linear evolution limit we assume the subhorizon approximation
of the DM and DE perturbation equations to be valid.
Even in this limit the DE and DM perturbations are coupled.
Fortunately,  the effect of DE perturbations is small
compared with that of the DM \cite{He_09cf}. Using
the linear perturbation equations for DM and DE  together
with eqs.~(\ref{zetamd},~\ref{dymd}), in the spherical model with an inhomogeneous
DE distribution we can obtain the linearly extrapolated density threshold
$\delta_c(z)=\delta_c(z=z_{coll})$ above which the structure collapses.

\subsection{Press-Schechter formalism and Galaxy cluster number counts.}
\label{sec:cluster_counts}

The variation of the cluster number counts with redshift
has been studies as a promising tool to discriminate different DE models
\cite{Multamaki_03f,Solevi_06f,Mota_04f,Nunes_06f,Wang_09f,Brax_10f,
Wintergerst_10f,Abramo_07f,Abramo_09f} and to
test coupled quintessence models \cite{Nunes_06bf,Manera_06f,Mota_08f}.
In these studies, DM and DE were assumed to be conserved separately
at the background level; the models only included energy loss due to the DE
inside a DM halo and mass conservation implied that the DM density evolved
as $\rho_c\sim R^{-3}$. By contrast, in our study we will consider
that the interaction within the dark sector exists at all scales.

Press and Schechter \cite{Press_74f} designed a formalism
to predict the number density of collapsed objects using the spherical
collapse model. Although this formalism is a crude approximation
and it is not precise enough to predict the exact number of clusters
\cite{Jenlins_01f}, it can be useful to understand
how the interaction and the clustering of DE influence
the threshold density of cluster collapse and, consequently,
the cluster number counts as a function of mass and redshift.

In the Press-Schechter formalism, the comoving number density of
collapsed  DM halos of mass $[M,M+dM]$ at redshift $[z,z+dz]$ is \cite{Press_74f}
\begin{equation}
\frac{dn(M,z)}{dM}=\sqrt{\frac{2}{\pi}}\frac{\rho_m}{3M^2}
\frac{\delta_c}{\sigma}e^{-\delta_c^2/2\sigma^2}
\left[-\frac{R}{\sigma}\frac{d\sigma}{dR}\right] ,
\label{probabilities}
\end{equation}
where $\rho_m=\rho_c+\rho_b$ is the comoving mean
matter density at each particular redshift. In most cases,
it is a constant and equals the present mean
matter density, but this is not true when DE
interacts with DM \cite{Manera_06f}. The quantity
$\sigma=\sigma(R,z)$ represents  the root mean square density fluctuation on a sphere of  radius $R$. It has the explicit form \cite{Viana_96f},
\begin{equation}
\sigma(R,z)=\sigma_8\left(\frac{R}{8h^{-1}{\rm Mpc}}\right)^{-\gamma(R)}D(z),
\end{equation}
where $\sigma_8$ has been evaluated on a sphere of
radius $R=8h^{-1}{\rm Mpc}$ and $D(z)$ is
the growth function defined by
$D(z)=\delta_c(z)/\delta_c(0)$. The index
$\gamma$ is a function of the mass scale and the
shape parameter $\Gamma$ of the matter power
spectrum \cite{Viana_96f},
\begin{equation}
\gamma(R)=(0.3\Gamma+0.2)\left[2.92+\log_{10}\left(\frac{R}{8h^{-1}{\rm Mpc}}\right)\right].
\end{equation}
To simplify, we will use the value of $\Gamma = 0.3$ throughout the analysis.
The radius $R$ at given $M$ can be calculated by the relation \cite{Viana_96f},
\begin{equation}
R=0.951h^{-1}{\rm Mpc}\left(\frac{M}{\Omega_m10^{12}h^{-1}M_\odot}\right)^{1/3},
\end{equation}
where the mass is given in $h^{-1}M_\odot$.
The Press-Schechter formalism gives the comoving number density
of halos, which can be compared with astronomical data. To this purpose,
we calculate the all sky number of halos per unit of redshift
\begin{equation}
\frac{dN}{dz}=\int d\Omega\frac{dV}{dzd\Omega}\int n(M)dM,
\end{equation}
where the comoving volume element per unit
redshift is $dV/dzd\Omega=r^2(z)/H(z)$ and $r(z)$
is the comoving distance $r(z)=\int_0^z\frac{dz'}{H(z)}$.

In the subsequent discussion we do not aim for a precise
comparison with data, that would require computationally
expensive N-body simulations, but to see in which
direction the interaction modifies the cluster number counts.
We will study the effect of a homogeneous and an inhomogeneous
DE distribution. We will show that when the DE is distributed
inhomogeneously,  it plays an important role in the gravitational instability and
the collapse of halos and in structure formation.

\subsubsection{Interaction proportional to the
energy density of DE ($\xi_1=0,\xi_2 \neq 0$).}

\begin{figure}
\begin{center}
  \begin{tabular}{cc}
\includegraphics[width=3in,height=3in]{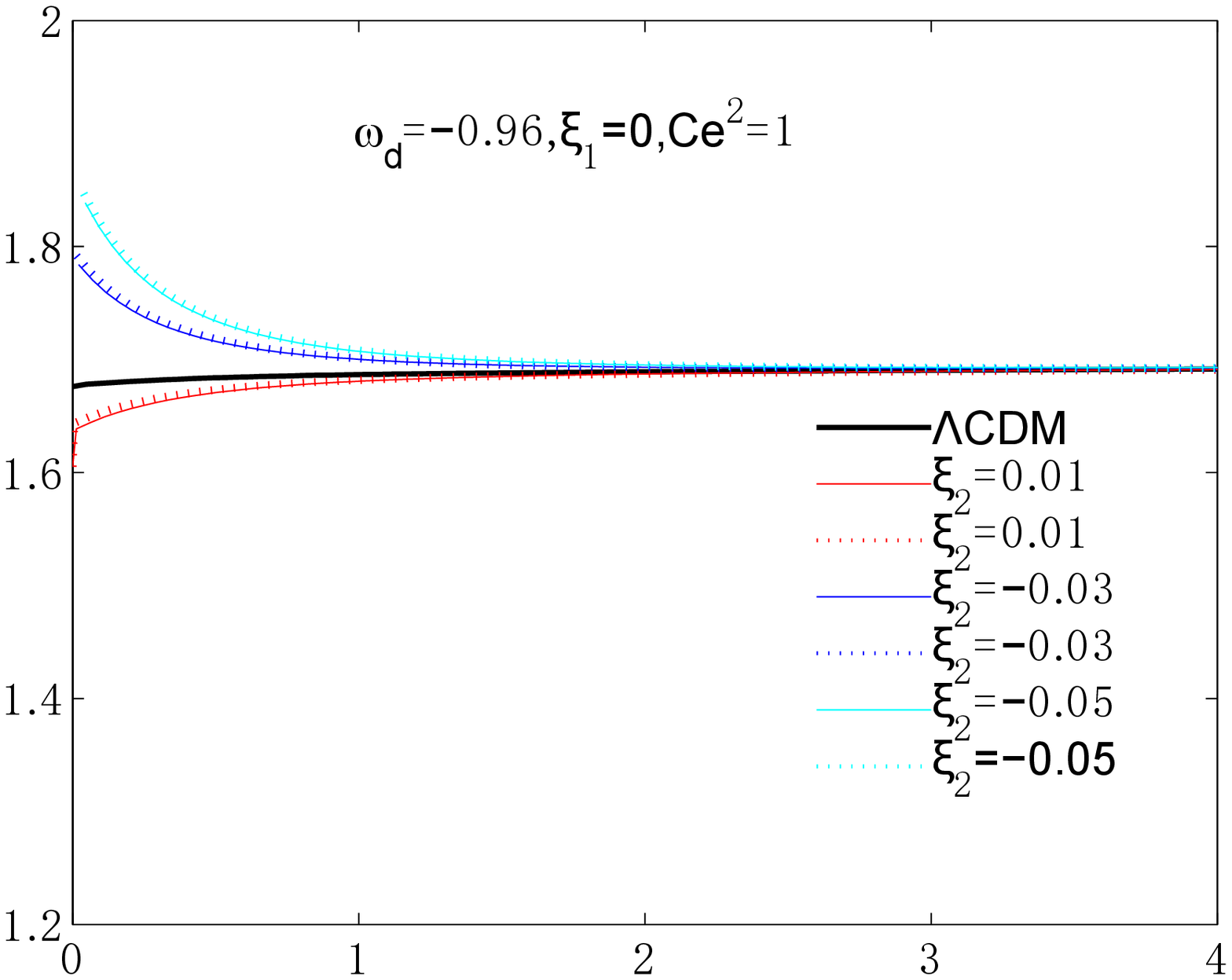}&
\includegraphics[width=3in,height=3in]{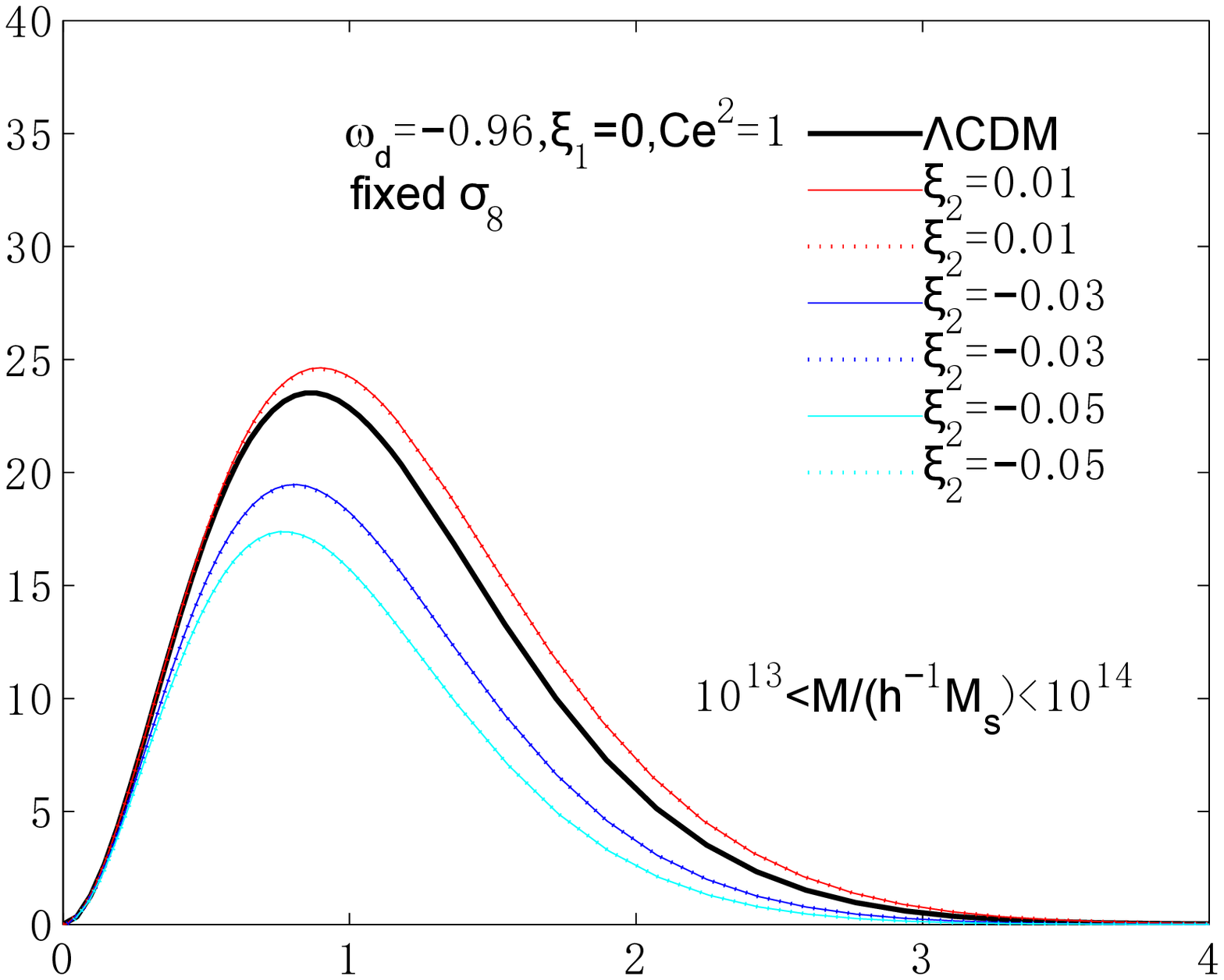} \\
    (a)&(b) \\
\includegraphics[width=3in,height=3in]{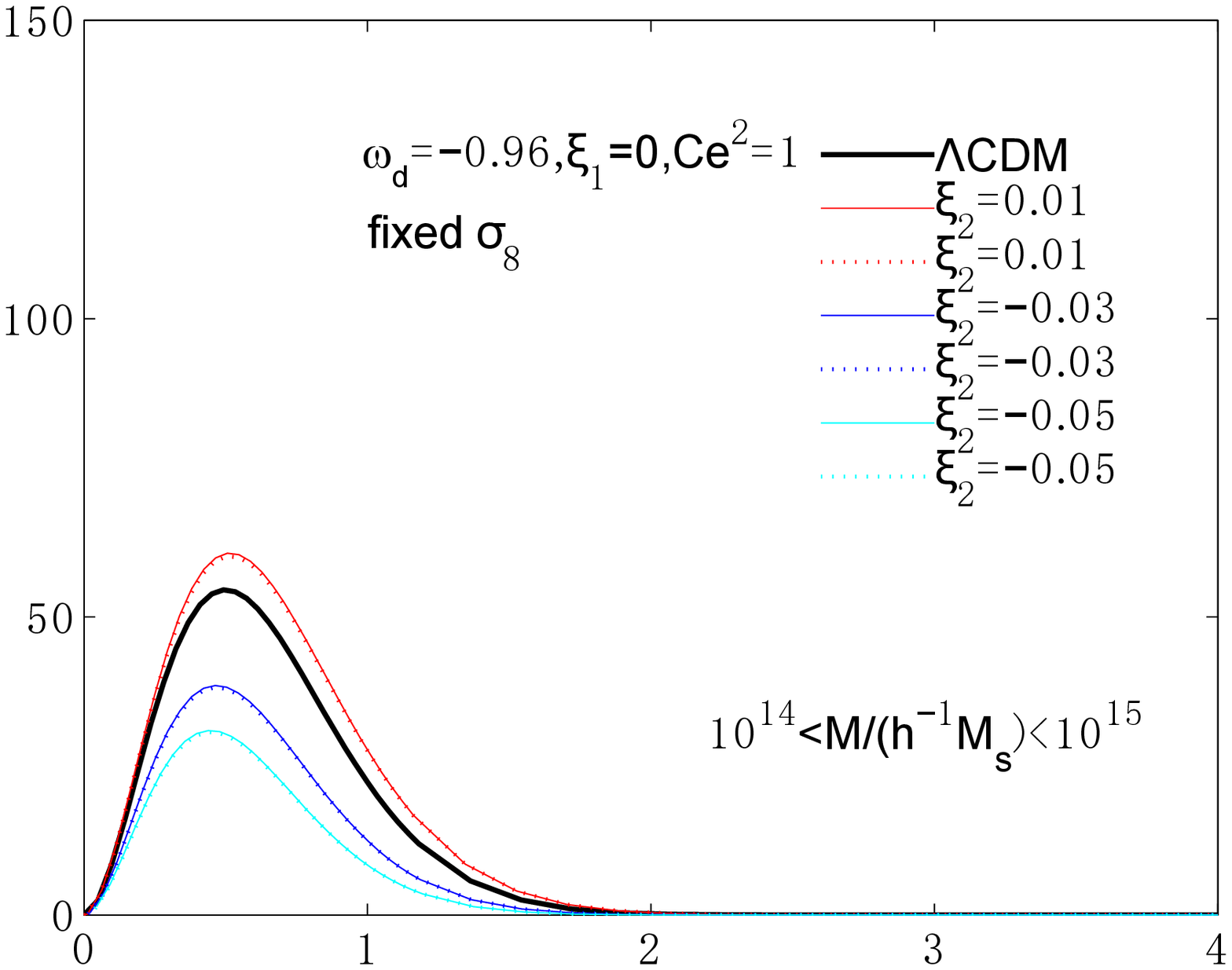}&
\includegraphics[width=3in,height=3in]{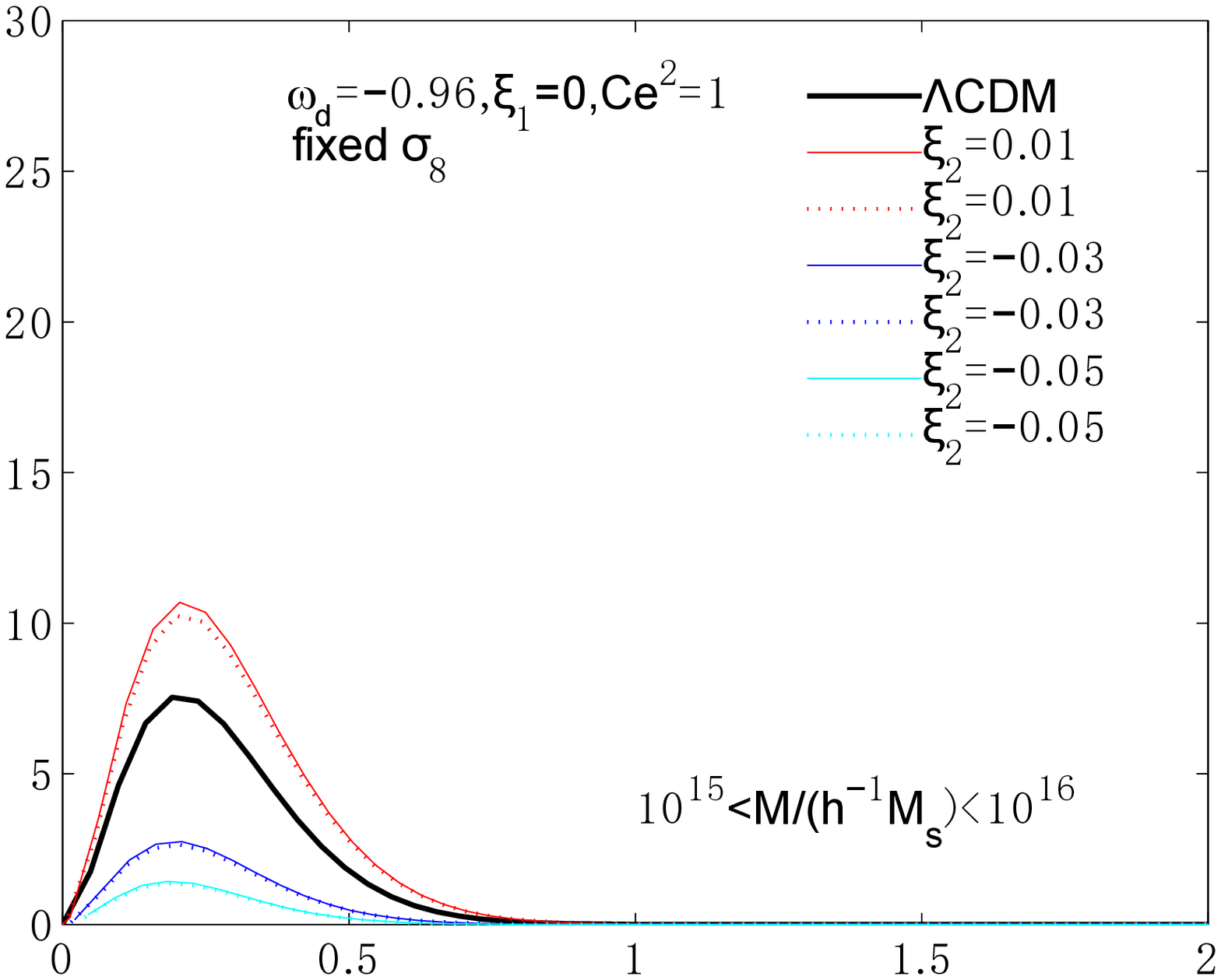}\\
    (c)&(d)\\
  \end{tabular}
\end{center}
\caption{Cluster number counts for an interaction kernel
$Q=\xi_2\rho_d$ and
DE EoS with $\omega_d>-1$. Solid lines correspond to a
DE distributed homogeneously and dotted lines to an
inhomogeneous distribution.}\label{dequint}
\end{figure}

When the interaction is proportional to the DE energy
density, we have shown in Sec.~\ref{sec:sec4} that the curvature
perturbation is always stable for both
quintessence and phantom DE EoS.
In Fig.~\ref{dequint} we present the results with
a constant DE EoS with $\omega_d>-1$. Solid lines
correspond to the case when the DE is distributed
homogeneously. Fig.~\ref{dequint}$a$ corresponds
to $\xi_2>0$, i.e., DE decays into DM). In this case,
the critical mass density decreases compared with the $\Lambda$CDM
model; clusters collapse more rapidly than in the concordance model.
When DM decays into DE, the threshold of collapse
$\delta_c$ is higher than in the concordance model,
so it is more difficult for an overdensity  to collapse.
The effect of the coupling is significant and
changes cluster number counts. This can be seen in
Figs~\ref{dequint}b-\ref{dequint}d. For a positive coupling
($\xi_2>0$) cluster number counts are
larger than in the ${\rm \Lambda}$CDM model but the opposite is true when the coupling is negative
($\xi_2<0$).

When the DE is distributed inhomogeneously, we
need to consider DE perturbations and their
effect in structure formation. In Fig.~\ref{dequint}
these results are shown as dotted lines. In this case,
the critical threshold $\delta_c$ is slightly larger than
the results without DE perturbations. The
differences due to the effect of DE inhomogeneities is
negligible compared with that of the interaction.

When the DE EoS is $\omega_d<-1$ and the DE
distribution is homogeneous, the effect of
the interaction on the evolution of critical threshold
$\delta_c$ and the galaxy number counts is similar to
when $\omega_d>-1$. However, when the DE clusters
and $\omega_d<-1$, the number counts are slightly larger
than in the homogeneous case, although the differences
are negligible, especially at low masses.

\begin{figure}
\begin{center}
  \begin{tabular}{cc}
\includegraphics[width=3in,height=3in]{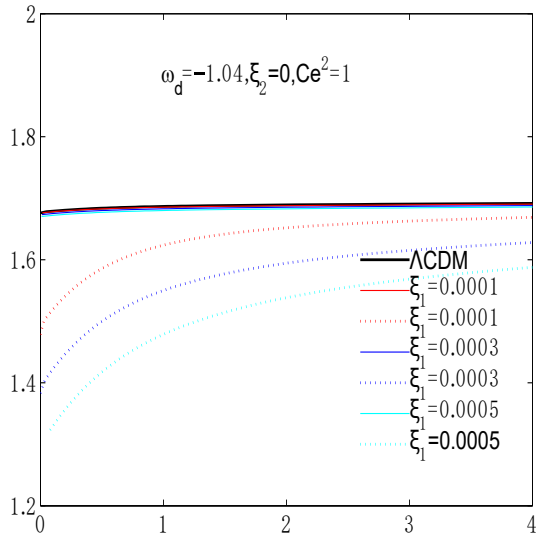}&
\includegraphics[width=3in,height=3in]{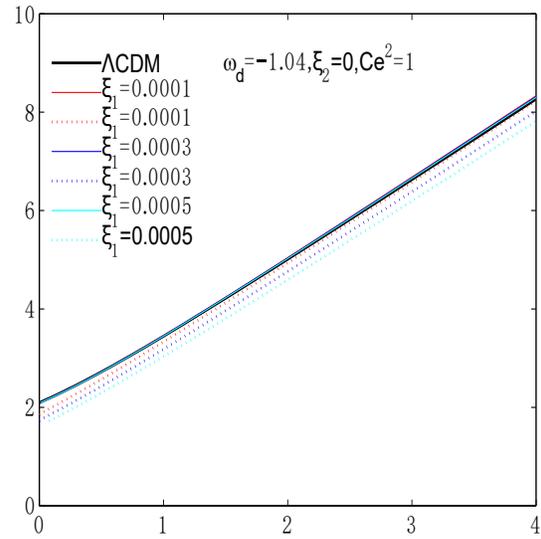}\nonumber \\
    (a)&(b)\nonumber \\
\includegraphics[width=3in,height=3in]{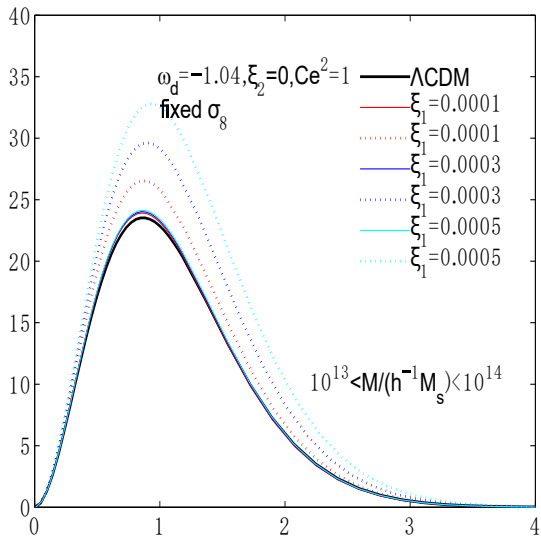}&
\includegraphics[width=3in,height=3in]{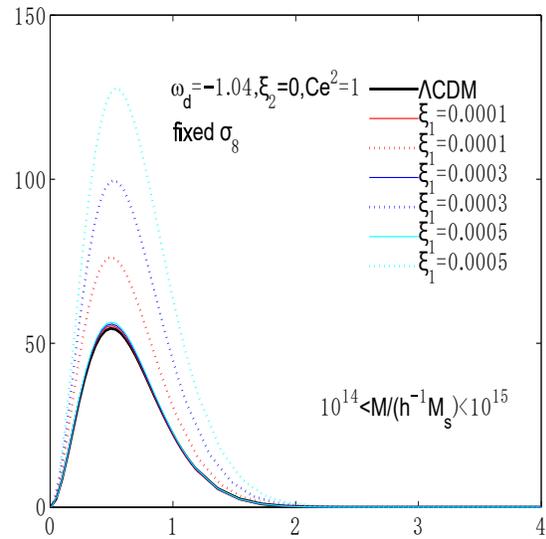}\\
    (c)&(d)\nonumber \\
  \end{tabular}
\end{center}
\caption{Cluster number counts with an interaction kernel $Q=\xi_1\rho_c$.
Solid lines correspond to a homogeneous DE distribution and dotted lines
to the inhomogeneous case.}\label{dmm}
\end{figure}

Compared with the effect of the interaction, the
clustering properties of the DE have a negligible
effect in the threshold of collapse $\delta_c$ and
consequently also in cluster number counts. This is an
indication that the fluctuation of DE field $\sigma_d$ is
small. In eq.~(\ref{DEvv}), a small value of $\sigma_d$
leads to $U_{md}\sim U_{dm}\sim U_{dd}\sim T_{d}\rightarrow 0$,
implying that the DE plays a very small role in the virialization of
the structure.  The small effect is a  consequence of  subhorizon DE perturbations being
much smaller than those of the DM
\cite{He_09cf}.

\subsubsection{The interaction proportional to the
energy density of DM ($\xi_1>0,\xi_2=0$).}\label{DM}

In this case, we will only consider stable perturbations, $\omega_d<-1$
since perturbations are unstable if $\omega_d>-1$
(see Sec.~\ref{sec:sec4}).
We also restrict the coupling to be
positive to avoid having DE with a negative energy
density at early times \cite{He_08f}.
In Fig.~\ref{dmm} we present our  results.
Solid lines correspond to homogeneous
and dotted lines to inhomogeneous DE distributions.
The figure shows that the results for small couplings
are similar to ${\rm \Lambda}$CDM;
Fig~\ref{dmm}$c$ shows that larger positive coupling leads to
higher number of clusters, as expected since
$\delta_c$ and the ratio $\delta_c(z)/\sigma_8D(z)$ are
smaller at high redshift for a larger (positive) $\xi_1$
than for the concordance model. The behavior is
consistent with the case when the interaction is
proportional to the DE.

When the DE clusters, we need to
consider the effect of its perturbations. In this case
$\delta_c$ is more suppressed at low redshift
than in the case of homogeneous DE.
The ratio $\delta_c(z)/\sigma_8D(z)$, shown in
Fig~\ref{dmm}b, is also smaller than in the case of
homogeneous DE, leading to a larger number of clusters
as shown in Fig~\ref{dmm}c.

\section{Observational Tests.} \label{sec:sec6}

As discussed in Sec.~\ref{sec:challenges}, general constraints on
the form of the interaction can be obtained by imposing that at the background
level the Universe underwent a radiation and a matter dominated
periods that lasted long enough as to allow the formation of CMB temperature anisotropies
and the emergence of large scale structure, followed by the present period of
accelerated expansion. Similar restrictions apply to models of modified gravity
\cite{amendola_08af}. It is more informative to test
theoretical predictions with observations.
In models without interaction, the effects of DE are more significant at
redshifts $z\le 2$ \cite{frieman_08fb}, while the interaction extends
the effect to larger redshifts and could even alter the sequence of
cosmological eras. In this section we will
summarize the magnitudes and data most commonly used to constraint DM/DE interacting
models.

\begin{table}
\begin{indented}
\item[]
\begin{tabular}{@{}ll}
\br
Acronym & Meaning \& url address \\
\mr
2MASS   & 2-micron All Sky Survey\\
	& http://www.ipac.caltech.edu/2mass\\
6dFGS	& 6-degree Field Galaxy Survey \\
	& http://www-wfau.roe.ac.uk/6dFGS\\
ACT	& Atacama Cosmology Telescope \\
	& http://www.princeton.edu/act/ \\
BOSS 	& Baryon Oscillation Spectroscopic Survey \\
	& https://www.sdss3.org/surveys/boss.php\\
CMBPol	& Cosmic Microwave Background Polarization\\
	& http://cmbpol.uchicago.edu/\\
COrE	& Cosmic Origins Explorer \\
	& http://www.core-mission.org/\\
CFHT	& Canada-France-Hawaii Telescope \\
	& http://www.cfht.hawaii.edu/\\
DES	& Dark Energy Survey \\
	& http://www.darkenergysurvey.org/\\
eBOSS 	& Extended Baryon Oscillation Spectroscopic Survey \\
	& eBOSS  http://www.sdss.org/sdss-surveys/eboss/\\
Euclid 	& Euclid Satellite, Euclid Consortium\\
	& http://www.cosmos.esa.int/web/euclid/home\\
HETDEX 	& Hobby-Eberly Telescope Dark Energy Experiment  \\
	& http://hetdex.org/\\
JDEM	& Joint Dark Energy Mission  \\
	& http://jdem.lbl.gov/\\
J-PAS	& Javalambre Physics of the accelerating universe Astronomical Survey \\
	& http://j-pas.org/\\
LSST	& Large Synoptic Survey Telescope \\
	& http://www.lsst.org/lsst/\\
Pan-STARSS &Panoramic Survey Telescope and  Rapid Response System  \\
	& http://www.ps1sc.org/\\
Planck  & Planck Satellite, Planck Collaboration\\
	& http://www.cosmos.esa.int/web/planck/pla\\
PRISM	& Polarized Radiation Imaging and Spectroscopy Mission \\
	& http://www.prism-mission.org/\\
SDSS 	& Sloan Digital Sky Survey  \\
	& https://www.sdss3.org/\\
SKA 	& Square Kilometre Array  \\
	& https://www.skatelescope.org/\\
SNAP 	& Super Nova Acceleration Probe  \\
	& http://http://snap.lbl.gov/\\
SPT 	& South Pole Telescope  \\
	& https://pole.uchicago.edu/\\
WFIRST	& Wide Field Infrared Survey Telescope  \\
	& http://http://wfirst.gsfc.nasa.gov/\\
WMAP	& Wilkinson Microwave Anisotropy Probe \\
	& http://lambda.gsfc.nasa.gov/\\
\br
\end{tabular}
\end{indented}
\caption{\label{table:facilities} Observational facilities.}
\end{table}

\subsection{Data on the Expansion History.}

Most models are usually too complex to be studied in detail and their
viability is first established at the background level. Even at zero
order one could expect significant differences on the Hubble expansion between
DM/DE interacting models and the concordance model. In interacting models, the DM
has an effective equation of state different from zero; the effect of the
interaction is not equivalent to an EoS varying arbitrarily
with time and the differences on the background evolution is usually first
tested to verify the viability of any given model. In this section we will briefly
summarize the most common observables, data sets used to this purpose and
the constraints imposed on the models described in Secs.~\ref{sec:sec2}
and \ref{sec:sec3}. In Table~\ref{table:facilities} we list the acronyms of
the current and future observational facilities that provide the most commonly
used data sets.

\subsubsection{Luminosity Distance Tests: Constraints from Supernovae.}

\begin{figure}
\centering
\epsfxsize=0.7\textwidth\epsfbox{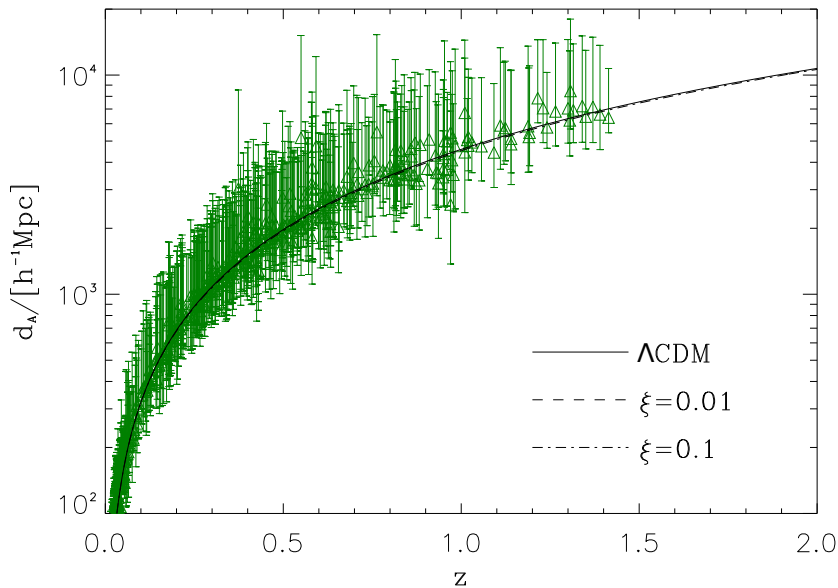}
\caption{Luminosity distances derived from SNIa Union 2 sample data.
The sample can be downloaded from {\it http://supernova.lbl.gov/Union/}.
As illustration, solid, dashed and dot-dashed lines represent
the luminosity distance of the concordance model with the cosmological
parameters measured by the Planck Collaboration and a
fluid DM/DE model with $\omega=-0.9$ and different interaction parameter.}
\label{fig:dl_data}
\end{figure}

Supernovae Type Ia  (SN Ia) are still the most direct
probe of the expansion and of the existence of DE.
They are accurate standard candles \cite{sandage_01f} and
have been used to establish that the Universe is accelerating at
present \cite{perlmutter_98f, Riess_1998f}. By comparing their
intrinsic luminosity to the measured flux one obtains a direct
estimate of the physical distance to the object.  Luminosity
distances are the first test of many of the DE
and modified gravity models that have been proposed
in the literature. Many
SN samples such as ESSENCE \cite{wood-vasey_07f}
SDSS-II \cite{kessler_09f}, CfA3 \cite{hicken_09f}
Union-2 \cite{amanullah_10f} are publicly available
and continue to be updated. The data tests models
at $z\le 1$, where most SNIa have been found. Nevertheless,
future surveys will provide SNIa to higher redshifts, like
the Wilson SN, the furtherest SNIa to date with a redshift $z=1.914$,
providing stronger constraints on model parameters.
In Fig.~\ref{fig:dl_data} we represent the luminosity distances
obtained from Union-2 sample with their error bars. For comparison,
we show the standard $\Lambda$CDM model
with the cosmological parameters measured by the Planck Collaboration
(see Table~~\ref{table:parameters}) and two fluid models
with the same cosmological parameters except an EoS and interaction parameters
$\omega=-0.9$ and  $\xi=0.01,0.1$ (see the definition in
eq.~\ref{eq:kernels}).
The differences between models are rather small, a clear indication
of the difficulty of determining the interaction using only luminosity distances.

\subsubsection{Angular Diameter Distance Tests.}

Angular diameter distances, $D_A=D_L(1+z)^{-2}$, have also been used
to constrain the rate of expansion. This distance is computed by
measuring the angular size subtended by a standard ruler of known size.
In models where the fraction of DE is negligible at recombination,
the particle and sound horizons at last scattering \cite{white_94f} are
such rulers. The pressure waves that propagate in the pre-recombination
photo-baryon plasma imprint oscillations in both the matter and radiation
power spectra. The angular scale subtended by the first acoustic peak
in the radiation power spectrum has been used to determine the spatial
flatness of the Universe \cite{hu_95f,hu_96f}. The effect of this
Baryon Acoustic Oscillations (BAOs) of the matter power
spectrum is to imprint a characteristic scale in the clustering of
matter and galaxies, which appears in the galaxy correlation function as
a localized peak at the sound horizon scale $r_s$.
The angular scale of the comoving sound horizon is
$r_s(z_*)=\int_{z_*}^\infty c_S H^{-1}dz$ where $c_S(z)$ is the sound speed
of the photon-baryon plasma. The CMB acoustic scale is $l_A=\pi (1+z_*)D_A(z_*)/r_s(z_*)$,
but this distance prior is applicable only when the model in question is based
on the standard FRW \cite{Komatsu_2011}. The interaction changes the expansion
history and the distance to the last scattering surface and the angular scale
of the acoustic peaks and the application of the BAO scale to interacting DM/DE models
is not straight-forward.

\begin{figure}
\centering
\epsfxsize=0.7\textwidth\epsfbox{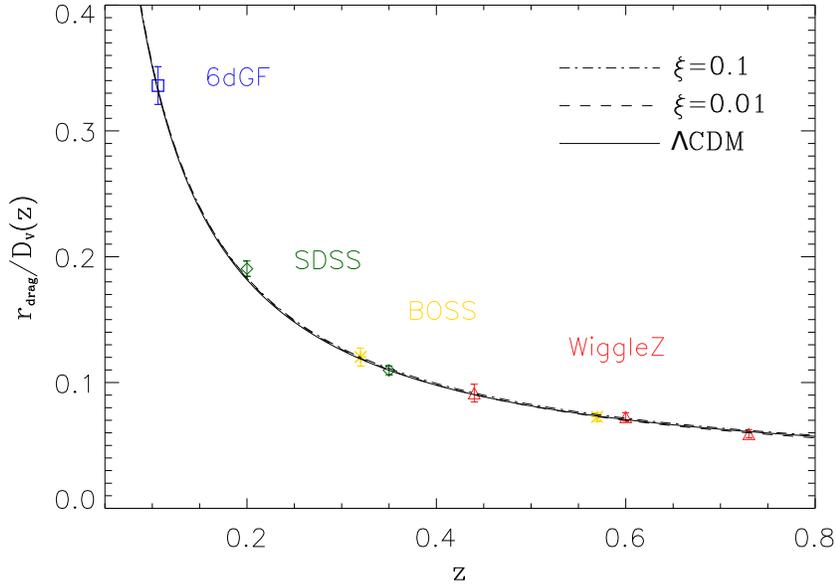}
\caption{Ratio of the BAO drag scale to the angle-averaged dilation scale $D_V(z)$.
The 6dFGS is from \cite{beutler_11f} (blue),
SDSS-III from \cite{percival_07f} (missing), the SDSS MGS \cite{ross_15f} (green),
the BOSS from \cite{anderson_14f} (gold) and WiggleZ \cite{kazin_14f} (red).
Like in Fig.~\ref{fig:dl_data}
solid, dashed and dot-dashed lines corresponds to the theoretical prediction
of the concordance model and the interacting DM/DE model IV for different interaction
parameters.}
\label{fig:bao_data}
\end{figure}

Assuming a $\Lambda$CDM cosmology,
the Planck Collaboration has determined  the sound horizon at
the drag scale $(z_d=1020)$, the scale relevant for the BAO which is
slightly different from the decoupling of the photon-baryon plasma $(z_*=1090)$,
is $r_{drag}=147.60\pm 0.43$Mpc \cite{Planck_XVI_14f,Planck_I_15f}
(Notice that this magnitude is independent of the Hubble constant).
This scale can be measured in the correlation function using galaxy
surveys of large volume like the Sloan Digital Sky Survey (SDSS)
\cite{Eisenstein_2005a}. Since the observed galaxy coordinates
are angles and redshifts, the conversion of coordinate separations to comoving
distances will depend on the angular diameter distance $d_A(z)$ and the
expansion rate $H(z)$. Errors on the two quantities are correlated, and
in the existing galaxy surveys the best determined combination is approximately
an average of the radial and angular dilation scale, $D_V(z)$, defined in
\cite{Eisenstein_2005a} as $D_V=[(1+z)^2D_A^2zc/H(z)]^{1/3}$.
BAO data have been used by  \cite{He_11f} to constrain DM/DE models.
In Fig~\ref{fig:bao_data} we represent the distance scale $r_{drag}/D_V(z)$ obtained
from different data sets: 2dFGRS \cite{2dF},
6dFGRS \cite{beutler_11f,6dF}, SDSS-III \cite{SDSS,percival_07f},
SDSS MGS \cite{ross_15f}, BOSS ``low-z'' and CMASS surveys \cite{anderson_14f,carvalho_15f}
and WiggleZ \cite{Wiggle,kazin_14f,beutler_15f}. In addition, we represent the concordance
LCDM model and two interacting DM/DE fluid models as in Fig~\ref{fig:dl_data}.

As an alternative to the sound horizon, the CMB shift parameter
$R=\sqrt{\Omega_m}\int_0^{z_{ls}}E(z')^{-1}dz'$ has been used a standard ruler.
In this expression $z_{ls}$ is the redshift of the Last Scattering surface and $E(z)$
is given in eq.~(\ref{Eofz}).
This parameter has been measure to be $R= 1.725 \pm 0.018$
\cite{WangYMukherjee_2007a} from  WMAP \cite{Hinshaw_2007,Komatsu_2011}
and has been applied in the literature to test interacting models.
\cite{Feng_08f,Feng_07f,He_08f,Micheletti_2009a,Micheletti_2009b}.

\subsubsection{Expansion Rate, look-back time and age.}

\begin{figure}
\centering
\epsfxsize=0.7\textwidth\epsfbox{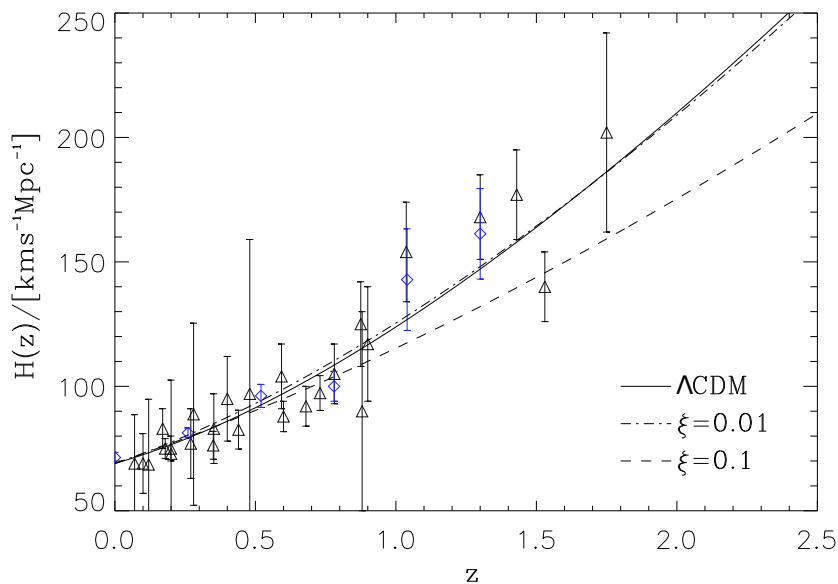}
\caption{Hubble factor as a function of redshift. The data has been
taken from \cite{heavens_14f}, \cite{oka_14f},
\cite{chuang_13f} and \cite{samushia_14f}
from the Baryon Oscillation Spectroscopic Survey Data Release  7, 9
and 11, respectively, \cite{blake_12f} (at $z = 0.44,0.6,0.73$) using data
from WiggleZ DE Survey. Black triangles  corresponds to the data from
the compilation of \cite{farooq_13f} and \cite{Font-Ribera:2013wce}
while diamonds are given in \cite{heavens_14f}.
Lines correspond to model predictions
and follow the same conventions than in Fig.~\ref{fig:dl_data}.}
\label{fig:hubble}
\end{figure}

The interaction modifies the rate of expansion of the Universe and changes
the Hubble parameter and  the age of astronomical
objects as a function of redshift compared with the concordance model.
Data on the Hubble expansion has been derived from  the differential ages of old,
passively evolving, galaxies \cite{Simon_2005a} or from the spectra of red
galaxies \cite{stern_10f}. The data provides a measurement of the expansion
rate of the Universe in an almost model independent way \cite{heavens_14f}.
A direct measurement of $H(z)$ was obtained by
\cite{gaztanaga_09f} using the 2-point correlation of Sloan Digital Sky Survey
luminous red galaxies and taking the BAO peak position as a standard ruler
in the radial direction and is independent of the BAO measurements of
\cite{percival_07f}. In \cite{hasinger_02f} the age of a quasar at $z=3.91$ was estimated
to be $t_{quasar}\le 2.1$Gyr and this value has been used to constrain cosmological models
\cite{friaca_05f}. In \cite{Wang_07f} and \cite{duran_11f} it was argued the age of
this quasar favors the existence of an interaction.
In Fig~\ref{fig:hubble} we represent the value of the Hubble expansion as
a function of redshift
measured from different data from BOSS \cite{oka_14f,chuang_13f,samushia_14f}
and WiggleZ \cite{blake_12f}. These data has been compiled by \cite{farooq_13f}
and \cite{Font-Ribera:2013wce} and are represented with triangles.
Also included is data from \cite{heavens_14f} (diamonds)
derived using SNIa, BAO data from Fig~\ref{fig:bao_data}
and expansion data from \cite{verde_14f}.
Lines show the prediction for the models  of Fig.~\ref{fig:dl_data}.

The look-back time (see Sec~\ref{sec:distances}) can also be used
to constrain models. In \cite{Jimenez_2003a}
and \cite{Simon_2005a} the ages of 35 and 32 red galaxies are respectively given.
For the age of the universe one can adopt $t_0^{obs} = 13.73 \pm 0.12$ Gyr
\cite{Komatsu_2009w} from e.g. 5 years WMAP data, the seven year data
\cite{Komatsu_2011} or else the most recent data from Planck, gives
$t_0^{obs}= 13.799\pm 0.013$Gyr \cite{AdePAR_2015a} (actually, we cannot
go that far in precision, considering the interaction, the above values
being model dependent).
Although this estimate for $t_0^{obs}$ has been obtained assuming
a $\Lambda CDM$ universe, it does not introduce systematical
errors in our calculation: any systematical error eventually
introduced here would be compensated by adjusting $df$, in
(\ref{eq:look-back}). On the other hand, this estimate is in perfect
agreement with other estimates, which are independent of the
cosmological model, as for example $t_0^{obs}=12.6^{+3.4}_{-2.4}$Gyr,
obtained from globular cluster ages \cite{Krauss_2003} and
$t_0^{obs}=12.5\pm3.0 $Gyr, obtained from radioisotopes studies
\cite{Cayrel_2001}.

\subsubsection{Cluster Number Counts.}

The cluster mass function and cluster redshift distribution probe the
late time evolution of the Universe and the associated DE effects
(see Sec.~\ref{sec:cluster_counts}).
These observables provide specific signatures of DM/DE
interaction and the existence of DE inhomogeneities \cite{basilakos_10f,He_10f}.
Also, high resolution simulations of LSS
formation show significant differences in the mass function between DM/DE
models and $\Lambda$CDM \cite{sutter_08f}.
Cluster counts are exponentially sensitive to the properties of
the DE but their effectiveness relies in obtaining a good estimators
of the cluster mass. They could be used to constrain the properties of DE
if the evolution in the relationship between observable quantities and the
cluster mass can be calibrated \cite{lima_05f}. Cluster surveys such
as those of DES, SPT, WFirst, Euclid or e-Rosita, detecting about $10^4$ clusters
out to redshifts $z=2$ would provide enough statistics to constrain
measure $\Omega_{DE},w_{DE}$ with 3\% and 6\% accuracy \cite{lima_04f}.
The dependence of cluster abundances and DM/DE coupling has been studied
in \cite{He_10f,Manera_06f} who found that increasing the coupling reduces significantly
the cluster number counts, and that DE inhomogeneities increases cluster
abundances. Wiggles in cluster number counts were shown to be a specific signature
of coupled DE models. The interaction can significantly enhance the
probability to observe very massive clusters at redshifts $z\ge 1.4$,
that are very unlikely in the concordance model \cite{baldi_11f}.

Observationally, cluster counts from the Planck Collaboration show the
data is not in full agreement with the concordance model.
In 2013, the Planck Collaboration described
a tension between the constraints on $\Omega_m$ and $\sigma_8$
from cluster counts and those from the primary CMB in the concordance model
\cite{Planck_XX_14f}, result confirmed by the analysis of the latest data
\cite{Planck_XXIV_15f}. At present, it is unclear if the tension arises from
low-level systematics in the data or is an indication of new physics;
unstable DM models help to ease the tension \cite{berezhiani_15f}
and could be more easily accommodated in interacting quintessence models.

\subsubsection{The Sandage-Loeb test.}

The Sandage-Loeb test provides a direct measurement of the expansion of the Universe
by measuring the redshift drift of extra-galactic sources. The test was first
proposed in \cite{sandage_62f} and it was shown to be feasible
using the spectra of distant quasars at the redshift interval $z=[2,5]$ \cite{loeb_98f}.
Measuring the redshift drift of quasar spectra would be useful to determine
the cosmic expansion history on a redshift interval where other DE
probes are unable to provide useful information \cite{corasaniti_07f}.
The data would be helpful to break the
degeneracy between the values of $\Omega_m$ and $H_0$ existing in BAO,
SNIa and CMB data \cite{geng_14fa}, to constrain the equation of state parameter of
DE and, consequently, also constrain any possible DM/DE interactions
at those redshifts \cite{geng_14fb}.
Monte Carlo simulations of data taken with high resolution spectrographs
using $\sim 40$m telescopes have been carried out to show how these
observations would constrain DM/DE interactions
\cite{geng_15f}.
The test has not been carried out yet due to the lack of data but as we
indicate in Sec.~\ref{sec:sec8},  it could
be within reach of the forthcoming generation of observational facilities.

\subsubsection{Equivalence Principle tests.}

The interaction between DM and DE produces a violation of the
equivalence principle between baryons and DM. The DE
modifies the gravity felt by the DM particles. At Newtonian
scales, the interaction simply renormalizes Newton gravitational constant
for the DM \cite{damour_90f,Olivares_06f}.
The skewness of the large scale structure is a probe of gravitational
clustering \cite{peebles_80f} and can be used to probe violations
of the equivalence principle \cite{amendola_04bf}. Moreover, if the DE
couples only to DM and not to baryons, as requested by the
constraints imposed by local gravity measurements, the baryon fluctuations
develop a constant, scale-independent, large-scale bias which is in principle
directly observable \cite{amendola_02f}.

\subsection{Constraints from large scale structure.}

The growth rate of large scale structure is another
very sensitive probe of the evolution of the gravitational
potential. The transition from a matter dominated Universe
to a period of accelerated expansion freezes the growth
of matter density perturbations \cite{peebles_84f, barrow_93f}.
The interaction modifies the scale-invariant Harrison-Zel'dovich
matter power spectrum \cite{harrison_70f,zeldovich_72f}
adding power at large scales \cite{duran_12f}
and leading to a mismatch between the
CMB-inferred amplitude of the fluctuations, late-time
measurements of $\sigma_8$ \cite{kunz_04f,baldi_11f}  and galaxy rotation curve \cite{Baldi2012P}.

\begin{figure}
\centering
\epsfxsize=0.7\textwidth \epsfbox{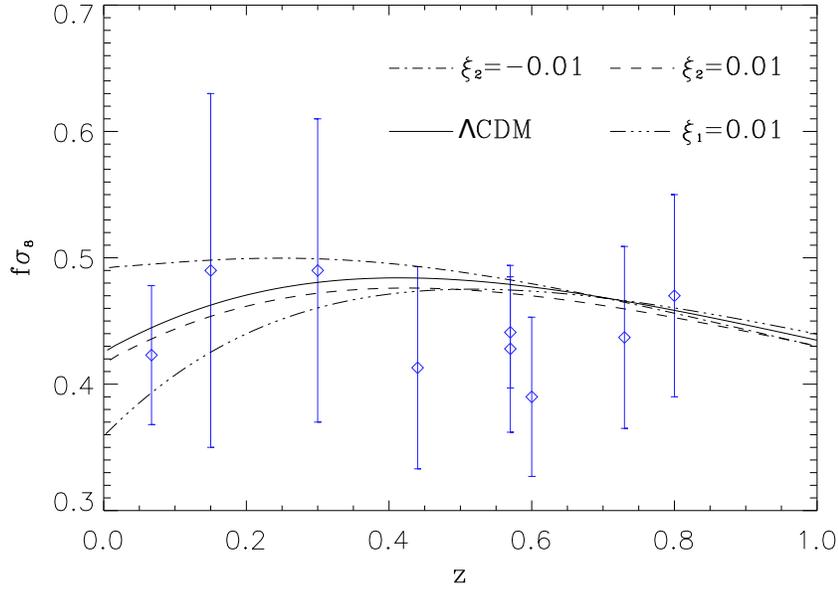}
\caption{Growth factor data in units of $\sigma_8$, obtained from different
surveys:  6dF \cite{beutler_11f} ($z=0.067$),
\cite{howlett_15f} ($z=0.15$), SDSS DR7 \cite{oka_14f} ($z=0.3$) and SDSS DR9,
\cite{chuang_13f} ($z=0.57$) and VIPERS \cite{torre_13f} ($z=0.8$).
Lines correspond to the concordance model and a DM/DE interacting fluid model
with EoS parameter $\omega_d=-1.1$ with different interaction parameters.
}
\label{fig:growth_data}
\end{figure}

\subsubsection{The growth rate of matter density perturbations.}

In the concordance model, the vacuum energy is
homogeneously distributed within the horizon and only perturbations
in the matter fluid are considered. If DE perturbations exist,
they affect the evolution of matter perturbations
through the gravitational field
and are themselves affected by the interaction.
The  dimensionless growth rate is defined as $f=d\ln D_+/da$ and $D_+$ is the growth factor; it is customary
to express it as $f(\Omega_m)\simeq\Omega_m^{\gamma(\Omega_m)}$ \cite{Peebles_93f}.
In most models of DE, structure formation stops
when the accelerated expansion begins. In contrast,
the coupling of DE to DM can induce the growth
of perturbations even in the accelerated regime \cite{amendola_02f,
Olivares_06f}. Depending on the parameters, the growth may be much faster than
in a standard matter-dominated era.  Then, the growth
index $\gamma$ probes the nature of DE and can  discriminate between models
\cite{linder_07f, tsujikawa_09f, sapone_10f, duran_11f, He_09cf}.
Data on the product $f(z)\sigma_8(z)$ of the  growth rate $f(z)$ of matter
density perturbations and the redshift-dependent rms fluctuation
of the linear matter density field $\sigma_8$ has been compiled by
\cite{basilakos_13f}. This estimator is (almost) model independent
\cite{song_09f}. The Euclid satellite is expected to measure
the growth factor within 1 to 2.5\% accuracy for each of 14 redshift
bins in the interval of redshifts $[0.5-2]$
\cite{amendola_13f} (see Sec~\ref{sec:sec9}).
A model independent test to probe possible departures from the
concordance model at perturbation level was proposed in \cite{nesseris_14f}
and found that if the data on the growth factor was free of systematics, it
was in conflict with the concordance model.

There are also effects at high redshifts. Due to the interaction
the growth rate of matter density perturbations during the radiation
dominated regime is slower compared to non-interacting models with
the same ratio of DM to DE today. This effect
introduces a damping on the power spectrum at small scales
proportional to the strength of the interaction similar to the
effect generated by ultra relativistic neutrinos \cite{Olivares_06f}.
The interaction also shifts matter-radiation equality to larger
scales \cite{duran_12f}.

In Fig~\ref{fig:growth_data} we present a compilation of growth factor
data from different surveys, measured in units of the amplitude of the
matter power spectrum at $8h^{-1}$Mpc, $\sigma_8$ \cite{beutler_11f,chuang_13f,
howlett_15f,kazin_14f,marin_15f,oka_14f,torre_13f}.
The different lines correspond to the
concordance model and a DM/DE interacting fluid model with EoS parameter
$\omega_d=-1.1$. The interaction parameters are given in the figure.

\subsubsection{Redshift Space Distortions.}

The growth of cosmological structure induces galaxy peculiar
velocities, i.e. coherent flows of galaxies towards matter
overdensities. When redshifts are used to map galaxy positions,
the reconstructed spatial distribution of objects is distorted
in the radial direction, effect known as Redshift Space Distortions (RSD).
On large scales galaxies trace the linear growth of cosmological
structures enhancing the amplitude of the 2-point correlation
function; on small scales virialized structures are elongated
along the line of sight. The amplitude of the former effect is
directly proportional to the logarithmic growth rate of density fluctuations
$f_m(z)=d\ln\delta(a)/d\ln a$ \cite{kaiser_87f,hamilton_92f}.
The recent RSD measurements
\cite{Beu,samushia_14f,Eriksen} indicate values smaller than the
$\Lambda$CDM prediction.
Analysis of the BOSS sample using RSD has yielded the tightest
constraints to date on the growth rate of structure \cite{reid_12f}.
In \cite{mohammad_15f} the cross-correlation between
galaxies and galaxy groups was used to measure RSD
as a probe of the growth rate of cosmological structure.

\subsubsection{The Alcock-Paczynski test.}

Alcock and Packzynki (A-P) test expresses the idea that if one assumes
the wrong cosmological model to convert redshift measurements to distances,
an intrinsically spherical object or pattern comoving with the Hubble flow
will appear ellipsoidal \cite{alcock_79f}. The distortion is proportional to $D_A(z)$
and $H(z)$. The BAO is an example of the A-P test but with an object of a known size
\cite{kim_14f}. The test is partially degenerate due to the
apparent ellipticity of clustering caused by RSD. Although the two effects
have a different dependence with scale,  with a sufficient large volume one
could expect to separate them \cite{matsubara_04f}.

\begin{figure}
\centering
\epsfxsize=0.7\textwidth \epsfbox{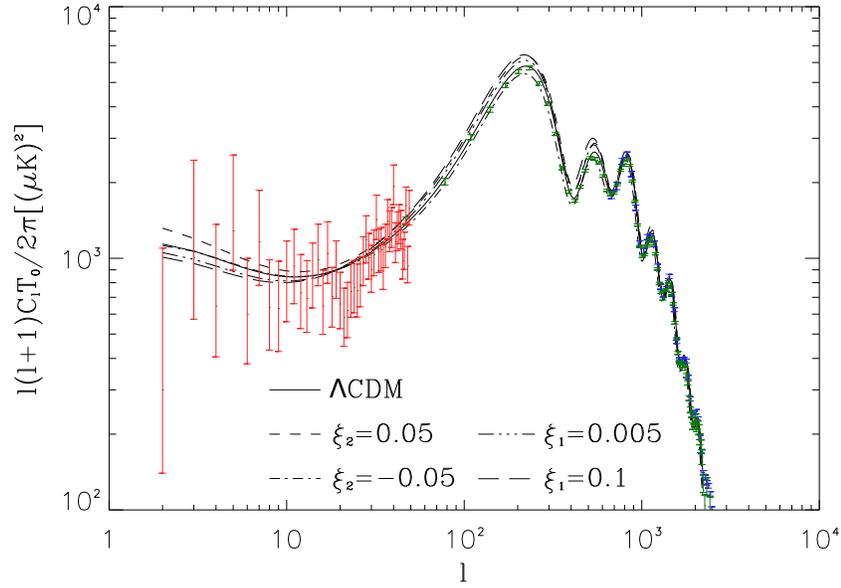}
\caption{Radiation Power Spectrum from Planck (low multipoles
red and high multipoles in green) and the South Pole Telescope (blue).
Planck data can be downloaded from the Planck Legacy Archive
{\it http://www.cosmos.esa.int/web/planck/pla} and
the latest compilation of SPT data (as of May 2014)
can be downloaded from {\it http://pole.uchicago.edu/public/data/story12/index.html}
Lines correspond to the theoretical model predictions for the
concordance $\Lambda$CDM model and the interacting DM/DE fluid model. In this case,
the EoS parameter is
$\omega_d=-0.9$ when $\xi_2\ne 0$ and $\omega_d=-1.1$ when $\xi_1\ne 0$.
}
\label{fig:cmb_data}
\end{figure}

\subsection{Constraints from CMB temperature anisotropies.}

The Cosmic Microwave Background is the main source of information
on the physics of the early Universe.
The different time evolution of the gravitational potentials in interacting
models induces several effects that change the radiation power
spectrum with respect to the concordance model.
Baryons and DM  evolve differently, affecting
the ratio of height between the odd and even peaks \cite{amendola_12f}.
The lensing potential changes modify the lensing B-mode contribution
\cite{amendola_14f}.  The sound speed of gravitational waves is
also modified and it affects the amplitude of the primordial B-mode \cite{raveri_14f}.
The damping tail of the radiation power spectrum varies, providing  a measurement of the abundance of DE at different redshifts
\cite{calabrese_11f,reichardt_12f}. Comparing models with data
requires to follow the evolution of all energy density components by
solving the perturbation equations from some early time
\cite{amendola_04af,Amendola_03bf,Olivares_05f,Olivares_08af}.

CMB temperature anisotropies have been measured by several
experiments out to $\ell=3000$ \cite{bennett_13f,sievers_13f,story_13f,Planck_XV_14f}.
Earlier studies on DM/DE interaction used WMAP, and additionally other
data sets on CMB temperature anisotropies, to set upper limits
on the strength of the DM/DE coupling
\cite{Olivares_05f,Olivares_08af,Gavela_09f,Vacca_09f,Xia_09f,Valiviita_10f,
pettorino_12f,Salvatelli_13f,Costa_14f,Costa_2014b, He_09bf,He_11f,HeJH_2011f} or
to constrain the cross-section of DM/DE interactions \cite{Xu_12f}.
More recently, Planck CMB data have been used to constrain interacting models
with different kernels discussed in the literature:
$Q=H\xi_1\rho_c$ \cite{Costa_14f,valiviita_15f},
$Q=H\xi_2\rho_{d}$ \cite{Costa_14f,Salvatelli_13f,yang_14fa,yang_14fb,bolotin_15f},
$Q=H(\xi_1\rho_c+\xi_2\rho_d)$ \cite{Costa_14f,AbdallaE_2015a,Pu_14f}.
Other couplings have also been discussed; \cite{Salvatelli_14f}
has shown that the data provides a moderate Bayesian evidence in
favor of an interacting vacuum model. Nevertheless, since
CMB temperature anisotropy data probes the DM density at the
time of last scattering, i.e., at redshift $z\sim 1100$ while the interaction
preferentially modifies the evolution between the last scattering surface and
today, the data is degenerate with respect to $\Omega_ch^2$ and the interaction
rate and complementary data sets are required to break the degeneracy such as Baryon
Acoustic Oscillations, supernova data, the growth of matter
density perturbations and the ISW
discussed below \cite{Valiviita_10f,He_10f,Xu_11f,clemson_12f}.

Data on temperature anisotropies, covariance matrices, window functions
likelihoods and many other resources are publicly available\footnote{All
data products can be downloaded from {\it http://lambda.gsfc.nasa.gov/product/}}.
As an illustration, in Fig~\ref{fig:cmb_data}
we plot the radiation power spectrum measured by the Planck Collaboration
(red and green)
\cite{Planck_I_15f} and the South Pole Telescope \cite{story_13f} (blue).
For comparison, we also superpose  the model predictions for the
concordance $\Lambda$CDM model and the interacting DM/DE fluid model. In Model I the EoS parameter is
The EoS parameter is $\omega=-0.9$ when the kernel is proportional to the
$\omega_d=-0.9$ and in Model III is $\omega_d=-1.1$.

\subsubsection{The ISW Effect.}

The decay of the gravitational potential
 affects the low-multipoles of the radiation
power spectrum by generating ISW temperature
anisotropies \cite{sachs_67f,kofman_85f,hu_95f}. These
anisotropies are generated by evolving gravitational potential
at the onset of the accelerated expansion and change with
the interaction \cite{He_09bf,HeJH_2011f}.
The contribution to the radiation power spectrum due to the evolution
of the gravitational potential at low redshifts is given by eq.~(\ref{eq:isw_auto}).
The ISW effect at low redshifts is
generated by local structures, the spatial pattern of ISW anisotropies
will correlate with the distribution of the large scale structure and
the cross-correlation with a template of the matter distribution is
given by eqs.~(\ref{eq:isw_auto},\ref{eq:isw_cross}) \cite{Crittenden_96f}.
Several groups looked for evidence
of ISW effect by cross-correlating the WMAP data with templates built
from different catalogs \cite{Boughn_04f,Nolta_04f}. The ISW signal
at the position of superclusters is larger than expected in the
$\Lambda$CDM model \cite{granett_08f}. An update of these
earlier results have been carried out by the Planck Collaboration that reported
a $\sim 4\sigma$ detection of the ISW effect \cite{Planck_XIX_14f} and
confirmed the previously mentioned ISW anomaly. Since the DM/DE
interaction damps the growing mode of the Newtonian potential faster
compared to models with no interaction so it enhances the ISW effect
\cite{giannantonio_08f}. Comparison of the fraction of DE measured in CMB
temperature maps which obtained from the ISW effect
would provide a direct evidence of DM/DE interaction \cite{Olivares_08bf}.
The Planck Collaboration constrained the vacuum energy density to be
$0.49<\Omega_\Lambda<0.78$ at the 68\% confidence level compatible with the
concordance model but also in agreement with interaction models with $\xi\le 0.01$
\cite{Olivares_08bf}. Similar constraints but on a different coupling were
obtained by \cite{Schaefer_08f,Xia_09f,mainini_12f}.

\subsubsection{Constraints from Peculiar Velocities.}

The interaction does not only change the time evolution of
the gravitational potentials. It also changes the potentials themselves
\cite{Olivares_06f}. Since the gravitational potential gives rise to
peculiar motions, the interaction, in turn, will modulate peculiar
velocities of baryons and their effect on the CMB temperature anisotropies.
The correlation of spatial variations in the distribution of galaxy
luminosities with the peculiar velocity field has been shown to be a
powerful test of gravity and dark energy theories on cosmological scales
\cite{feix_15f}.
In addition, while in the concordance model baryons trace the DM
distribution in the linear regime, in DM/DE interacting models
baryons do not follow DM as closely due to the DE inertial drag on the DM \cite{He_10f}.
Peculiar velocities modify temperature fluctuations on the CMB at
small scales via the conventional Kinetic Sunyaev-Zel'dovich (kSZ)
effect \cite{sunyaev_72f} generated by the ionized gas in the diffuse
intergalactic medium and in the potential wells of galaxy clusters. Large scale
flows have been detected in WMAP \cite{kashlinsky_08f,kashlinsky_10f}
and Planck \cite{atrio_14f} data, the pair-wise velocity dispersion
of the unresolved cluster population has been measured \cite{hand_12f}
and the kSZ radiation power spectrum has been constrained using the
data on CMB temperature anisotropies \cite{george_15f,Planck_XIII_15f}.
In \cite{xu_13f} it was determined that peculiar velocities could
be from five times smaller to two times larger than in the concordance
model and showed that peculiar velocities could provide constraints
stronger than those derived from the ISW effect. The evidence of interaction
between dark sectors in the kSZ observations were discussed in \cite{xu_13f}.

\subsection{Gravitational lensing.}

The effect of lensing is to remap the CMB fluctuations. The effect is
parameterized in terms of the lensing potential $\varphi$ \cite{lewis_06f},
\begin{equation}
\varphi(\hat{n})=-2\int_0^{\varsigma_{rec}}d\varsigma\frac{\varsigma_{rec}-\varsigma}{\varsigma_{rec}\;\varsigma}
\Psi(\varsigma\hat{n};\tau_0-\tau),
\end{equation}
itself a function of a gravitational potential $\Psi$.
The observed anisotropy in direction $\hat{n}$ is the unlensed (primordial)
anisotropy in the direction $\hat{n}+\nabla\varphi(\hat{n})$.
In this expression we have assumed that the Universe is spatially flat;
$\varsigma$ is the conformal distance, $\tau$ the conformal time,
$\tau_0$ the conformal time at present and the integration is carried
out up to recombination.  The change of the gravitational  potential changes the lensing of the CMB radiation power spectrum; therefore the lensing potential power spectrum  probes
the matter power spectrum up to recombination \cite{acquaviva_06f}. It is also sensitive
to the growth index \cite{smith_09f}. The {\it Planck}
satellite has measured the lensing potential using data on temperature
and polarization anisotropies with a statistical significance of $40\sigma$
and has released an estimate of the lensing potential over approximately 70\%
of the sky,  in band powers
for the multipole range $40\le \ell\le 400$ and with an associated likelihood for
cosmological parameter constraints \cite{Planck_XV_15f}. The data reveals
some tension with the concordance $\Lambda$CDM model: the lensing amplitude is $A_L=1.22\pm 0.10$, which
is in $2\sigma$ tension with the amplitude of the CMB spectrum reconstructed from lensing
deflection angle spectrum, $A^{\phi\phi}_L=0.95\pm 0.04$ while both quantities
should be unity in the concordance model. In \cite{hu_15f} it was argued that some models
whose effective Newton gravitational constant is larger than that in $\Lambda$CDM could
explain the discrepancy. However, in order
to lens the CMB anisotropy, the growth of matter density perturbation needs to be
enhanced, giving a value of $\sigma_8$ larger than the measured value.
A full non-linear study of gravitational lensing on different cosmological
models is still lacking.
In \cite{carbone_13f,pace_14f} the impact of the DM/DE coupling  on weak
lensing statistics was analyzed by constructing realistic simulated weak-lensing
maps using ray-tracing techniques through a suite of N-body cosmological simulations.
Model-independent constraints on the growth function of structure and the
evolution of the DE density can be obtained from the reconstruction through
lensing tomography \cite{hu_02bf} and further information
can be obtained from the power spectra and cross-correlation measurements of
the weak gravitational lensing in the CMB and from the
cosmic shearing of faint galaxies images \cite{hu_02af}.

\subsection{Model Selection Statistics.}

The data sets described above are not exclusively tests of interacting
DE models. Fitting data to the model predictions to determine/constrain
its parameters is not sufficient to establish the validity of the model.
The concordance model requires only the energy density
associated to the cosmological constant to explain the current period
of accelerated expansion. In contrast, interacting DE models require at
least two extra parameters: the EoS and
the DM/DE coupling constant. It could seem that further improvement
on the available data sets will help as to constrain/measure the cosmological
parameters with higher accuracy. Equally important is the inference problem of
allowing the data to decide the set of parameters needed to explain the
data, known as model selection \cite{liddle_07f}. Adding extra parameters
necessarily improves the fit to the data at the expense of reducing
the predictive power of the model \cite{liddle_04f}. The purpose
of model selection statistics is to address whether the improved fit
favors the introduction of extra parameters. The most commonly used
criteria id the Akaike Information Criterion (AIC) \cite{akaike_74f} and
the Bayesian Information Criterion (BIC) \cite{schwarz_78f}.
If $\hat{\cal L}(\theta)$ is the maximum of a likelihood function of
a model of $\theta$ parameters, $N$ the number of data points and $k$ the number
of parameters to be estimated from the data, then
\begin{equation}
AIC=2k-2\ln\hat{\cal L}(\theta);\qquad
BIC=-2\ln\hat{\cal L}(\theta) + k\ln N .
\end{equation}
Information criteria penalize the introduction of new parameters that do not significantly
improve the quality of the fit. For instance, if adding an extra parameter reduces
the BIC by 2-6 units, the data shows a positive evidence in favor of the new parameter
being required to explain the data; if the decrement is 6-10, the evidence is
called strong and if it is larger than 10, very strong \cite{liddle_04f}.
The BIC generally penalizes the number of parameters more strongly than the AIC,
although it depends on the number of data points $N$ and parameters $k$.

\section{Observational Constraints on Specific Models.}\label{sec:sec7}

Determining the properties of interacting DE/DM models requires the use of all the
available observational data, combining different probing techniques
described in Sec.~\ref{sec:sec6} and has been extensively considered in
the literature \cite{Amendola_2000a,Amendola_2007a,Feng_07f,Bean_08bf,Micheletti_2009a}.
Comparisons with the CMB data using WMAP and Planck results have also
been carried out \cite{Olivares_05f,Olivares_08af,Costa_14f,Salvatelli_13f}.
Further constraints, relying on cluster properties \cite{Abdalla_09f,Abdalla_10f}
as well as structure formation \cite{He_10f}, age constraints and other properties
\cite{Shen_05a,Wang_07f,Abdalla_2007PLB,valiviita_15f} have also been discussed.

In this section, we will review the constraints that observations
have imposed on some specific models. Since the data is constantly evolving,
not all the models have been tested using the most recent data.

We will first go over the constraints obtained by using the observational data on the universe expansion history. Those data are not only got from distance based methods such as the SN data, but also obtained based on time-dependent observables for instance the age estimates of galaxies. We will combine four fundamental observables  including
the new 182 Gold SNIa samples \cite{Riess_2006}, the shift parameter of the CMB
given by the three-year WMAP observations \cite{WangYMukherjee_2007a}, the BAO measurement from the SDSS \cite{Eisenstein_2005a}and age estimates of
35 galaxies provided in \cite{Simon_2005a, santos2007} to perform the joint
systematic analysis of the coupling between dark sectors.

Furthermore we will review the constraints on the interacting DE models by employing the data from the CMB temperature anisotropies together with some other external data described as follows:
\begin{enumerate}
\item{} CMB temperature anisotropies:
From the 2013 Planck data release we use
the high-$\ell$ TT likelihood, which includes
measurements up to a maximum multipole
$\ell_{max} = 2500$, combined with the low-$\ell$ TT
likelihood, which includes measurements in the range $\ell= 2-49$
\cite{Planck_I_14f,Planck_XIII_15f} (Fig.~\ref{fig:cmb_data}).
We also include the polarization measurements from WMAP 9yr \cite{bennett_13f},
in particular the the low-$\ell$ ($\ell<32$) TE, EE and BB likelihoods.

\item{} BAO: We combine the results from three
data sets of BAO: the 6DF at redshift $z = 0.106$
\cite{beutler_11f}, the SDSS DR7 at redshift $z = 0.35$
\cite{padmanabhan_12f} and the SDSS
DR9 at $z =0.57$ \cite{anderson_14f} (Fig~\ref{fig:bao_data})

\item{} SNIa data: The Supernova Cosmology Project (SCP) Union 2.1
compilation \cite{Suzuki:2011hu} with 580 measured luminosity distances
(Fig.~\ref{fig:dl_data})

\item{} Hubble constant data: Finally we also include the
Hubble constant $H_0 = 73.8 \pm 2.4 km s^{-1}Mpc^{-1}$,
measured by \cite{Riess:2011yx}. This value is in tension with the
result of the Planck Collaboration \cite{Planck_I_14f,Planck_I_15f},
so the constraints derived using $H_0$ could be shifted slightly
if a different value is used.
\end{enumerate}
We will limit the study to models with $\Omega_k=0$; our numerical
calculations have been performed using the CMBEASY \cite{Doran 05f} and CAMB
\cite{Lewis 1999bs} codes. We modified the codes to include the effect
of the DM/DE coupling at the background and perturbation level.
To compare the theoretical predictions with observations, we perform
Monte Carlo Markov Chain (MCMC), using a modified
version of the CosmoMC program \cite{Lewis:2002ah, lewis_06f}.

\begin{figure*}[htbp!]
\centering
\includegraphics[scale=0.55]{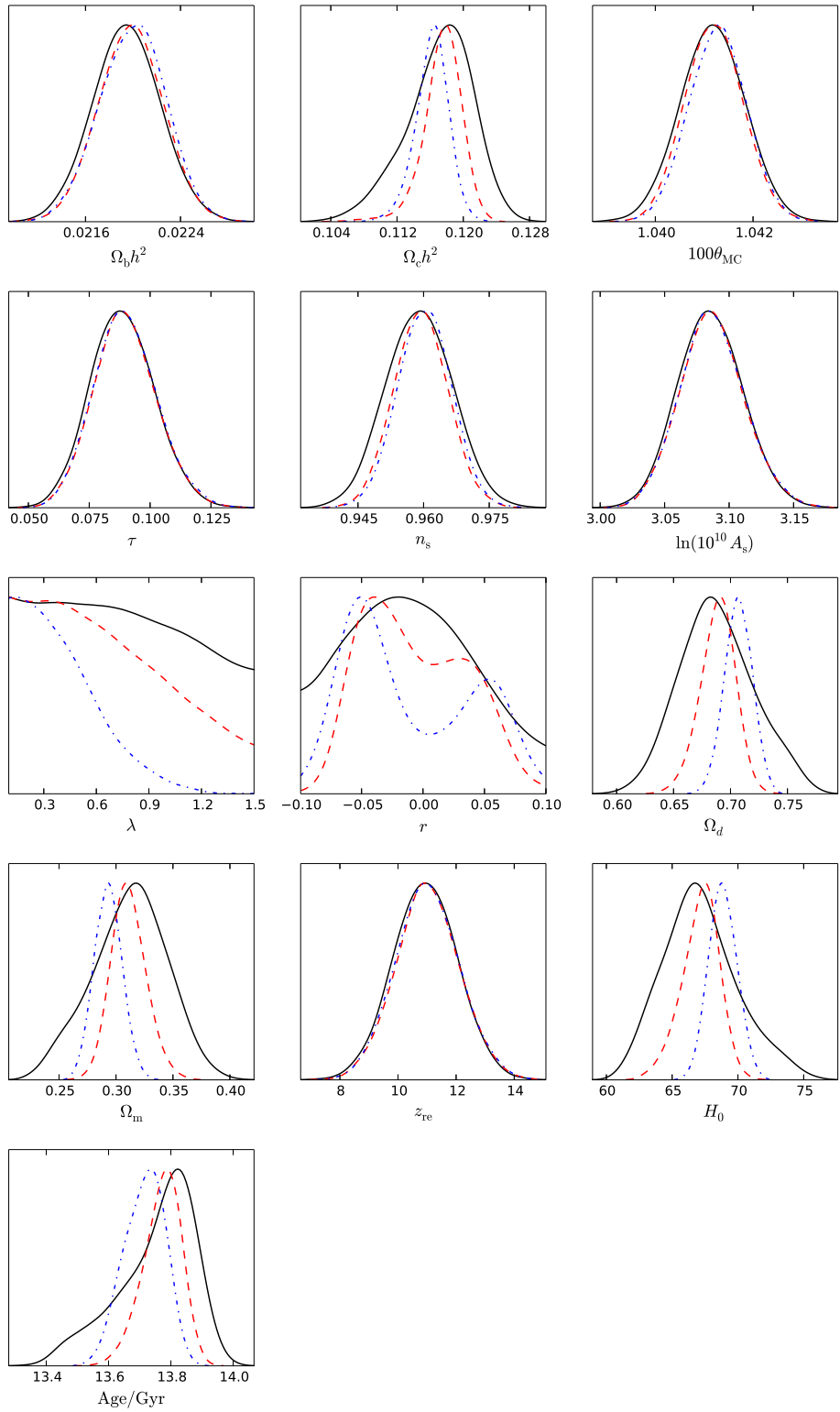}
\caption{Marginalized likelihoods for the parameters of the Yukawa model. The black
solid lines correspond to the Planck constraints, the red dashed lines
correspond to Planck+BAO and the blue dot-dashed lines correspond to
Planck+BAO+SNIa+$H_0$.}\label{figure1_sec8}
\end{figure*}

\subsection{Constraints On The Phenomenological Interaction From The Universe Expansion History}

In this subsection we examine different phenomenological interaction models between DE and DM by performing statistical joint analysis
with observational data arising from the 182 Gold type Ia
SN samples, the shift parameter of the CMB given by the three-year WMAP observations, the BAO measurement
from the SDSS and age estimates of 35
galaxies.  We compare the compatibility of these data sets. Especially we find that including the time-dependent observable, we can add
sensitivity of measurement and give complementary results for the
fitting.  The complementary effect of adding the time observable was also observed by using the lookback time observation in \cite{pires2006}.    Detailed analysis of our result can be found in \cite{Feng_08f}.

We do not specify any special model for DE, but consider as an example the commonly used parametrization of the EoS of DE in the form $\omega_{d}(z)=w_0+\frac{w_1z}{(1+z)^2}$. After employing the MCMC method to explore the parameter space, we obtain the constraints on the model parameters in the following table when the coupling between DE and DM is taken proportional to energy densities of DM, DE and total dark sectors (T), respectively.

\tabcolsep0.2in

\begin{small}PARAMETERS AT $68.3\%$ CONFIDENCE LEVEL\\
\begin{tabular}{cccccc}
\hline \hline
Coupling&EoS & $w_0$ & $w_1$ & $\Omega_m$ & $\xi/\xi_1/\xi_2$\\
\hline
T&$\omega_{d}$&$-1.50^{+0.31}_{-0.30}$&$3.90^{+2.09}_{-2.31}$&$0.26^{+0.02}_{-0.02}$&
$0.01^{+0.03}_{-0.03}$\\
\hline

DM&$\omega_{d}$&$-1.50^{+0.32}_{-0.31}$&$3.91^{+2.12}_{-2.34}$&$0.25^{+0.02}_{-0.02}$&$0.01^{+0.04}_{-0.04}$\\
\hline

DE&$\omega_{d}$&$-1.49^{+0.31}_{-0.30}$&$3.78^{+2.13}_{-2.39}$&$0.26^{+0.02}_{-0.02}$&$0.05^{+0.06}_{-0.10}$\\
\hline
\end{tabular} \\
\end{small}
\begin{scriptsize}For the chosen EoS, the priors on the model parameters are $-10<w_0<10$,$-15<w_1<15$,
$0<\Omega_m<0.8$, $-1<\xi_i<1$.
\end{scriptsize}\\

In \cite{Feng_08f} it was argued that adding the age constraint, the coupling between DE and DM tends to be a small positive value which gives more compatibility among different data sets.  We will see below that this result is consistent with the constraints obtained by using the CMB temperature anisotropies and other external observations.

\subsection{Field Description of The Interaction Between DE and DM.}

In Sec.~\ref{sec:sec2} we presented a  field description of
DM, DE and their interaction. In Fig.~\ref{figure1_sec8} we present
the results of the likelihood analysis for a DE described as
an scalar field with an exponential potential
$V(\varphi)=Ae^{-\lambda\varphi/M_{pl}}$ and an interaction
of the linear type: $F(\varphi)=M-\beta\varphi$ (see Sec.~\ref{sec:k-essence}).
The influence of the fermionic mass $M$ and the Yukawa coupling $\beta$
is degenerate and both parameters can not be fit simultaneously.
Let us define $\varrho=\beta /M$ to represent the amplitude of the Yukawa
coupling in units of the fermionic mass. Only this parameter will be
constrained by the observations. This approach has the advantage of
decreasing one degree of freedom.
The cosmological parameters have similar values to those in the
concordance model. The constraint is much weaker on the parameter of the DE
self-interaction potential $\lambda$. For Planck data alone the likelihood of
interaction parameter $r$ is almost symmetric around zero. Adding low redshift
data, $\lambda$ tends to its lower limit, while $r$ breaks the symmetry around
zero. Including BAO, SNIa and $H_0$, a null interaction is disfavored and
the likelihood of $\varrho$ shows a preference for negative values.

Furthermore it was found \cite{Costa_2014b} that if we can determine the
scalar potential parameter $\lambda$, for example if we have a
theoretical model to fix it as a large value,  we observe that the
Yukawa interaction between DE and DM can be preferred by the
cosmological data. This shows that the field description of the
interaction between DE and DM is compatible with observations.
In addition, the best fit value of the cosmological parameter that we have obtained helps to alleviate the coincidence problem, since there will be
 more time for the DM and DE energy densities to become comparable.
This will be discussed further in the end of this section.

For the DE described by the scalar field as $k$-essence, more
discussions on breaking the degeneracy of the model parameters
can be found in \cite{Micheletti_2009a}. For more general discussion
on the constraints of the non-minimally coupled k-essence type DE
models, the readers can refer to \cite{Planck_XIV_15f}.

\begin{figure}
\begin{center}
\begin{tabular}{c}
Model I \nonumber\\
\includegraphics[width=5in,height=1.7in]{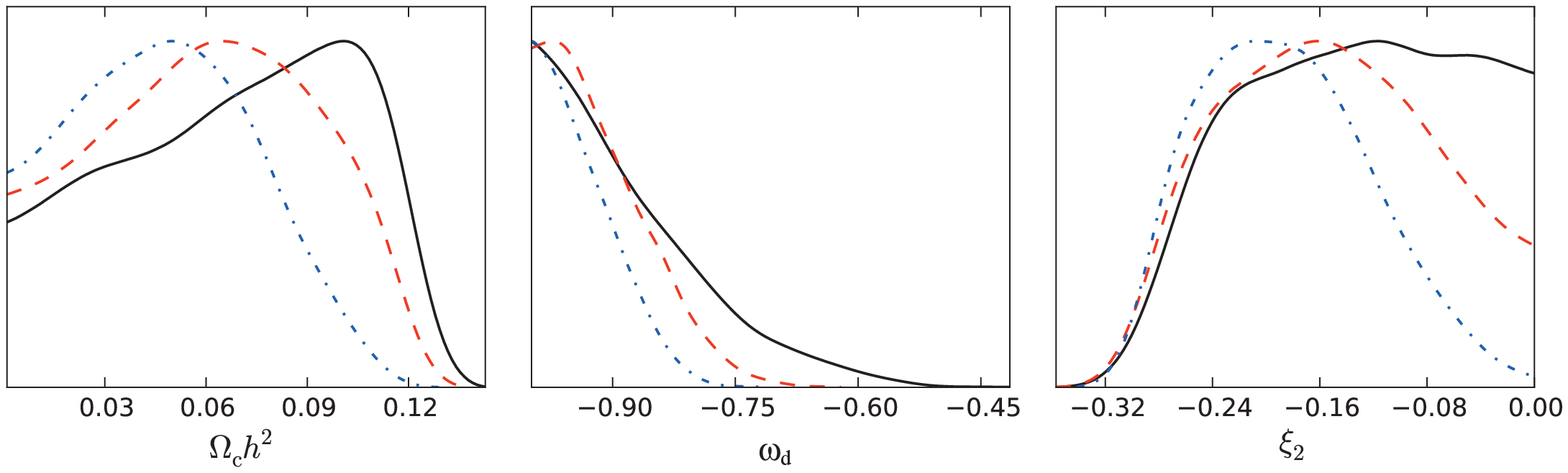}\nonumber\\
Model II \nonumber\\
\includegraphics[width=5in,height=1.7in]{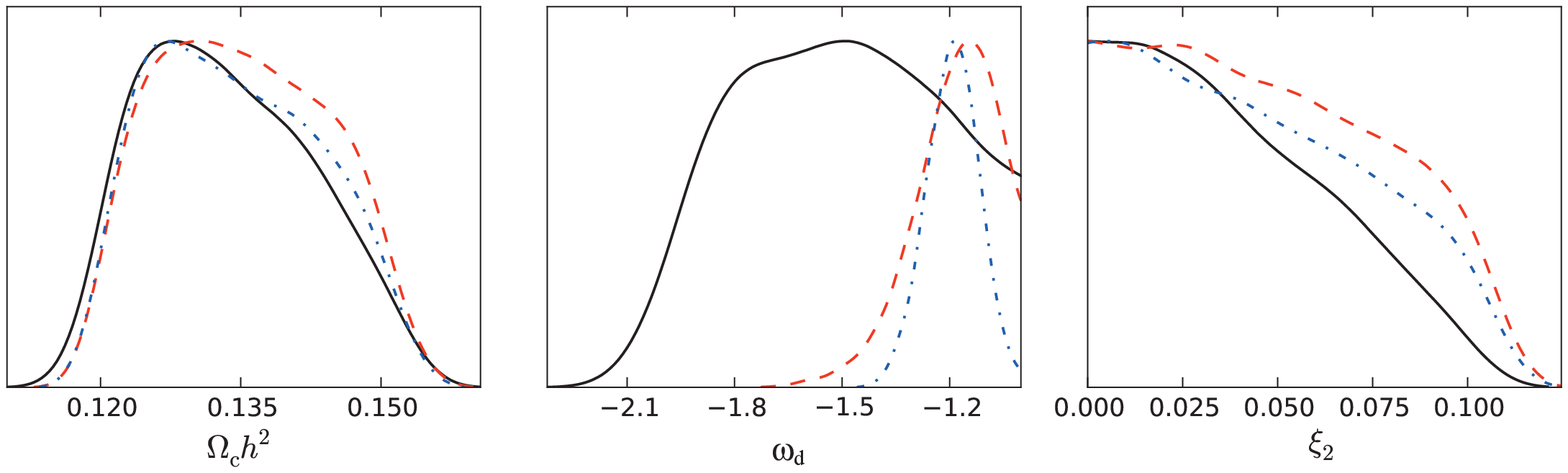}\nonumber\\
Model III \nonumber\\
\includegraphics[width=5in,height=1.7in]{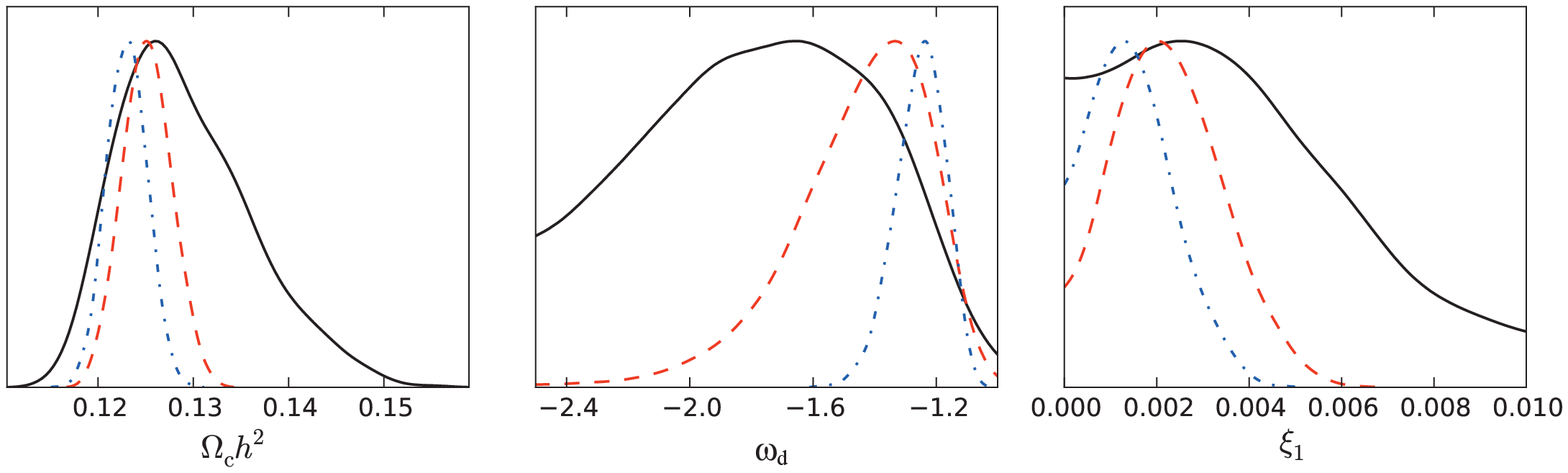}\nonumber\\
Model IV \nonumber\\
\includegraphics[width=5in,height=1.7in]{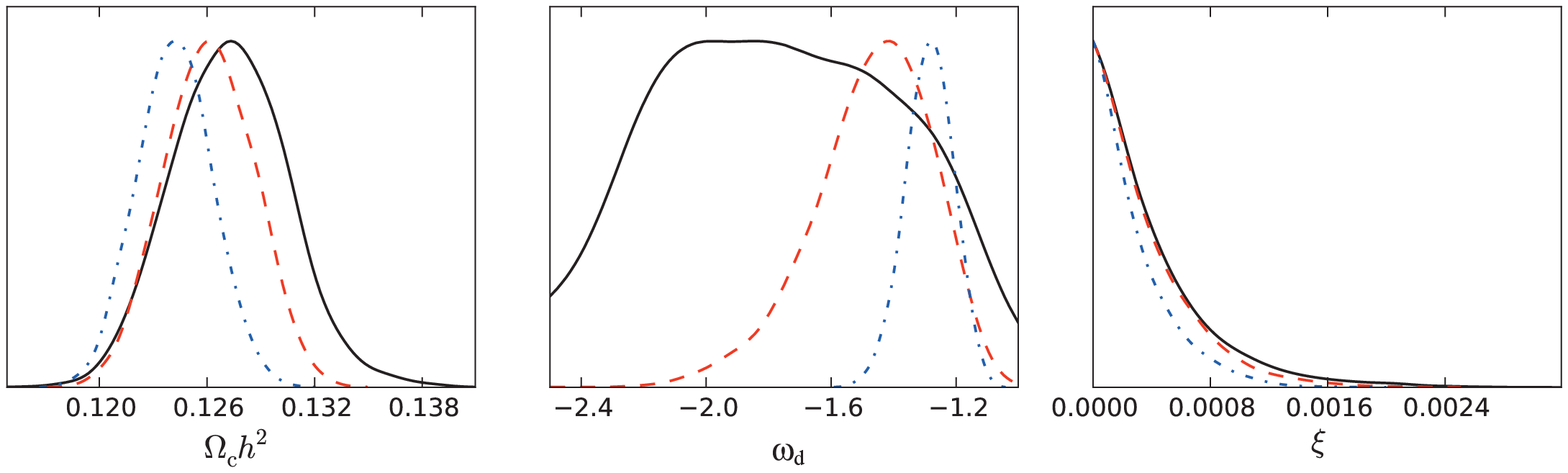}\\
\end{tabular}
\end{center}
\caption{Marginalized likelihoods for the parameters describing interacting fluid
models of Table~\ref{table:models}. Black solid lines correspond to the
constraints from Planck data alone, red dashed lines to Planck+BAO and blue
dot-dashed lines to Planck+BAO+SNIa+$H_0$.}
\label{figure2_sec8}
\end{figure}

\subsection{Phenomenological Description of The Interaction Between DE and DM.}

The phenomenological description of the interaction between
dark sectors was introduced in Sec.~\ref{sec:sec2} and the linear
perturbation theory of the model was discussed in Sec.~\ref{sec:sec4}.
For the sake of simplicity in our subsequent discussion, we will review
only models with a DE EoS parameter $\omega_d=const$. The results for
a variable EoS have been reported in \cite{Pu_14f,Xu_11f}.
We restrict our analysis to the models that satisfy the
stability condition (Sec.~\ref{sec:sec4.stability}). The interaction kernels
were summarized in Table~\ref{table:models}.

The results of the MCMC analysis using different data sets
are shown in Fig.~\ref{figure2_sec8}. When
the coupling is proportional to the energy
density of the DE, the data constraints the value of the interaction parameter
to be in the range $-0.3\le\xi_2\le 0.1$.
When the coupling is proportional to the energy density of the DM or
the total energy density of the dark sector, the constraint is much tighter and the coupling are positive at the 68\%
confidence level (CL).  Including additional data
tightens the constraints on the cosmological parameters compared with the CMB data alone.
Results for different  models can be
found in \cite{ Amendola_03bf,
Wang_07f,Guo_07f,Feng_07f,Feng_08f,Xia_09f,
Martinelli_10f,Valiviita_10f,Fabris_10f,Honorez_10f,
He_09bf,He_11f,Xu_11f,Xu_12f,Salvatelli_13f,Xia_13f,
Costa_14f,Salvatelli_14f,Xu lixin14, Li_14f, Li_14fb,Bertolami_07f,Abdalla_09f,Abdalla_10f, Murgia2016}.
The conclusion of these studies shows that interacting DM/DE models  are compatible with observations. In \cite{Salvatelli_14f} it was found
evidence of the existence of interaction but with a low CL.
More recently the Planck Collaboration also found that  coupled
DM/DE were in agreement with observations
\cite{Planck_XIV_15f}. In the next section
we will further discuss the evidence in favor of an interactions.

\begin{figure}[!htp]
\centering
\includegraphics[width=\textwidth]{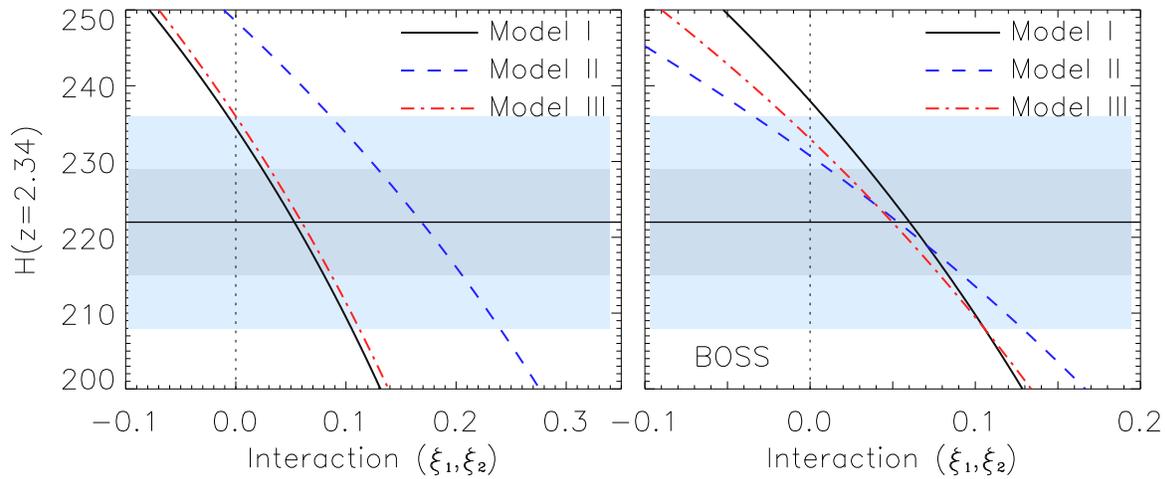}
\vspace*{-5cm}
\caption{Hubble function at redshift $z=2.34$ as a function of the
interaction parameters $\xi_1,\,\xi_2$ for Models III and I, II, respectively.
In the left panel we used the cosmological parameters from Table V, XI and XII
of \cite{Costa_14f}. In the  right panel, labeled BOSS,
the cosmological parameters are those of Table~\ref{table:BOSS_parameters}.
The horizontal line corresponds to the BOSS measured value $H(2.34)=222
\pm 7 ~\mathrm{km~s^{-1}~Mpc^{-1}}$
with the shaded areas representing $1\sigma$ and $2\sigma$ CL.}
\label{fig:all_models}
\end{figure}

\subsection{High Redshift Evidence for Interacting DE.}\label{sec:sec7.evidence}

Together with CMB data, low redshift observations like luminosity distances
from SNIa have been used to test the DM/DE interaction.
At low redshifts the deviations from $\Lambda$CDM are not very
pronounced and it is easier to establish the existence of a dynamical DE
component or even an interacting one using high redshift observations.
Recently, the analysis of BOSS data presented
evidence against the concordance model by measuring the BAO in the
redshift range $2.1 \leq z \leq 3.5$ from the correlation
function of the  Ly$\alpha$ forest from high redshift quasars
\cite{Delubac:2014aqe}.
Their result indicates a $1.8\sigma$ deviation (from \textit{Planck}+WMAP)
and $1.6\sigma$ deviation (from ACT/SPT) from $\Lambda$CDM at $z=2.34$. While the
statistical evidence is still not significant, if confirmed, this
result cannot be explained by a dynamical DE component, and it suggests a
more exotic type of DE.
An interacting DE appears as a simple and efficient solution to explain the
BOSS result. If DE and DM interact and the former transfers energy to the
latter, as required to alleviate the coincidence problem (see sec~\ref{sec:sec2.direction})
and indicated by the data \cite{Costa_14f}, it would explain the value
of the Hubble parameter, $H(z=2.34) = 222 \pm 7 \mathrm{km s}^{-1}
\mathrm{Mpc}^{-1}$, obtained by the BOSS collaboration,  value that is smaller
than the expectation from $\Lambda$CDM \cite{AbdallaE_2015a}. It would
explain the discrepancy of the angular diameter distance at high redshifts.
Let us now briefly summarize which of the models given in Table~\ref{table:models}
can explain the BOSS results.

Let us consider a universe filled only with DM, DE and baryons. We can
use the Hubble parameter obtained from the Friedmann equation and compare
it with the value obtained by the BOSS collaboration for different sets
of adjusted cosmological parameters. We can also compare the constraints
for $H(z=2.34)$ and $D_A(z=2.34)$ given by the BOSS experiment with
constraints from CMB adjusted data using $\Lambda$CDM and the interaction model.
To carry out this analysis we need to establish first the evolution with redshift of the
energy densities of each component, specially DE and DM since due to the interaction
they are not independently conserved. For the Models I and II, they behave as \cite{He_08f}
\begin{eqnarray}
\rho_d&=(1+z)^{3(-1+\omega_d+\xi_2)}\rho_d^0 , \nonumber \\
\rho_c&=(1+z)^3\left\{\frac{\xi_2\left[1-(1+z)^{3(\xi_2+\omega_d)}\right]
\rho_d^0}{\xi_2+\omega_d}+\rho_c^0\right\} . \label{int_DE}
\end{eqnarray}
The baryonic density is given by the usual expression $\rho_b=\rho_b^0(1+z)^3$.
For the Model III, the evolution is
\begin{eqnarray}
\rho_d&=(1+z)^{3(1+\omega_d)}\left(\rho_d^0+\frac{\xi_1\rho_c^0}{\xi_1+\omega_d}\right)
-\frac{\xi_1}{\xi_1+\omega_d}(1+z)^{3(1-\xi_1)}\rho_c^0 , \nonumber\\
\rho_c&= \rho_c^0(1+z)^{3(1- \xi_1)} .\label{int_DM}
\end{eqnarray}
From these solutions, it is easy to establish that when the energy is
transferred from the DE to the DM, the energy density of the DM is
always smaller than what it would have been in the standard $\Lambda$CDM model.
Since $\rho_c$ is the dominant component at redshifts $z\ge 1$ and it is
smaller than in the concordance model, so it would be the Hubble parameter,
as indicated by the BOSS data.

\begin{table}[htb]
\centering \caption{Cosmological parameters
used by the BOSS collaboration
\cite{Delubac:2014aqe}.}
\begin{tabular}{ccc}
\hline
Parameter & Bestfit & $\sigma$ \\
\hline
$h$ & $0.706$ & $0.032$ \\
$\Omega_{c}^{0} h^{2}$ & $0.143$ & $0.003$ \\
$\Omega_{d}^{0}$ & $0.714$ & $0.020$ \\
$\Omega_{b}^{0} h^2$ & $0.02207$ & $0.00033$ \\
\lasthline
\end{tabular}
\label{table:BOSS_parameters}
\end{table}

To make the previous statement more quantitative we took two sets of values
for the cosmological parameters $\Omega_d^0,\Omega_c^0,\Omega_b^0$ and $H_0$:
(1) the values used by BOSS collaboration, obtained from the Planck Collaboration
analysis of the $\Lambda$CDM model and listed in Table~\ref{table:BOSS_parameters}, and
(2) the values derived by \cite{Costa_14f}
by fitting DM/DE interacting fluid models to the \textit{Planck}, BAO, SnIa and
$H_0$ indicated above.  Using both data sets, the Hubble parameter
at $z=2.34$ has been computed using eqs.~(\ref{int_DM},\ref{int_DE})
for the cosmological models listed in Table~\ref{table:models}.
The results are shown in Fig.~(\ref{fig:all_models}). The right panel corresponds
to the cosmological parameters of the BOSS collaboration and the left panel to
those of \cite{Costa_14f}. The figure shows the measured value $H(z=2.34)$
and its $1\sigma$ and $2\sigma$ contours. In both panels, the $\Lambda$CDM model
that corresponds to the case of no interaction is always outside the
$1\sigma$CL. While still not significant, it does show that the data
prefers an DM/DE interacting model with positive interaction.
Further improvements on the data could help to establish the existence of
an interaction.

\begin{figure}
\centering
\includegraphics[width=5in,height=4.5in]{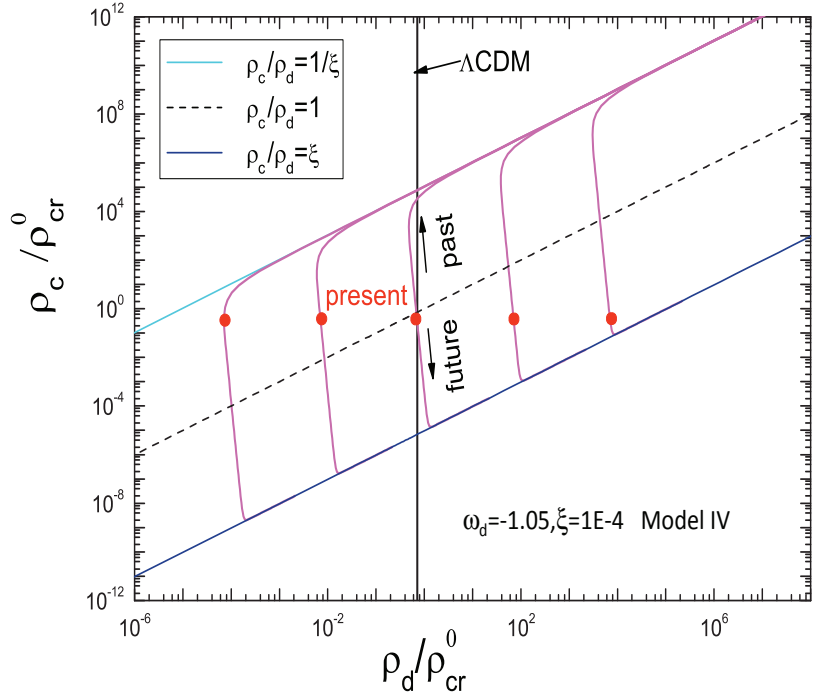}
\caption{$\rho_c^0$ is the critical energy density today. The
attractor solutions  of $r$ does not depend on the initial conditions at the early
time of the universe. The purple lines represent the density evolution of
cosmological model with different initial conditions. Solid circles represent
the density contrast $r$ today. Values change with the initial  conditions
but they are bounded in two  attractor solutions
$r_s^-\sim\xi,\,r_s^+\sim1/\xi$ in $\rho_c-\rho_d$ plane.}
\label{figure5_sec7}
\end{figure}

\subsection{The Coincidence Problem.}

One motivation to study interacting DM/DE models is to alleviate
the coincidence problem. Only at the present time the DM/DE ratio in the   $\Lambda{\rm CDM}$ model
is of order unity, demanding a fine-tuning
on the initial conditions at the Planck scale of 90 orders of magnitude
(see Sec.~\ref{sec:sec1.coincidence}). Let us examine if once interacting DE models
are confronted with observations, the goal can be satisfied. We
will concentrate on the analysis of the phenomenological fluid model.
The results shown in Fig.~\ref{figure2_sec8} suggest that positive
coupling parameters are compatible with the data and positive values
work in the direction of solving the cosmological coincidence problem
\cite{He_09bf,He_08f} and similar conclusions have also been reached for
the field description of the DE \cite{Costa_2014b}.
Let us particularize our analysis for fluid models, and in
particular for Model IV.
In Sec.~\ref{sec:sec2.solution} we have demonstrated how this model
solves the coincidence problem. As illustrated in Fig.~\ref{fig:ratios}(a)
the ratio has two attractor solutions $r_s^{\pm}$; the past solution
and future ratios are given in eq.~(\ref{r+-}). When the coupling
is $0<\xi\ll 1$, the ratios behave asymptotically as
\begin{equation}
r_s^+\sim\frac{1}{\xi} \quad {\rm and}\quad  r_s^-\sim\xi ,
\end{equation}
i.e., the behavior of the attractor solutions of the ratio $r$ only
depends on the coupling constant $\xi$ and does not depend on the
initial conditions at the Planck scale. This solution of the coincidence
problem is illustrated in Fig.~\ref{figure5_sec7}.
Purple solid lines represent the evolution of the energy densities
in units of the critical energy density today, $\rho_{cr}^0$,
with different initial conditions. The density
contrast $r$ at present is different for different initial conditions but
all the curves are bounded by the two attractors solutions $r_s^+$ and $r_s^-$.
In this particular case we fixed $\xi=10^{-4}$ and $\omega_d=-1.05$.
During the whole thermal history of the Universe, the
DM to DE ratio takes values within the range $10^{-4}<r<10^{4}$;
it changes much less than that in the $\Lambda{\rm CDM}$ model, thus
the cosmological coincidence problem is greatly alleviated.

\section{Current and future observational prospects.}\label{sec:sec8}

The discovery that the expansion of the Universe is accelerating
has led to large observational programs being carried out to
understand its origin. New facilities are being designed and
built aiming to measure the expansion history and the growth of structure in
the Universe with increasing precision out to greater redshifts.
Since the interaction in the
dark sectors changes the expansion history of the Universe and the evolution
of matter and radiation density perturbations, peculiar velocities and
gravitational fields, these new facilities will not only test
the current period of accelerated expansion but also explain
the nature on the interaction between dark sectors.
Thus, current and future observations could and will be used
to set up constraints on interactions between dark sectors and
clarify the nature of DM/DE interactions.
Observations of type Ia supernova, Baryon Acoustic
Oscillations (BAO), gravitational lensing, redshift-space distortions and the
growth of cosmic structure probe the evolution of the Universe at $z\le 2-3$.
In parallel, the physics of DM/DE interactions at recombination can be
probed by the CMB radiation power spectrum while the ISW effect and
lensing pattern of the CMB sky are sensitive to the growth of matter
at lower redshifts.

The DE Task Force (DETF) was established to advise the different
U.S. funding agencies on future DE research.
Their report categorized different experimental approaches by introducing
a quantitative ``figure of merit'' that is sensitive to the
properties of DE, including its evolution with time \cite{albrecht_06f}.
Using this figure of merit, they evaluated ongoing and future DE studies
based on observations of Baryon Acoustic Oscillations, Galaxy Clusters, Supernova
and Weak Lensing. The DETF categorized the different experiments by their different
degree of development. Stage I referred to the discovery experiments, Stage II to
the on-going experiments at that time when the report was elaborated
(circa 2006), Stage III
was defined as the next generation that are currently in full operation. They
also looked forward to a Stage IV generation of more capable experiments.
Examples of Stage II surveys are the Canada-France-Hawaii Telescope (CFHT)
Legacy Survey, with observations of SNIa \cite{conley_11f} and weak lensing \cite{heymans_12f}
and that ended in 2009, the ESSENCE \cite{wood-vasey_07f} and SDSS-II \cite{frieman_08f}
supernova surveys and BAO measurements from the SDSS
\cite{Eisenstein_2005a,Percival_2010a,padmanabhan_12f}. While new observations
continue to be expanded and improved with more recent instruments,
the CHFT Lensing survey remains the largest weak lensing survey to date.

In this section we will briefly review projects that are currently operating
or under construction (Stage III and IV).
All of these facilities share the common feature of surveying wide areas
to collect large samples of galaxies, clusters, and/or supernovae and
they will help clarify the nature of the interaction between dark sectors.
More details can be found in \cite{weinberg_13fb}.

\subsection{Ground Based Observations.}

The existing and planned ground based DE experiments collect data on
SNIa, galaxy clustering and gravitational lensing.  Wide-field imaging is used  to measure weak
gravitational lensing and clustering of galaxies in bins of photometrically
estimated redshifts and wide-field spectroscopy, to map the clustering of galaxies,
quasars and the Ly-$\alpha$ forest and measure distances and expansion rates
with BAO and the history of structure growth with redshift-space distortions (RSD).
Type Ia supernovae are searched to determine the distance-redshift relation.

\subsubsection{Stage III: 6dFGS, BOSS, HETDEX, Pan-STARRS, WiggleZ.}

The 6-degree Field Galaxy Survey (6dFGS) has mapped the nearby universe over
$\sim 17,000$ deg$^2$ of
the southern sky with galactic latitude $|b|>10^0$. The median redshift of
the survey is $z=0.053$. It is the largest redshift survey of the nearby
Universe, reaching out to $z\sim 0.15$. The survey data includes images,
spectra, photometry, redshifts and a peculiar velocity survey of a
subsample of 15,000 galaxies. The final release of redshift data is given
in \cite{jones_04f}.

The Baryon Oscillation Spectroscopic Survey (BOSS) is currently the largest
spectroscopic redshift survey worldwide, mapping $10^4$ deg$^2$ up to $z=0.7$.
BOSS is the largest of the four surveys that comprise SDSS-III
and has been in operation for 5 years since 2009. Its goals are to measure angular
diameter distance and expansion rate using BAO, using 1.5 million galaxies
\cite{anderson_14f}. Using Ly-$\alpha$ lines towards a dense grid
of high-redshift quasars, it has pioneered a method to measure BAO at redshifts
$z=[2,3.5]$. The analysis of SDSS Data Release 9 has provided a measurement of the
BAO scale at $z\sim 2.5$ with a precision of 2-3\% \cite{busca_13f,slosar_13f}.
This survey will be followed by the extended BOSS (eBOSS) that will be
operating for six years and will extend the BOSS survey to higher redshifts.

Similar to BOSS, the Hobby-Eberly Telescope DE Experiment (HETDEX)
at the Austin McDonald observatory has the goal of
providing percent-level constraints on the Hubble parameter and
angular diameter distance on the redshift range $z=[1.9,3.5]$
by using a combination of BAO and power spectrum shape information.
It will be achieved by surveying 0.8 million $Ly-\alpha$ emitting galaxies
on a field of view of 420deg$^2$ \cite{hetdex_08f}.

The Javalambre Physics of the accelerating universe Astronomical Survey (J-PAS)
is a new astronomical facility dedicated to mapping the observable
Universe in 56 colors \cite{Benitez:2014ibt}.
The starting date for this multi-purpose astrophysical
survey is 2015.  In five years, J-PAS will cover $\sim 8000$deg$^2$
using a system of 54 narrow band and 2 broad-band
filters in the range $300-1000$nm. The filter system was optimized to accurately
measure photometric redshifts for galaxies up to $z\sim 1$.
The main instruments are a 2.5m telescope
located at El Pico del Buitre (Teruel, Spain) and
a 1.2 Giga-pixel camera. The main goals of the survey are to
measure angular and radial components of BAO from the galaxy clustering,
determine the evolution of the cosmic volume from cluster counts
and luminosity distances from SNIa. The filter system will permit to determine
the redshifts of the observed supernovae.  The camera is not optimized
to measure galaxy elipticities so weak lensing studies would require
ellipticity measurements obtained from other surveys.
The JPAS telescope will measure BAO from high redshift
quasars to achieve a better precision than BOSS \cite{Benitez:2014ibt},
open the possibility of using the test described in Sec.~\ref{sec:sec7.evidence}
to disproof the concordance model.

The Panoramic Survey Telescope and Rapid Response System (Pan-STARRS)
describes a facility with a cosmological survey among its major goals.
The final goal is to use four coordinated telescopes to carry out survey of
the full sky above DEC=$-45^0$ \cite{morgan_14f} that will go a factor $\sim 10$
deeper than the SDSS imaging survey. The survey will provide data on high
redshift SN, galaxy clustering and gravitational lensing. For that purpose,
in addition to the wide survey, an ultra-deep field of $1200$deg$^2$ will
be observed down to magnitude 27 in the $g$ band with photometric
redshifts to measure the growth galaxy clustering. Data from this facility
has already been used to constrain the equation of state parameter \cite{zheng_14f}.

The WiggleZ DE Survey is a large-scale redshift survey carried out at
the Anglo-Australian Telescope and is now complete. It has measured redshifts
for $\sim 240,000$ galaxies over 1000deg$^2$ in the sky. It combines measurements
of cosmic distance using BAO with measurements of the growth of structure from
redshift-space distortions out to redshift $z=1$ \cite{kazin_14f}.

\subsubsection{CMB Experiments: ACT, SPT.}

The Atacama Cosmology Telescope (ACT) operates at 148, 218 and 277 GHz
with Full-Width at Half Maximum
angular resolutions of $(1.4^\prime,1.0^\prime,0.9^\prime)$ \cite{swetz_10f}.
ACT observes the sky by scanning the telescope in azimuth at a
constant elevation of $50.5^0$ as the sky moves across the
field of view in time, resulting in a stripe-shaped observation area.
The collaboration has released two observed areas of $850$deg$^2$
and $280$deg$^2$ \cite{dunner_13f}.
Sky maps, analysis software, data products and model templates
are available  through NASA Legacy Archive for Microwave Background Data
Analysis (LAMBDA).

The South Pole Telescope (SPT) is a 10m telescope designed
to map primary and secondary anisotropies in the CMB, currently
operating at 95, 150, 220 GHz with a resolution
with resolution (1.7, 1.2, 1.0)$^\prime$. The
noise levels are 18$\mu$K at 150GHz and $\sim\sqrt{2}$ larger for the
other two channels \cite{george_15f}. It has observed a region
of $2,540$deg$^2$. Data in the three frequencies were used to produce
a radiation power spectra covering the multipole range $2000<\ell<11,000$.
At present is the most precise measurement of the radiation
power spectrum at $\ell>2,500$ at those frequencies; at those angular
scales the signal is dominated by  the SZ effect and is not so relevant
to constrain models of DM/DE interaction.

A polarization-sensitive receiver have been installed on the SPT; data at
95 and 150 GHz has provided a measurement of the $B$-mode polarization
power spectrum from an area of $100$deg$^2$, spanning the range
$300\le\ell\le 2300$. The resulting power spectra was consistent with
predictions for the spectrum arising from the gravitational lensing
of $E$-mode polarization \cite{keisler_15f}.

\subsubsection{Stage IV: DES, eBOSS, JPAS, LSST, SKA, WFMOS, BINGO.}

The DE Survey (DES) is a wide-field imaging and supernova survey
on the Blanco 4m telecope at Cerro Tololo (Chile)
using the DE Camera. It has started operations
and it will continue for five years.
The DE Spectroscopic Survey Instrument (DESI) is a wide field
spectroscopic instrument intended to start in 2018 and operate also for five years
in the nearly twin Mayall telescope at Kitt Peak (Arizona).
DESI will obtain spectra and redshifts for at least 18 million emission-line
galaxies, 4 million luminous red galaxies and 3 million quasi-stellar objects,
to probe the effects of DE on the expansion history
BAO and measure the gravitational growth history through RSD. The resulting
3-D galaxy maps at redshift $z<2$ and Ly-$\alpha$ forest at $z>2$ are expected
to provide the distance scale in 35 redshift bins with a one-percent precision
\cite{levi_13f}. The imaging survey will detect 300 million galaxies, with
approximately 200 million WL shape measurements, almost a two-order
of magnitude improvement over the CFHTLens Weak Lensing survey.

Approved as a major cosmology survey in SDSS-IV (2014-2020),
eBOSS will capitalize on this premier facility with spectroscopy on
a massive sample of galaxies and quasars in the relatively uncharted
redshift range that lies between the BOSS galaxy sample and the BOSS
Ly-$\alpha$ sample. Compared with BOSS, this new survey
will focus on a smaller patch of 7500 deg$^2$ but it will reach higher
magnitudes.  It will measure both the distance-redshift relation
and the evolution of the Hubble parameter using different density
tracers; the clustering of Luminous Red Galaxies (LGRs) and Emission Line Galaxies (ELGs),
quasars and Ly-$\alpha$ systems to probe the BAO scale in the redshift
ranges [0.6,0.8], [1,2.2] and [2.2,3.5] respectively and
it will achieve 1-2\% accuracy in distance
measurements from BAOs between $0.6<z<2.5$.

The Javalambre Physics of the accelerating universe Astronomical Survey (JPAS)
is a new astronomical facility dedicated to mapping the observable
Universe in 56 colors \cite{Benitez:2014ibt}.
The starting date for this multi-purpose astrophysical
survey is 2015.  In five years, JPAS will cover $\sim 8000$deg$^2$
using a system of 54 narrow band and 2 broad-band
filters in the range $300-1000$nm. The filter system was optimized to accurately
measure photometric redshifts for galaxies up to $z\sim 1$.
The main instruments are a 2.5m telescope
located at El Pico del Buitre (Teruel, Spain) and
a 1.2 Giga-pixel camera. The main goals of the survey are to
measure angular and radial components of BAO from the galaxy clustering,
determine the evolution of the cosmic volume from cluster counts
and luminosity distances from SNIa. The filter system will permit to determine
the redshifts of the observed supernovae.  The camera is not optimized
to measure galaxy elipticities so weak lensing studies would require
ellipticity measurements obtained from other surveys.
The JPAS telescope will measure BAO from high redshift
quasars to achieve a better precision than BOSS \cite{Benitez:2014ibt},
open the possibility of using the test described in Sec.~\ref{sec:sec7.evidence}
to disproof the concordance model.

The Large Synoptic Survey Telescope (LSST) is a wide-field, ground-based
telescope, designed to image $\sim 20,000$ deg$^2$ in six optical
bands from 320nm to 1050nm. The telescope will be located on Cerro Pach\'on (Chile)
and it will operate for a decade allowing
to detect galaxies to redshifts well beyond unity. Its science goals are
to measure weak and strong gravitational lensing, BAO, SNIa and
the spatial density, distribution, and masses of galaxy clusters
as a function of redshift. Its first light is expected on 2019.

The Square Kilometre Array (SKA) is a radio-facility which
is scheduled to begin construction in 2018. The  HI galaxy redshift survey
can provide us with accurate redshifts (using the 21cm line) of millions
of sources over a wide range of redshifts, making it an ideal redshift
survey for cosmological studies \cite{AbdallaE_2015a, bacon_14f,bull_14f,kitching_15f,
raccanelli_14f,santos_14f}. Although technically challenging, the SKA
could measure the expansion rate of the Universe
in real time by observing the neutral hydrogen (HI) signal of galaxies at
two different epochs \cite{darling_12f,klockner_15f}.

Wide-Field Multi-Object Spectrograph (WFMOS) is a camera specially devoted
to Galaxy Surveys. It will be mounted atop the 8.2m
Subaru Telescope on Mauna Kea (Hawaii). One of
the science goals of the WFMOS camera is high precision measurements
of BAO. The WFMOS DE survey comprises two parts: a 2,000 deg$^2$
survey of two million galaxies at redshifts $z<1.3$ and a
high redshift survey of about half a million
Lyman Break Galaxies (LBGs) at redshifts $2.5<z<3.5$ that
would probe distances and the Hubble rate beyond $z=2$
(see \cite{bassett_05f} for more details).

BINGO is a radio telescope designed to detect BAO at radio frequencies
by measuring the distribution of neutral hydrogen at cosmological distances
using a technique called Intensity Mapping. The telescope will be located
in a disused, open cast, gold mine in Northern Uruguay. It will operate in
the range [0.96,1.26]~GHz to observe the redshifted 21 cm Hydrogen line. It
will consist of a two-mirror compact range design with a 40m diameter
primary and it will have no moving parts to provide an excellent polarization
performance and very low side-lobe levels required for intensity mapping.

\subsubsection{Stage IV: CMB experiments.}

Currently, the interest on CMB ground experiments is centered on polarization.
For a cosmic variance limited experiment polarization alone places
stronger constraints on cosmological parameters than CMB temperature \cite{Galli_14f}.
Experiments like SPTpol \cite{keisler_15f} and Quixote \cite{genova-santos_15f}
are currently taken data aiming to characterize the polarization of the CMB
and of the Galactic and extragalactic sources. CMB experiments devoted
to measuring polarization from the ground are also being proposed;
the scientific capabilities of a CMB polarization experiment like
CMB-S4 have been considered that in combination with low redshift data
would be able to constrain, among other parameters, the DE equation
of state and dark matter annihilation \cite{wu_14f,abazajian_14f}.

\subsection{Space based surveys.}
Satellite surveys usually require a dedicated facility and, consequently, are
more expensive than those carried out from the ground. Their significant
advantage is that, by observing outside the atmosphere, the data usually
contains a lower level of systematic errors.

\subsubsection{Stage III: WMAP, Planck.}

The Wilkinson Microwave Anisotropy Probe (WMAP) was a satellite mission
devoted to measure CMB temperature fluctuations at frequencies
operating between 23 and 94GHz. Launched on June 30, 2001
and operated for 9 years up to the end of September 2010. The main
results and data products of the nine years of operation are described
in \cite{bennett_13f}. The final 9yr data released was soon followed
by those of the Planck Collaboration.
The Planck satellite observed the microwave and sub-millimeter sky
from August 12th, 2009 to Oct 23rd, 2013  in nine frequencies between 30 and 857 GHz,
with angular resolution between 33' and 5'.  Its goal was to produce CMB maps
both in temperature and polarization. The Planck Collaboration has released
data on  CMB temperature anisotropies, Thermal Sunyaev-Zledovich (TSZ) effect. The
measured  temperature and polarization, a catalog of Sunyaev-Zeldovich (SZ)
clusters and likelihood codes to assess cosmological models against the Planck data
\cite{Planck_I_14f,Planck_I_15f} and other data products  can be downloaded from the
Planck Legacy Archive \cite{planck}. The Temperature-Temperature, Temperature-E mode
and E mode-E mode power spectra are measured up to $\ell\sim 2000$
\cite{Planck_XV_14f}, allowing the CMB lensing potential \cite{Planck_XVII_14f}
and the constraint on cosmological models beyond the $\Lambda$CDM model \cite{Planck_XIV_15f}.

\subsubsection{Stage IV: eRosita, Euclid and WFIRST.} \label{sec:sec9}

eROSITA will be a X-ray satellite that will be launched in 2016.
It will perform the first imaging all-sky survey in energy range
0.3-10 KeV \cite{merloni_12f}. The goal of eROSITA is the detection
of $\sim 10^5$ galaxy clusters out to redshifts $z>1$, in order
to study the large scale structure in the Universe and test and
characterize cosmological models including DE.
In the soft X-ray band (0.5-2 keV), it will be about 20 times more
sensitive than the ROSAT all sky survey, while in the hard
band (2-10 keV) it will provide the first ever true imaging survey
of the sky at those energies.

Euclid is a European Space Agency DE satellite mission scheduled
for launch in 2020. This mission
is designed to perform two surveys: a wide 15,000 deg$^2$ survey in the optical
and near-infrared and a deep survey on 40 deg$^2$ two magnitudes deeper.
These facilities are not independent between each other.
Euclid will map the extra-galactic sky with the resolution of the
Hubble Space Telescope, with optical and Near-Infrared (NIR) imaging and NIR spectroscopy.
Photometric redshifts for the galaxies in the wide survey will be provided
from ground photometry and from the NIR survey. In addition, 50 million spectroscopic
redshifts will be obtained. Euclid data will allow to measure the expansion history
and the growth of structure with great precision. A detailed quantitative forecast of Euclid
performance has been discussed in \cite{laureijs_11f}.
The data will allow to constrain many different cosmological models;
when the growth factor is parametrized as $f_g=\Omega_m(z)^\gamma$ the value
$\gamma\simeq 0.545$ corresponds to the $\Lambda$CDM model and Euclid
will measure this parameter with a precision of $\Delta\gamma=0.03$ \cite{heavens_07f}.
Forecasts for other parametrizations of the growth factor and for other magnitudes
such as the bias, DE sound speed, redshift space distortions are given in
\cite{amendola_13f}.

The Wide Field Infrared Survey Telescope (WFIRST) is an american satellite mission
that is currently being reviewed and expected to be launch in 2023.
This mission updates and expands earlier proposed
missions like the Super Nova Acceleration Probe (SNAP) and the Joint DE
Mission (JDEM). Like Euclid, one of its primary science goals is to determine the
properties of DE and in many respects complements EUCLID. WFIRST strategy
is to construct a narrow and deep galaxy redshift survey of 2000 deg$^2$.
Both satellites will measure the redshift for a similar number of galaxies
and will obtain a comparable precision for the Baryon Acoustic Oscillations
derived angular diameter distances and Hubble constant redshift evolution
\cite{spergel_13f}. Nevertheless, due to their different observing strategy
will allow cross-checks that will help to identify and eliminate systematics.
The combination of both data sets will significantly improve the constraints
on the dark energy parameters.

Many synergies will come from cross-correlating data from different observations
For instance, Euclid, WFIRST and SKA have similar scientific aims but will carry
observations at different wavelengths. Euclid and WFIRST probe the low redshift Universe,
through weak lensing and galaxy clustering measurements. The SKA has the
potential to probe a higher redshift regime and a different range in scales
of the matter power spectrum, which are linear scales rather than the quasi-non-linear
scales to which Euclid and WFIRST will be sensitive. The  combination of different observations will particularly sensitive to signatures of modified gravity.
Cross-correlation of different data sets will
help to control systematics for the primary science.
The SKA, WFIRST and Euclid will be commissioned on similar timescales
offering an exciting opportunity to exploit synergies between these facilities.
\cite{kitching_15f}

\subsubsection{CMB experiments: CMBpol, COrE, PRISM.}

The Cosmic Origins Explorer (COrE) is a Stage IV full-sky, microwave-band
satellite proposed to ESA within Cosmic Vision 2015-2025. COrE will provide
maps of the microwave sky in polarization and temperature in 15 frequency bands,
ranging from 45 GHz to 795 GHz, with angular resolutions from $23^\prime$
at 45 GHz and $1.3^\prime$  at 795 GHz, with sensitivities roughly
10 to 30 times better than Planck \cite{armitage-caplan_11f}.

The Polarized Radiation Imaging and Spectroscopy Mission (PRISM)
is a large-class mission proposed to ESA in May 2013 within the framework of
the ESA Cosmic Vision program. Its main goal is to survey the CMB sky
both intensity and polarization in order to precisely measure the
absolute sky brightness and polarization. The mission will detect
approximately $10^6$ clusters using the thermal SZ effect and a peculiar velocity
survey using the kinetic SZ effect that comprises our entire Hubble volume \cite{andre_14f}.
NASA is carrying similar efforts through
the Primordial Polarization Program Definition Team (PPPDT)
that converge towards a satellite dedicated to the study of
CMB polarization (CMBPol) \cite{baumann_09f}.

Combing these complementary ground based and space based observations,
we would hopefully achieve a better understanding of the nature
of DM, DE and the interaction within the dark sectors.

\subsubsection*{Acknowledgements}
We thank S. Tsujikawa for comments and suggestions.
E. A. wishes to thank FAPESP and CNPq (Brazil) for support and
A. A. Costa, E. Ferreira and R. Landim for discussions and suggestions.
F. A. B. acknowledges financial support from the Ministerio de Ciencia
e Innovaci\'on, grant FIS2012-30926 and
the ``Programa de Profesores Visitantes Severo Ochoa'' of the Instituto
de Astrof{\'\i}sica de Canarias. B. W. would like to acknowledge the support by National Basic Research Program of China (973 Program 2013CB834900)
and National Natural Science Foundation of China and he wishes to thank J. H. He and X. D. Xu for helpful discussions.

\end{document}